\documentclass{aastex62}
\usepackage{natbib}
\usepackage{color}
\usepackage{amsmath}

\received{--}
\revised{--}
\accepted{--}

\shortauthors{Dong et al.}

\begin{document}

\title{Propagation of very-high-energy $\gamma$-rays from distant blazars}

\correspondingauthor{Y.G. Zheng}\email{ynzyg@ynu.edu.cn}

\author{L.J. Dong}
\affiliation{Department of Physics, Yunnan Normal University, Kunming, Yunnan, 650092, People's Republic of China}
%
%

\author[0000-0003-0170-9065]{Y.G. Zheng}
\affiliation{Department of Physics, Yunnan Normal University, Kunming, Yunnan, 650092, People's Republic of China}

\author[0000-0002-9071-5469]{S.J. Kang}
\affiliation{School of Physics and Electrical Engineering, Liupanshui Normal University, Liupanshui, Guizhou, 553004, People's Republic of China}




\begin{abstract}
We re-derive the possible dependence of the redshift with very high energy (VHE) $\gamma$-ray photon index. The results suggest that the universe to VHE $\gamma$-rays is becoming more transparent than usually expected. We introduce the extragalactic background light (EBL) plus the photon to axion-like particle (ALP) oscillations to explain this phenomenon. We concentrate our analysis on 70 blazars up to redshift $z \simeq 1$. Assuming this correlation is solely the result of photon-photon absorption of VHE photons with the EBL, which finds the deviations between the predictions and observations, especially at redshifts $0.2 < z < 1$. We then discuss the implications of photon-ALP oscillations for the VHE $\gamma$-ray spectra of blazars.  A strong evidence shows that: 1) the EBL attenuation results that the VHE $\gamma$-ray photon index increases non-linearly at the ranges of redshift, $0.03 < z < 0.2$; 2) the photon-ALP oscillation results in a attractive characteristic in the VHE $\gamma$-ray photon index at the ranges of redshift, $0.2 < z < 1$. We suggest that both the EBL absorption and photon-ALP oscillation can influence on the propagation of VHE $\gamma$-rays from distant blazars.
\end{abstract}



\keywords{Blazars (164) --- BL Lacertae objects (158) --- Flat-spectrum radio quasars (2163) }

\section{Introduction}\label{sec:intro}
TeV $\rm \gamma$-ray sources play a crucial role in exploring the universe's most violent non-thermal phenomena and the cosmic background radiation. The observation of these $\rm \gamma$-rays depends on Imaging Atmospheric Cherenkov Telescopes (IACTs) like HESS, MAGIC, and VERITAS, which collect data in the interval 100 $GeV$ $< {E_\gamma } < 100$ $TeV$ \citep{2008ICRC....3.1341W}. These very-high-energy (VHE) $\gamma$-ray photons may be absorbed by the extragalactic background light or converted to axion-like particles \citep{2015Advantages,2020MNRAS.493.1553G}. These interactions affect the VHE $\gamma$-ray spectra of blazars, which result in fascinating astrophysical effects.

Extragalactic background light (EBL) is the accumulated and diffuse radiation from the formation and evolution of stellar objects and galaxies \citep{2001ARA&A..39..249H}. It is a basic observation of cosmology whose significant contributions to the EBL spectrum are the stellar emission (peaking at optical-UV) and the dust emission (peaking in the IR)\citep{2014ApJ...795...91S}. The VHE $\gamma$-rays from distant blazars are expected to be strongly attenuated across the universe by the possible interaction with the EBL, which affects the VHE $\gamma$-ray spectra of blazars in the $ \sim $  10 $GeV$ to 10 $TeV$ energy regime  \citep{1998A&A...334L..85S,2013APh....43..112D}. Direct measurement of EBL is difficult due to strong foreground contamination by the Galactic and zodiacal light. Therefore, indirect estimation of the cosmological model is used. Throughout the paper, we use the optical depth which is deduced by the EBL model in  \cite{2017A&A...603A..34F,2018A&A...614C...1F}.

It is assumed that there is a large-scale magnetic field in the nano-Gauss range in the universe. The existence of the cosmic magnetic field has been confirmed by the AUGER observatory \citep{2007Constraints}.  Photons can oscillate into a new, very light, spin-zero particle${-}$named Axion-Like Particle (ALPs) \citep{2011PhRvD..84j5030D}, which are characterized by coupling to two photons, the Lagrangian in Equation (1) gives a mathematical description of it \citep{2007PhRvL..99w1102H,2018PhRvD..98d3018G}:

\begin{equation}
{{\cal L}_{a\gamma \gamma }} =  - \frac{1}{4}{g_{a\gamma \gamma }}{F_{\mu \nu }}{{\tilde F}^{\mu \nu }}a = {g_{a\gamma \gamma }}E \cdot Ba.
\end{equation}										
Where ${g_{a\gamma \gamma }}$ is the two-photon coupling constant (which has the dimension of an inverse energy), ${F^{\mu \nu }}$  is the usual electromagnetic field strength, ${\tilde F_{\mu \nu }}$ is its dual, E and B are the electric and magnetic components of the electromagnetic field, and $a$ is the ALP field.
In the presence of an external magnetic field B, the same $\gamma$-$\gamma$-a vertex produces an off-diagonal element in the mass matrix of the photon-ALP system. As a result, the interaction eigenstates are different from the propagation eigenstates. This phenomenon causes photon-ALP oscillations \citep{2008LNP...741..115M,2009Evidence,2011PhRvD..84j5030D,2022MNRAS.516..216C}. It allows for a reduction of the absorption of EBL, thereby reducing the cosmic opacity.
ALP can be directly detected by the GammeV experiment at FERMILAB and the planned photon regeneration experiment ALPS at DESY, in addition to indirectly detecting astrophysical effects in high-energy laboratories. \citep{2008arXiv0812.3495E,2010PhRvD..82k5018A}.

Generally, we stress that those nearby blazars at redshift ${\rm{z}} < 0.03$ do not suffer EBL absorption. The observed data for $0.03 < z < 0.2$ display a highly significant positive correlation, mainly attributed to the absorption of EBL, because there is no such correlation in other bands \citep{2014ApJ...795...91S,2022MNRAS.511..994M}. VHE $\gamma$-ray photons at higher redshifts are expected to have obvious EBL absorption \citep{2004A&A...413..807K,2009Intrinsic}. However, we found $\Gamma _{obs}$ is independent of $0.2 < z < 1$. A possible explanation relies upon the previously mentioned possible existence of ALPs \citep{2008PhRvD..77f3001S,2009Evidence}. Lorentz Invariance Violation (LIV) may be the reason for the lack of absorption \citep{2000PhLB..493....1P} and the emission of cosmic rays from blazars whose energy is smaller than 50 $EeV$. These cosmic rays can travel large distances across the universe, where they can interact with the EBL before reaching our galaxy, secondary photons are generated in these interactions \citep{2012ApJ...751L..11E,2013ApJ...764..113Z}. Based on their analysis, the reason for the change of the VHE $\gamma $-rays photon index of the $TeV$ blazar at the $z = 0.2$ could be clearly understood.

We find that EBL and ALPs deeply influence VHE $\gamma$-rays from distant blazars in interstellar space. We compile the redshift of 70 blazars from the TeVCat\footnote{http://tevcat.uchicago.edu}, we first use the mathematical model ${\Gamma _{obs}}(z) = \alpha z + \beta $ to analyze the correlation between ${\Gamma _{obs}}(z)$ and $z$. In addition, we compared this experimental VHE $\gamma$-ray photon index: $\Gamma _{obs}^{CP}$ is calculated theoretically in conventional physics and $\Gamma _{obs}$ is collected by TeVCat. A vital characteristic of these VHE $\gamma$-ray photon index are the presence of a break at $z = 0.2$, which can be successfully explained by considering the photon-ALP conversion during $\gamma$-ray propagation.

This paper is structured as follows. In Section 2, we describe the sample. In Section 3, we deduce a relationship between the VHE $\gamma$-ray photon index and the redshift of VHE blazars. In Section 4, we estimate the absorption of EBL. Section 5 applies photon-ALP oscillations to explain the independent phenomena between the VHE $\gamma $-rays photon index. We discuss the results in Section 6. In this paper, we assume the Hubble constant $H_0=70km{s^{ - 1}}Mp{c^{ - 1}}$, the dimensionless cosmological constant $\Omega_{\Lambda}=0.7$, matter energy density $\Omega_{M}=0.3$, and radiation energy density $\Omega_{R}=0$.

\section{Sample description} \label{sec:method}
This paper selects from the TeVCat catalog seventy blazars for analysis. If the physical information are not present in TeVCat, they can be obtained from other literatures. \cite{2009MNRAS.394L..21D} systematically explained for the first time the relationship between the VHE $\gamma $-rays photon index and the redshift for these blazars. They considered nineteen blazars with known redshift, VHE $\gamma$-ray flux and spectrum. Eighteen blazars were considered as BL Lac, and the other was classified as FSRQ. We updated the results of \cite{2009MNRAS.394L..21D}, including all blazars detected by IACTs in the VHE range.

The non-thermal radiation of the blazars has a broadband range extending from the radio band to the $\gamma$-ray band. It can emit $\gamma$-ray photons up to the $TeV$ energy range \citep{1995PASP..107..803U}. Blazars are usually classified as BL Lacertae objects (BL Lacs) \citep{2010ApJ...716...30A} and flat spectrum radio quasars (FSRQs) \citep{2011ApJ...743..171A}. BL Lacs are characterized by weak emission rays or no emission lines (equivalent width $<$ 5{\AA}), whereas FSRQs (equivalent width $>$ 5{\AA}) show broad, strong emission lines. According to the calculations of the peak frequency of the synchrotron SED component. BL Lac objects are divided into the following four categories: low-synchrotron-peaked (LBL, ${\nu _{{\rm{syn}}}} < {10^{14}} Hz$), intermediate-synchrotron-peaked (IBL, ${10^{14}} < {\nu _{{\rm{syn}}}} < {10^{15}} Hz$), high-synchrotron-peaked
(HBL, ${\nu _{{\rm{syn}}}} > {10^{15}} Hz$), or extreme high-frequency
peaked blazars (EHBL, ${\nu _{{\rm{syn}}}} > {10^{17}} Hz$) \citep{2019ICRC...36..676F}. The blazar samples analyzed in this paper consist of 70 VHE $\gamma$-ray photon spectral indices. From these sources, forty-six are cataloged in the TeVCat as HBL, seven as IBL and LBL, seven as EHBL, three simply as "Blazar" and seven as FSRQ. Most of these blazars are listed in Table 1, and the results are displayed in Figure 1.

\section{The correlation between redshift and VHE $\gamma$-ray photon index} \label{sec:method}
The absorption of VHE $\gamma$-ray photons interacts with EBL through the pair production process, the result that the VHE $\gamma$-ray spectra of blazars to steepen \citep{2013A&A...554A..75S}. It is expected that there is a correlation between the VHE $\gamma$-ray observed photon index and the blazar's redshift, if the blazar has similar intrinsic photon spectral index. To examine this correlation, we created a study between the VHE $\gamma$-ray photon index ($\Gamma _{obs}$) and the source redshift (${\rm{z}}$) for the blazars listed in Table 1.

\begin{equation}
\Gamma _{obs} = \alpha z + \beta.
\end{equation}

First, the relationship between the VHE $\gamma$-ray photon index and redshift was obtained by linear fitting. Secondly, we performed a correlation study between these quantities for the ensemble of blazars.

\begin{figure}
\centering
\includegraphics[height=8cm,width=11cm]{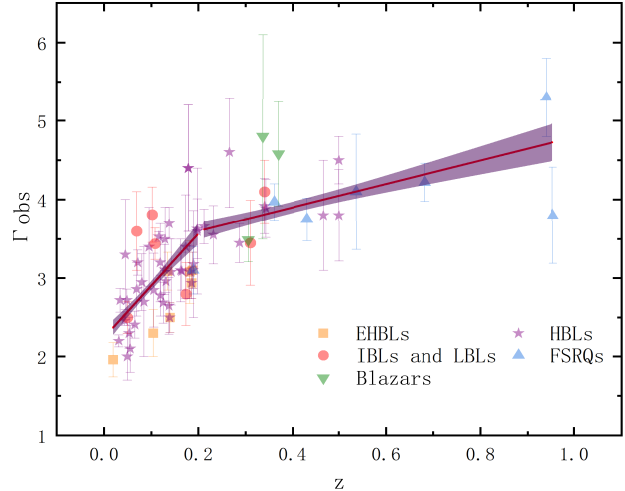}
\caption{Scatter plot between the VHE $\gamma $-rays photon index of the elected blazars with redshift. Different colors/symbols depict blazars, as mentioned in the labels. The solid red line represents the linear regression, and the purple area indicates the $1\sigma$ confidence bands.}
\label{sub:fig1}
\end{figure}

In Figure 1, we show the best fit line for $\Gamma _{obs}$ and ${\rm{z}}$, providing the confidence range of $1\sigma$. We found the best straight line between $\Gamma _{obs}$ and $0.2 < z < 1$ to be $\Gamma _{obs} = (1.49 \pm 0.39)z + (3.30 \pm 0.17)$. The Spearman rank correlation analysis for the blazars with $0.2 < z < 1$  resulted in a rank correlation coefficient, ${r_s} = 0.40$ with a null hypothesis probability ${P_{rs}} = 0.07$. In the same way, the Pearson's correlation analysis resulted in a linear correlation coefficient $\rho  =  0.41$  with a null hypothesis probability ${P_\rho } = 0.07$ and ${R^2} = 0.44$, and adjusted ${R^2} = 0.41$. For blazars with $0.03 < z < 0.2$ the dependence between $\Gamma _{obs}$ and ${\rm{z}}$ in the panel is $\Gamma _{obs} = (6.61 \pm 1.09)z + (2.25 \pm 0.11)$, ${r_s} = 0.49$ and ${P_{rs}} < 0.05$, $\rho  = 0.53$ and ${P_\rho } < 0.05$, while ${R^2} = 0.43$ and the adjusted ${R^2} = 0.42$.

 We use the second ROSAT all-sky survey (2RXS) source catalog containing 29 HBLs \citep{2016A&A...588A.103B}, six years of the Beppo-SAX catalog consisting of 31 HBLs \citep{2005A&A...433.1163D}, the seventy months of the Swift-BAT catalog consisting of twenty-five HBL \citep{2013ApJS..207...19B}, the low-energy $\gamma$-ray ($GeV$) spectral index with redshift for the ninety-nine HBLs listed in the fourth catalog of Fermi-LAT \citep{2020ApJS..247...33A} (Figure 2). The results of the Pearson and Spearman rank correlation study are: ${r_s} = 0.11$, ${P_{rs}} = 0.57$; $\rho  = 0.08$, ${P_\rho } = 0.69$ for 2RXS, ${r_s} = 0.14$, ${P_{rs}} = 0.45$; $\rho  = 0.09$, ${P_\rho } = 0.62$ for Beppo-SAX, ${r_s} = -0.07$, ${P_{rs}} = 0.75$; $\rho  = 0.07$, ${P_\rho } =0.73$ for Swift-BAT, ${r_s} = -0.04$, ${P_{rs}} = 0.66$; $\rho  = 0.01$, ${P_\rho } = 0.96$ for 4LAC-HSP. We found no evidence of any correlation of the X-ray spectral index and the low-energy $\gamma$-ray ($GeV$) spectral index with redshift. \cite{2014ApJ...795...91S} and \cite{2020MNRAS.493.1553G} confirm that luminosity is not correlated with the VHE $\gamma$-ray observed photon index. Therefore, the possibility of a selection effect or the Malmquist bias in the correlation between the VHE $\gamma$-ray photon index and redshift can be ruled out.

On the bias of above analysis, it is strong evidence evident that the steepening of the VHE $\gamma$-ray spectra of blazars results from EBL-induced absorption. Our statistical results show that a correlation between the VHE $\gamma$-ray photon index and the redshift ($0.03 < z < 0.2$), mainly because of EBL absorption. We verify this observation by studying the correlation between X-rays or low-energy spectral indices with the redshift for HBL. The details of EBL absorption are described in the next chapter.

\begin{figure}
\centering
\includegraphics[height=8cm,width=11cm]{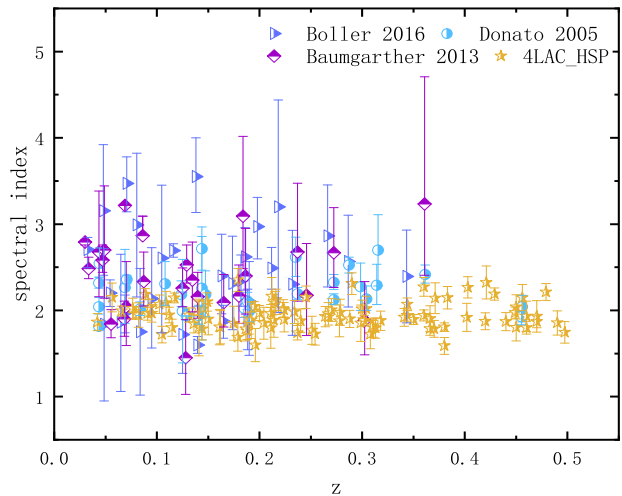}
\caption{Scatter plot of the observed X-ray and low-energy $\gamma$-ray ($GeV$) spectral index of HBL with redshift. The filled violet triangle is from the ROSAT catalog \citep{2016A&A...588A.103B}, the filled purple diamonds are from the Swift-BAT catalog \citep{2013ApJS..207...19B}, and the blue circles are from the Beppo-SAX catalog \citep{2005A&A...433.1163D}. The golden pentagons are from the fourth catalog of Fermi-LAT \citep{2020ApJS..247...33A}.}
\label{sub:fig2}
\end{figure}

\section{Extragalactic background light} \label{sec:method}
When the VHE $\gamma$-ray hard photons energy $E$ of blazars propagating scatter off soft background photons energy $\epsilon$ permeating the universe and produce ${{\rm{e}}^ + }{e^ - }$ pairs through the standard $\gamma_{VHE}+\gamma_{EBL}\longrightarrow e^{+}+e^{-}$ process \citep{2012Sci...338.1190A}. This process produces cosmic opacity. The minimum threshold energy ${E_{th}}$ at which the background photon interacts with the $\gamma$-ray photon is given by the expression:
\begin{equation}
{E_{th}}(E,\varphi ) \equiv \frac{{2m_e^2{c^4}}}{{E(1 - \cos \varphi )}} ,
\end{equation}
where $\varphi$ is the angle between the interacting photons and ${m_e}$ denotes the electron mass.

The nature of the propagation of photons in the universe is significant. The corresponding cross section is \citep{1934PhRv...46.1087B,1954qtr..book.....H}
\begin{equation}
{\sigma _{\gamma \gamma }}(E,\epsilon,\varphi ) \simeq 1.25 \times {10^{ - 25}}(1 - {\beta ^2}) \times [2\beta ({\beta ^2} - 2) + (3 - {\beta ^4})\ln (\frac{{1 + \beta }}{{1 - \beta }})]c{m^2}.
\end{equation}
Assuming head-on collisions, the corresponding cross-section ${\sigma _{\gamma \gamma }}({E},\epsilon,\varphi )$ reaches its maximum where the VHE $\gamma$-ray photon energy $E$ and the background photon energy $\epsilon$ are related by $\epsilon({E}) \simeq ({{0.5 TeV} \mathord{\left/
 {\vphantom {{0.5 TeV} {{E}}}} \right.
 \kern-\nulldelimiterspace} {{E}}})$ eV.
Therefore, according to the energy range explored by IACTs, the cosmic opacity is dominated by the interaction with diffuse background photons with 0.005 $eV$ $<{\epsilon}<$ 5 $eV$.
The cross-section peaks provide a relation between the EBL photon interacting with a VHE $\gamma$-ray photon, given by:
\begin{equation}
E(TeV) = \frac{{0.86{\lambda _{EBL}}(\mu m)}}{{(1 - \cos \varphi )}}.
\end{equation}
Therefore, VHE $\gamma$-rays at a rest frame energy above 100 $GeV$ are sensitive to the optical and infrared bands of the EBL-SED, especially at a few $TeV$. Below 100 $GeV$, it is mainly the EBL photon in the UV part that causes the attenuation \citep{2013IJMPD..2230025C,2020JAsGe...9..309S}.

We carefully studied the propagation of VHE $\gamma$-rays photon beam emitted of $TeV$ blazars at redshift $z$ and detected at energy ${E_\gamma }$ in the standard $\Lambda$CDM cosmological model. The relation between the observed VHE $\gamma$-ray spectra ${\Phi _{{\rm{obs}}}}({E_\gamma },z)$ and the emitted one ${\Phi _{em}}(E)$ can be expressed as
 \begin{equation}
{\Phi _{{\rm{obs}}}}({E_\gamma },z) = {{\rm{e}}^{ - {\tau _\gamma }({E_\gamma },z)}}{\Phi _{em}}({E_\gamma }(1 + z)) ,
 \end{equation}
${\Phi _{{\rm{obs}}}}({E_\gamma },z)$ is exponentially suppressed as the distance and energy increase, so entailing the sources that are far enough are often invisible.

The energy emitted due to the universe's expansion is ${E_\gamma }(1 + z)$. The VHE $\gamma$-ray photon index of the selected $TeV$ blazars are all close to a single power-law, and they have this form ${\Phi _{obs}}({E_\gamma },z) \propto K_{obs}{E_\gamma }^{ - \Gamma _{obs}^{CP}}$, where $K_{obs}$ denotes the normalization constant. The  intrinsic photon spectra can be well predicted by the Synchrotron-Self-Compton (SSC) mechanism and the Hadronic Pion
Production (HPP) in proton-proton scattering \citep{2004vhec.book.....A}. So the intrinsic photon spectra of blazars also has a single power-law behavior ${\Phi _{em}}({E_\gamma }(1 + z)) \propto K_{em}^{CP}{({E_\gamma }(1 + z))^{ - \Gamma _{em}^{CP}}}$, where $K_{em}^{CP}$ denotes the normalization constant and ${\Gamma _{em}^{CP}}$  is the intrinsic photon spectral index. The theoretical calculation of VHE $\gamma$-ray photon index $\Gamma _{obs}^{CP}$ of $TeV$ blazars yields
\begin{equation}
\Gamma _{obs}^{CP}(z) = \frac{{\Gamma _{em}^{CP}\ln [\frac{{{E_\gamma }(1 + z)}}{{{E_0}}}]}}{{\ln (\frac{{{E_\gamma }}}{{{E_0}}})}} + \frac{{{\tau _\gamma }({E_\gamma },z)}}{{\ln (\frac{{{E_\gamma }}}{{{E_0}}})}} + \frac{{\ln \frac{{{K_{obs}}}}{{K_{em}^{CP}}}}}{{\ln (\frac{{{E_\gamma }}}{{{E_0}}})}},
\end{equation}
where ${{E_0}}$ = 100 $GeV$ is a fiducial energy.

Within the framework of standard cosmic geometry, the optical depth ${\tau _\gamma }({E_\gamma },z)$ arises by a high-energy photon ${E_\gamma }$ travelling through a cosmic medium filled with low-energy photons with the spectral number density ${n_\gamma }({\epsilon _\gamma },z)$ of EBL photons from a source at z to an observer. The optical depth of the EBL finally obtains  \citep{2011MNRAS.410.2556D,2012MNRAS.422.3189G,2013MNRAS.432.3245D,2017A&A...603A..34F}:
  \begin{equation}
 \begin{aligned}
{\tau _\gamma }({E_\gamma },z){\rm{ = }}\int_0^{\rm{z}} {dz\frac{{dl(z)}}{{dz}}} \int_{ - 1}^1 {d(\cos \varphi )\frac{{(1 - \cos \varphi )}}{2}}  \times \int_{{E_{th}}}^\infty  {d\epsilon(z){n_\gamma }} (\epsilon(z),z){\sigma _{\gamma \gamma }}(E(z),\epsilon(z),\varphi ) ,
\end{aligned}
\end{equation}
the distance a photon travels per unit of redshift z is
\begin{equation}
\frac{{dl(z)}}{{dz}} = \frac{c}{{{H_0}}}\frac{1}{{{{(1 + z)[{\Omega _\Lambda } + {\Omega _{\rm M}}{{(1 + z)}^3}]}^{{1 \mathord{\left/
 {\vphantom {1 2}} \right.
 \kern-\nulldelimiterspace} 2}}}}}  .
\end{equation}

It will be challenging for us to study the influence of EBL on the spectral distribution of broadband SED of blazars. We restrict our attention to the energy range 0.1 $TeV$ $< {E_\gamma } <$ 10 $TeV$ energy range covered by IACT, the corresponding diffuse EBL energy range is 0.05 $eV$ $< {\epsilon _0} < $5 $eV$. The EBL spectral number density is a key element in evaluating the optical depth within this range. Adopting the EBL photon density of \cite{2018A&A...614C...1F}, we can estimate the optical depth from Eq.(8). which is a monotonically increasing function of $z$ and ${E_\gamma }$. The farther the source is, the greater the probability that the beam photons will be absorbed.
The evolutionary effects of galaxy evolution are included in the effects of cosmic expansion \citep{2008IJMPD..17.1515R}.

We emphasize that EBL photon absorption is expected to be negligible for nearby blazars ($z < 0.03$). This means that the VHE $\gamma$-ray spectra of blazars should have the same shape as the intrinsic photon spectra. We assume that the selected blazars have the same  intrinsic photon spectral index, we further set ${\Gamma _{em}} \simeq 2.3$. According to Eq. (5), we use the lower limit (near infrared) and upper limit (far infrared) of the extragalactic infrared background radiation field to obtain the predicted energy of about 1 to 33 $TeV$, and the resulting value of $\Gamma _{obs}^{CP}$ is represented by the light gray area in Figure 3.

\section{Effects of a possible photon-ALP oscillation} \label{sec:method}
\begin{figure}
\centering
\includegraphics[height=20cm,width=18cm]{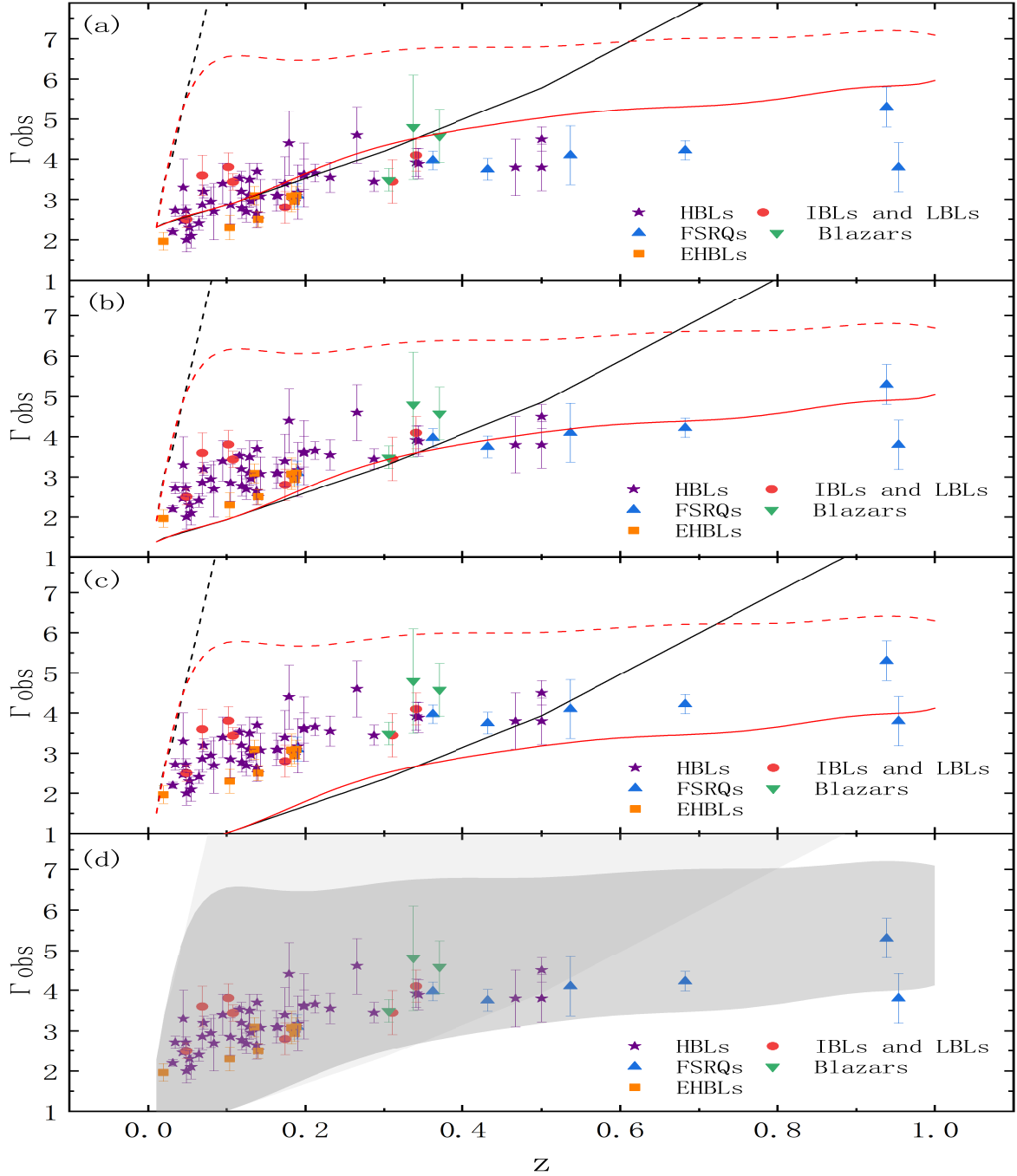}
\caption{Comparisons of predicted VHE $\gamma$-ray photon index with observed data at different values of redshift. Panel (a) shows the redshift dependent VHE $\gamma$-ray photon index with $K_{\rm em}=K_{\rm obs}$ for two different scenarios. The black curve corresponds to conventional physics, but the red curve corresponds to photon-ALP oscillations. Both the black and red dashed curves are plotted by adopting ${E_\gamma } = 33 ~\rm TeV$ and ${\Gamma _{em}} = 2.3$, and both the black and red solid curves are plotted by adopting ${E_\gamma }=1~\rm TeV$ and ${\Gamma _{em}} = 2.3$. Panel (b) and panel (c) show the same case as panel (a), but with $K_{em} = 10{K_{obs}}$ and $K_{em} = 100{K_{obs}}$,  respectively.
Panel (d) shows the expected areas when we adopt the parameters in the range of $0.01 < {{{K_{obs}}} \mathord{\left/
 {\vphantom {{{K_{obs}}} {{K_{em}}}}} \right.
 \kern-\nulldelimiterspace} {{K_{em}}}} < 1$ for two different scenarios. The light grey area $\Gamma _{obs}^{CP}$ is plotted by using conventional physics, and the dark grey area $\Gamma _{obs}^{ALP}$ is evaluated in the cases of proposed photon-ALP oscillations mechanism.}
\label{sub:fig3}
\end{figure}

A simple trend is easily recognized when comparing the conventional calculation with the actually VHE $\gamma$-ray photon index in Figure 3, $\Gamma _{obs}^{CP}$ increases more fastly than $\Gamma _{obs}$ for $0.2 < z < 1$. We are led to the conclusion that such behavior of $\Gamma _{obs}^{CP}$ is at odd with $\Gamma _{obs}$, calling for a departure from the conventional view to discuss the conflict.

One possible way to solve this difficulty is to modify the emission mechanism.
On the one hand, it is based on strong relativistic shock \citep{2007ApJ...667L..29S}. On the other hand, depending on the photon absorption inside the blazar \citep{2008MNRAS.387.1206A}. Simultaneously succeeded in significantly hardening the emission spectrum. Unfortunately, these methods fail to explain why only the most distant blazars have such a drastic departure from the conventional view show up.

The EBL model of FRV cannot explain the distribution of VHE $\gamma$-ray photon index for  $z > 0.2$. To explain this phenomenon reasonably, \cite{2011PhRvD..84j5030D} propose new physics in the form of ALPs. In the previous considerations, VHE $\gamma$-ray photons were assumed to travel freely throughout cosmological distances in a standard way. If there is a large-scale magnetic field in the universe, ALPs and VHE $\gamma$-ray photons can oscillate into each other, referred to as the photon-ALP oscillations. The extragalactic magnetic field ${B_{ext}}$ is usually supposed to originate from quasars and primeval galactic outflows \citep{1968Natur.217..326R,1969Natur.223..936H}, and it is modelled as a domain-like network. The extragalactic magnetic field has approximately the same strength in the domain with the coherence length ${L_{dom}}$ and changes direction randomly from one domain to another \citep{2001Magnetic}. \cite{2020MNRAS.493.1553G} have assumed the standard domain-like morphology for the ${B_{ext}}$ coherence length 1 $Mpc$ $\le {L_{dom}} \le 10$ $Mpc$.

Due to the lack of information about the situation of the extragalactic magnetic fields.  A variety of very different configurations of ${B_{ext}}$ have been proposed \citep{1994RPPh...57..325K, 2001PhR...348..163G, PhysRevD.96.043519}, the strength of ${B_{ext}}$ is limited to about lies in the range ${10^{ - 7}}$ $nG$ $< {B_{ext}} < 1.7$ $nG$ on the scale ${\rm O}(1)$ $Mpc$ by \cite{PhysRevLett.116.191302}.

It was first proposed by \cite{1968Natur.219..127R} and \cite{1969Natur.223..936H} in their studies of radio sources that ionized matter from galaxies gets ejected into extragalactic space. The associated magnetic field is frozen in due to the high conductivity, and amplified by turbulence, thereby magnetizing the surrounding space. Another more refined phenomenon was proposed by \cite{Kronberg_1999}. It's made up of the galactic super-winds emitted by primeval galaxies ${\rm{(z > 6)}}$ that magnetize the extragalactic space, producing fields lies in the $0.1$ $nG$ $< {B_{ext}} < 1$ $nG$ range on the $Mpc$ scale. A more concrete situation was further considered by \cite{Furlanetto_2001} relating to quasars outflows, which also predicts ${B_{ext}} = {\rm O}(1)$ $nG$ on the scale ${\rm O}(1)$ $Mpc$.

Typically, the VHE $\gamma$-ray photon emitted from blazars are quickly converted into axion-like particles, which travel unimpeded throughout the universe and are converted into photons before reaching the earth. Two concrete implementations of this idea have been studied. One of them - \cite{2007Constraints} proposed photon-ALP conversion in intergalactic magnetic fields. \cite{2012PhRvD..86g5024H} also suggested hardening the $TeV$ gamma spectrum of active galactic nuclei in galaxy clusters by converting photons into axion-like particles. The other is complementary in a sense because it premises the $\gamma  \to a$ conversion occurs inside the blazars, and $a \to \gamma $ conversion occurs in the Milky Way. \citep[e.g.,][]{1988PhRvD..37.1237R,2008PhRvD..77f3001S,2014JCAP...01..016W}. In addition, due to the effect of large-scale magnetic fields, the energy conservation oscillation between VHE $\gamma$-ray photon and ALP occurred. The VHE $\gamma$-ray photon are split into two forms, propagating as real photons over a period - which EBL absorbs - and for some time as ALP, in such a situation, they are free from any influence of the EBL \citep{2015Advantages}. It's main consequence is to reduce the EBL absorption, hence the effective optical depth $\tau _\gamma ^{{\rm{eff}}}({E_\gamma },z)$ during the ALP phase is smaller than the actual optical depth $\tau _\gamma ({E_\gamma },z)$ observed by the detector, and is still a monotonically increasing function of both ${E_\gamma }$ and ${\rm{z}}$. By taking into account the photon-ALP oscillation, Eq. (6) becomes
\begin{equation}
\Phi _{{\rm{obs}}}^{ALP}({E_\gamma },z) = P_{\gamma  \to \gamma }^{ALP}({E_\gamma },z)\Phi _{em}^{ALP}({E_\gamma }(1 + z)),
\end{equation}
where $P_{\gamma  \to \gamma }^{ALP}({E_\gamma },z)$ is the probability that a VHE photon is still a photon. Consequently, Eq. (7) is replaced by
\begin{equation}
\Gamma _{obs}^{ALP}(z) = \frac{{\Gamma _{em}^{ALP}\ln [\frac{{{E_\gamma }(1 + z)}}{{{E_0}}}]}}{{\ln (\frac{{{E_\gamma }}}{{{E_0}}})}} - \frac{{\ln P_{\gamma  \to \gamma }^{ALP}({E_\gamma },z)}}{{\ln (\frac{{{E_\gamma }}}{{{E_0}}})}} + \frac{{\ln \frac{{{K_{obs}}}}{{K_{em}^{ALP}}}}}{{\ln (\frac{{{E_\gamma }}}{{{E_0}}})}}.
\end{equation}

Since the photon-ALP oscillations greatly decrease the absorption effect of photons. In this view, the universe with energy $E \gtrsim 1$ $TeV$ is much more transparent than conventional physics \citep{2020MNRAS.493.1553G}. We demonstrate that the observed transparency can be naturally produced by the photon-ALP oscillation that can occur within the intergalactic magnetic field \citep{2007PhRvD..76l1301D}. Assuming an unpolarized beam, the propagation in the intergalactic magnetic field through a homogeneous magnetic field $B \sim 0.7$ $nG$ and coherence length of 6 $Mpc$ (both strength and coherence length as reference values at $z = 0$ ) and a constant electron density of the intergalactic medium of ${10^{ - 7}}$ $c{m^{ - 3}}$.
We take the ALP mass ${\rm{{{\rm{m}}_a}  \sim  1}}{{\rm{0}}^{ - 10}}$ $eV$ and the initial slope ${\Gamma _{em}} = 2.3$.
We note that the lower limit of parameter ${g_{a\gamma }} \sim 2.94 \times {10^{ - 12}}Ge{V^{ - 1}}$ in the case of two-photon coupling as reported in \cite{2020MNRAS.493.1553G}. The CAST experiment has placed limits on the axion-photon coupling strength at ${g_{a\gamma }} < 6.6 \times {10^{ - 11}}Ge{V^{ - 1}}$ \citep{2017NatPh..13..584A}. However, on the upper limit side, the stringent astrophysical limit with ${g_{a\gamma }} \sim 5.4 \times {10^{ - 12}}Ge{V^{ - 1}}$ is also suggested by \cite{2022PhRvD.105j3034D}. We adopt middle value ${g_{a\gamma }} \sim 4.17 \times {10^{ - 12}}Ge{V^{ - 1}}$ as the typical value.
We use the gammaALPs package \citep{2022icrc.confE.557M}, an open-source python code for computing the probability of photon-axion-like-particle oscillations in the intergalactic magnetic field\footnote{https://gammaalps.readthedocs.io}. It is based on numpy \citep{2020Natur.585..357H}, scipy \citep{2020NatMe..17..261V}, and astropy \citep{2013A&A...558A..33A} and uses numba\footnote{https://doi.org/10.1145/2833157.2833162} to speed up calculations \citep{2022icrc.confE.557M}. We only calculate the mixing in one random realization, showing the rapid-oscillatory phenomenon of the probability function, adopting the polynomial fitting method and try to fit the data, the best fitting curve of the photon-ALP oscillation probability was obtained.  Thus, we evaluated the probability of VHE $\gamma$-ray observed photon index, the resulting values of $\Gamma _{obs}^{ALP}(z)$ for 1 $TeV$ $< {E_\gamma } <$ 33 $TeV$ lies within the dark grey area of Figure 3 (d). Taking into account the photon-ALP oscillation and the absorption of EBL, the predicted results in Figure 3 (d) are consistent with the observed results of the samples.

Figure 3 shows the change of ${\Gamma _{obs}}$ with redshift z for different energy ${E_\gamma }$, where comparisons of predicted VHE $\gamma$-ray observed photon index with observed data for $K_{em} = {K_{obs}}$ (a), $K_{em} = 10{K_{obs}}$ (b), $K_{em} = 100{K_{obs}}$ (c) under EBL absorption and photon-Alp oscillation, respectively. We consider the case where ${K_{em}}$ and ${K_{obs}}$ can vary by more than zero magnitude and less than two orders of magnitude ($0.01 < {{{K_{obs}}} \mathord{\left/
 {\vphantom {{{K_{obs}}} {{K_{em}}}}} \right.
 \kern-\nulldelimiterspace} {{K_{em}}}} < 1$) with the same intrinsic photon spectral index ${\Gamma _{em}} = 2.3$ at the bottom (d) of Figure 3. We find that, the light grey area ${\Gamma _{obs}}$ increases non-linearly with redshift z due to the EBL absorption. The dark gray area (ALP scenario) in Figure 3 (d) at low redshift and low energy mimic conventional physics while it become differs from it as redshift and energy grows. It is shown that, the ALP scenario in the low redshift and low energy region, the probability of photon-ALP conversion is noneffective, the probability is dominated by the absorption of EBL. As the energy and redshift turns to the higher region, the photon-ALP conversion probability plays a very important role in the whole process, result in VHE $\gamma$-ray observed photon index is asymptotically independent of z for far-away sources.

\section{Discussion}\label{sec:intro}
It is issued that the VHE $\gamma$-ray photon acquire a split identity, where it propagates for some times as either a real photon or an ALP. The photons could suffer from EBL absorption, but the ALPs could produce photon-ALP oscillations effect. Phenomenologically, the issues offer a possible interpretation for the observed VHE $\gamma$-ray spectra. In order to test whether EBL absorption or photon-ALP oscillations effect existence, the early effort analyses a known redshift sample with 19 distant blazar sources \citep{2009MNRAS.394L..21D}. Their issue assumes that the mechanism results in the same intrinsic photon spectral index. Using the same EBL model, the calculated spectral index is used to predict the changes of the observed $\gamma$-ray spectra. However, a few sample sources result that the observed spectral indices of all blazars detected in the VHE band are not significantly in agreement with the theoretical predictions of the two schemes.

The purpose of this study is enlarging the sample to explore the potential implications on the propagation of VHE $\gamma$-rays from distant blazars. In the context, several interpretations for the emission mechanisms of the data are possible. Different sources possess different properties and may have different intrinsic spectral slopes (Persic $\&$ De Angelis 2008). In our analysis, we only considered the same intrinsic photon spectral index for all sources, and each source was not analyzed separately. We anticipate that the new generation of observatories will detect more sources in the future, and a detailed analysis of the intrinsic spectral slope for each type of source will be achieved. We compile a simple including 70 blazars to test the  VHE $\gamma$-ray spectra characteristics. In the paradigm, a simple mathematical model with ${\Gamma _{obs}}(z) = \alpha z + \beta $ is assumed. To test this hypothesis, we comparisons of predicted VHE $\gamma$-ray photon index with observed data at different values of redshift in this range (1 $< {E_\gamma } <$ 33) $TeV$. The predicted observed spectral index at 0.1 $TeV$ and 10 $TeV$ does not contain the source well. The main effect of photon-ALP oscillations is to reduce the EBL absorption, increasing the cosmic transparency for energies $E \gtrsim 1$ $TeV$. We consider 1 $TeV$ as the lower limit for two different scenarios. We focus on the energy range (0.1 $< {E_\gamma } <$ 10) $TeV$ for the observed energy of most sources. However, the observed energy of some sources may exceed this range, for example, the TeV BL Lac object Mkn 501 detected by the HAWC detector \citep{2019ICRC...36..654C}, the TeV BL Lac object 1ES 0229+200 detected by the HESS detector \cite{2007AA...470..475A}, etc. Therefore, we have assigned an appropriate energy upper limit of 33 TeV to enclose these sources. Additionally, we expect that the range (1 $< {E_\gamma } <$ 33) $TeV$ will include the detection of more high-energy blazars by future detectors. Our results show that the relationship between the VHE $\gamma $-rays photon index and the redshift clearly reveals the propagation properties of VHE $\gamma$-ray.

It is noted that, we take the parameters ${K_{em}}$ and ${K_{obs}}$ into account, the spectral slopes described by Eq. (7) and Eq. (11).
The redshift dependent values of ${K_{obs}}$, $K_{em}^{CP}$, $K_{em}^{ALP}$ in the ranges of redshift from 0 to 0.5 are given in \cite{2020MNRAS.493.1553G}, which shows that the relationship of ${K_{obs}} \lesssim K_{em}^{CP}$ and ${K_{obs}} \lesssim K_{em}^{ALP}$ are always correct at different redshift. In the context, we convenient to plot the parameters change for $K_{em} = {K_{obs}}$ (a), $K_{em} = 10{K_{obs}}$ (b), $K_{em} = 100{K_{obs}}$ (c) in Figure 3. It is evident that the additional term leads to diminish the spectral slope at a redshift z. We consider the case where ${K_{em}}$ and ${K_{obs}}$ can vary by more than zero magnitude and be smaller than two orders of magnitude ($0.01 < {{{K_{obs}}} \mathord{\left/
 {\vphantom {{{K_{obs}}} {{K_{em}}}}} \right.
 \kern-\nulldelimiterspace} {{K_{em}}}} < 1$) with the same intrinsic photon spectral index ${\Gamma _{em}} = 2.3$ at the bottom (d) of Figure 3. Since we adopt the ${E_0} = 100$ $GeV$, it is clear that there are two cases including the Eq. (7) and Eq. (11): 1) the ${\Gamma _{obs}}$ increases non-linearly at the ranges of $0.03 < z < 0.2$ because of the EBL absorption; 2) ${\Gamma _{obs}}$ is almost uniformly at $0.2 < z < 1$, we predict the potential effect of introducing photon-ALP oscillations to reduce the cosmic photon-photon opacity systematically. It can also be seen that the ALP scenario at low redshift and low energy can mimic conventional physics and ALP effect looks more evident as the redshift and energy grows. In these scenarios, it is shown that, if the ${K_{em}}$ and ${K_{obs}}$ vary by more than two orders of magnitude for more high redshift value, the model can expect a larger intrinsic photon index.

We note that, since the photon-ALP oscillation in the presence of an external magnetic field can effectively attenuate the EBL absorption of VHE photons, we can expect a diagnostic VHE $\gamma$-ray spectra of astrophysical sources. While the current VHE observations only provide limited dataset, the upcoming new generation Cherenkov observatories, such as LHAAASO, CTA, ASTRI, and SWGO, are in operation. The programme can provide abundant of distant blazar sources to test the intrinsic properties of VHE $\gamma$-ray spectra. We leave the open issues in the future.

\begin{table*}{}
\centering
\fontsize{8}{11}\selectfont
\caption{Blazars observed so far with the IACTs. The list gives the source name, type, redshift (z), the VHE $\gamma $-rays photon index ($\Gamma _{obs}$), and the reference. Statistical and systematic errors are added to the integral equation to yield the total error in the measured, observed spectral slope. When only statistical errors are cited, HESS has a systematic error of 0.1, VERITAS has a systematic error of 0.15, and MAGIC has a systematic error of 0.2.}\label{Tab1}
\begin{tabular}{ccccccccccccc}
\hline\hline
   Source & Type & z & $\Gamma _{obs}^{TC}$  & ref\\
\hline
    Mkn 421&HBL&0.031&2.20 $\pm$ 0.08&\cite{2007ApJ...663..125A}     \\
    Mkn 501&HBL&0.034&2.72 $\pm$ 0.15&\cite{2011ApJ...729....2A}     \\
    1ES 2344+514&HBL&0.044&2.46 $\pm$ 0.06&\cite{2017MNRAS.471.2117A}     \\
    Markarian 180&HBL&0.045&3.3 $\pm$ 0.7&\cite{2006ApJ...648L.105A}    \\
    1ES 1959+650&HBL&0.048&2.72 $\pm$ 0.14&\cite{2006ApJ...639..761A}     \\
    TXS 0210+515&HBL&0.049&2.0 $\pm$ 0.3&\cite{2020ApJS..247...16A}     \\
    1ES 2037+521&HBL&0.053&2.3 $\pm$ 0.2&\cite{2020ApJS..247...16A}    \\
    1ES 1727+502&HBL&0.055&2.1 $\pm$ 0.3&\cite{2015ApJ...808..110A}     \\
    PGC 2402248&HBL&0.065&2.41 $\pm$ 0.17&\cite{2019MNRAS.490.2284M}     \\
    PKS 0548-322&HBL&0.069&2.86 $\pm$ 0.34&\cite{2010AA...521A..69A}  \\
    1ES 1741+196&HBL&0.084&2.7 $\pm$ 0.7&\cite{2016MNRAS.459.2550A}\\
    SHBL J001355.9-185406&HBL&0.095&3.4 $\pm$ 0.5&\cite{2013AA...554A..72H}   \\
    1ES 1312-423&HBL&0.105&2.85 $\pm$ 0.47&\cite{2013MNRAS.434.1889H}     \\
    PKS 2155-304 & HBL&0.116&3.53$\pm$ 0.06&\cite{2010AA...520A..83H}\\
    B3 2247+381&HBL&0.1187&3.2 $\pm$ 0.5&\cite{2012AA...539A.118A}\\
    1RXS J195815.6-301119&HBL&0.119329&2.78 $\pm$ 0.26&\cite{2022icrc.confE.823B}     \\
    RGB J0710+591&HBL&0.125&2.69 $\pm$ 0.26&\cite{2010ApJ...715L..49A}     \\
    TXS 1515-273&HBL&0.1284&3.11 $\pm$ 0.32&\cite{2021MNRAS.507.1528A}     \\
    H 1426+428&HBL&0.129&3.5 $\pm$ 0.4&\cite{2002ApJ...580..104P}     \\
    1ES 1215+303&HBL&0.131&2.96 $\pm$ 0.14&\cite{2012AA...544A.142A}      \\
    PKS 1440-389&HBL&0.1385&3.7 $\pm$ 0.2&\cite{2020MNRAS.494.5590A}     \\
    1RXS J101015.9-311909&HBL&0.142639&3.08$\pm$ 0.42&\cite{2012AA...542A..94H}    \\
    1ES 1440+122&HBL&0.163058&3.1 $\pm$ 0.4&\cite{2016MNRAS.461..202A}     \\
    1ES 2322-409&HBL&0.1736&3.40 $\pm$ 0.66&\cite{2019MNRAS.482.3011A}    \\
    RX J0648.7+1516&HBL&0.179&4.4 $\pm$ 0.8&\cite{2011ApJ...742..127A}     \\
    1ES 1218+304&HBL&0.182&3.08$\pm$ 0.34&\cite{2009ApJ...695.1370A} \\
    1ES 1101-232&HBL&0.186 &2.94 $\pm$ 0.2&\cite{2007AA...470..475A}   \\
     RBS 0413&HBL&0.19&3.18 $\pm$ 0.68&\cite{2012ApJ...750...94A}     \\
    RBS 0723&HBL&0.198&3.6 $\pm$ 0.8&\cite{2020ApJS..247...16A}    \\
    MRC 0910-208&HBL&0.19802&3.63 $\pm$ 0.38&\cite{2022icrc.confE.823B}   \\
    1RXS J023832.6-311658&HBL&0.232&3.552 $\pm$ 0.371&\cite{2017ICRC...35..645G}     \\
    PKS 0301-243&HBL&0.2657&4.6 $\pm$ 0.7&\cite{2013AA...559A.136H}    \\
    1ES 0414+009&HBL&0.287&3.45 $\pm$ 0.25&\cite{2012AA...538A.103H}   \\
    1ES 0502+675&HBL&0.34&3.92 $\pm$ 0.35&\cite{2011ICRC....8...51B}  \\
    PKS 0447-439&HBL&0.343&3.89 $\pm$ 0.37&\cite{2013AA...552A.118H} \\
    1ES 0033+595&HBL&0.467&3.8 $\pm$ 0.7&\cite{2015MNRAS.446..217A}    \\
    PG 1553+113&HBL&0.5&4.50 $\pm$ 0.3& \cite{2008AA...477..481A} \\
    PKS 2005-489&HBL&0.071&3.2 $\pm$ 0.16&\cite{2010AA...511A..52H}  \\
    RGB J0152+017&HBL&0.08&2.95 $\pm$ 0.36&\cite{2008AA...481L.103A}     \\
    1ES 0806+524&HBL&0.138&2.65 $\pm$ 0.36&\cite{2015MNRAS.451..739A}  \\
    1ES 0229+200&HBL&0.1396&2.5 $\pm$ 0.19&\cite{2007AA...475L...9A} \\
    H 2356-309&HBL&0.165&3.09 $\pm$ 0.24&\cite{2006AA...455..461A}  \\
    1ES 0347-121&HBL&0.188&3.10 $\pm$ 0.23&\cite{2007AA...473L..25A}   \\
    1ES 1011+496&HBL&0.212&3.66 $\pm$ 0.22&\cite{2016AA...591A..10A}   \\
    PKS 1424+240&HBL&0.5&3.80 $\pm$ 0.58 &\cite{2010ApJ...708L.100A}  \\

\hline
\end{tabular}\\
\end{table*}

\begin{table*}{}
\centering
\fontsize{8}{11}\selectfont
\begin{tabular}{ccccccccccccc}
\hline\hline
   Source & Type & z & $\Gamma _{obs}^{TC}$  & ref\\
\hline
    RX J0648+1516&HBL&0.179&4.4 $\pm$ 0.8 &\cite{2011ApJ...742..127A}  \\
    BL Lacertae&IBL&0.069&3.6 $\pm$ 0.5&\cite{2007ApJ...666L..17A}     \\
    S5 0716+714&IBL&0.31&3.45 $\pm$ 0.54&\cite{2009ApJ...704L.129A}   \\
    W Comae&IBL&0.102&3.81 $\pm$ 0.35&\cite{2008ApJ...684L..73A}    \\
    MAGIC J2001+435&IBL&0.1739&2.8 $\pm$ 0.4&\cite{2014AA...572A.121A}      \\
    3C 66A&IBL&0.34&4.1 $\pm$ 0.4&\cite{2009ApJ...693L.104A}     \\
    VER J0521+211&IBL&0.108&3.44$\pm$ 0.2&\cite{2013ApJ...776...69A}    \\
    AP Librae&LBL&0.049&2.5 $\pm$ 0.2&\cite{2011ICRC....8..109C}   \\
    OJ 287&Blazar&0.3056&3.49 $\pm$ 0.28&\cite{2017arXiv170802160O}  \\
    TXS 0506+056& Blazar&0.3365&4.8 $\pm$ 1.3&\cite{2018ApJ...861L..20A}    \\
    S4 0954+65& Blazar&0.3694&4.58 $\pm$ 0.66&\cite{2018AA...617A..30M}      \\
    PKS 0736+017&FSRQ&0.18941&3.1 $\pm$ 0.3&\cite{2020AA...633A.162H}    \\
    PKS 1510-089&FSRQ&0.361&3.97 $\pm$ 0.23&\cite{2018AA...619A.159M}     \\
    4C +21.35&FSRQ&0.432&3.75 $\pm$ 0.27&\cite{2011ApJ...730L...8A}    \\
    3C 279&FSRQ&0.5362&4.1 $\pm$ 0.73&\cite{2008Sci...320.1752M}     \\
    B2 1420+32&FSRQ&0.682&4.22 $\pm$ 0.24&\cite{2008Sci...320.1752M}    \\
    PKS 1441+25&FSRQ&0.939&5.3 $\pm$ 0.5&\cite{2015ApJ...815L..22A}    \\
    S3 0218+35&FSRQ&0.954&3.8 $\pm$ 0.61& \cite{2016AA...595A..98A}    \\
    IC 310&EHBL&0.019&1.96 $\pm$ 0.22&\cite{2014AA...563A..91A}  \\
    RGB J2042+244&EHBL&0.104&2.3 $\pm$ 0.30&\cite{2020ApJS..247...16A}  \\
    1ES 0229+200&EHBL&0.14&2.50 $\pm$ 0.19&\cite{2007AA...475L...9A}    \\
    H 2356-309&EHBL&0.135&3.09 $\pm$ 0.24&\cite{2006AA...455..461A}    \\
    1ES 1218+304&EHBL&0.182&3.08 $\pm$ 0.40&\cite{2008AIPC.1085..565F}     \\
    1ES 1101-232&EHBL&0.186&2.94 $\pm$ 0.20&\cite{2007AA...470..475A}  \\
    1ES 0347-121&EHBL&0.188&3.10 $\pm$ 0.23&\cite{2007AA...473L..25A}     \\

\hline
\end{tabular}\\
\end{table*}

\section*{Acknowledgements}
We thank anonymous referee for constructive comments and suggestions. This work is partially supported by the National Natural Science Foundation of China (Grant Nos. 11873043 and 12163002). We also thank EditSprings\footnote{https://www.editsprings.cn/} for the expert linguistic services provided.
%

%
\bibliography{aastexbib}

\begin{thebibliography}{}
\expandafter\ifx\csname natexlab\endcsname\relax\def\natexlab#1{#1}\fi
\providecommand{\url}[1]{\href{#1}{#1}}
\providecommand{\dodoi}[1]{doi:~\href{http://doi.org/#1}{\nolinkurl{#1}}}
\providecommand{\doeprint}[1]{\href{http://ascl.net/#1}{\nolinkurl{http://ascl.net/#1}}}
\providecommand{\doarXiv}[1]{\href{https://arxiv.org/abs/#1}{\nolinkurl{https://arxiv.org/abs/#1}}}

\bibitem[{{Abdalla} {et~al.}(2019){Abdalla}, {Aharonian}, {Ait Benkhali},
  {Ang{\"u}ner}, {Arakawa}, {Arcaro}, {Armand}, {Arrieta}, {Backes}, {Barnard},
  {Becherini}, {Becker Tjus}, {Berge}, {Bernhard}, {Bernl{\"o}hr}, {Blackwell},
  {B{\"o}ttcher}, {Boisson}, {Bolmont}, {Bonnefoy}, {Bordas}, {Bregeon},
  {Brun}, {Brun}, {Bryan}, {B{\"u}chele}, {Bulik}, {Bylund}, {Capasso},
  {Caroff}, {Carosi}, {Casanova}, {Cerruti}, {Chakraborty}, {Chandra},
  {Chaves}, {Chen}, {Colafrancesco}, {Condon}, {Davids}, {Deil}, {Devin},
  {deWilt}, {Dirson}, {Djannati-Ata{\"\i}}, {Dmytriiev}, {Donath}, {Drury},
  {Dyks}, {Egberts}, {Emery}, {Ernenwein}, {Eschbach}, {Fegan}, {Fiasson},
  {Fontaine}, {Funk}, {F{\"u}{\ss}ling}, {Gabici}, {Gallant}, {Garrigoux},
  {Gat{\'e}}, {Giavitto}, {Glawion}, {Glicenstein}, {Gottschall}, {Grondin},
  {Hahn}, {Haupt}, {Heinzelmann}, {Henri}, {Hermann}, {Hinton}, {Hofmann},
  {Hoischen}, {Holch}, {Holler}, {Horns}, {Huber}, {Iwasaki}, {Jacholkowska},
  {Jamrozy}, {Jankowsky}, {Jankowsky}, {Jouvin}, {Jung-Richardt},
  {Kastendieck}, {Katarzy{\'n}ski}, {Katsuragawa}, {Katz}, {Kerszberg},
  {Khangulyan}, {Kh{\'e}lifi}, {King}, {Klepser}, {Klu{\'z}niak}, {Komin},
  {Kosack}, {Krakau}, {Kraus}, {Kr{\"u}ger}, {Lamanna}, {Lau}, {Lefaucheur},
  {Lemi{\`e}re}, {Lemoine-Goumard}, {Lenain}, {Leser}, {Lohse}, {Lorentz},
  {L{\'o}pez-Coto}, {Lypova}, {Malyshev}, {Marandon}, {Marcowith}, {Mariaud},
  {Mart{\'\i}-Devesa}, {Marx}, {Maurin}, {Meintjes}, {Mitchell}, {Moderski},
  {Mohamed}, {Mohrmann}, {Moulin}, {Murach}, {Nakashima}, {de Naurois},
  {Ndiyavala}, {Niederwanger}, {Niemiec}, {Oakes}, {O'Brien}, {Odaka}, {Ohm},
  {Ostrowski}, {Oya}, {Padovani}, {Panter}, {Parsons}, {Perennes}, {Petrucci},
  {Peyaud}, {Piel}, {Pita}, {Poireau}, {Priyana Noel}, {Prokhorov}, {Prokoph},
  {P{\"u}hlhofer}, {Punch}, {Quirrenbach}, {Raab}, {Rauth}, {Reimer}, {Reimer},
  {Renaud}, {Rieger}, {Rinchiuso}, {Romoli}, {Rowell}, {Rudak}, {Ruiz-Velasco},
  {Sahakian}, {Saito}, {Sanchez}, {Santangelo}, {Sasaki}, {Schlickeiser},
  {Sch{\"u}ssler}, {Schulz}, {Schwanke}, {Schwemmer}, {Seglar-Arroyo},
  {Senniappan}, {Seyffert}, {Shafi}, {Shilon}, {Shiningayamwe}, {Simoni},
  {Sinha}, {Sol}, {Spanier}, {Specovius}, {Spir-Jacob}, {Stawarz}, {Steenkamp},
  {Stegmann}, {Steppa}, {Sushch}, {Takahashi}, {Tavernet}, {Tavernier},
  {Taylor}, {Terrier}, {Tibaldo}, {Tiziani}, {Tluczykont}, {Trichard},
  {Tsirou}, {Tsuji}, {Tuffs}, {Uchiyama}, {van der Walt}, {van Eldik}, {van
  Rensburg}, {van Soelen}, {Vasileiadis}, {Veh}, {Venter}, {Viana}, {Vincent},
  {Vink}, {Voisin}, {V{\"o}lk}, {Vuillaume}, {Wadiasingh}, {Wagner}, {Wagner},
  {Wagner}, {White}, {Wierzcholska}, {W{\"o}rnlein}, {Yang}, {Zaborov},
  {Zacharias}, {Zanin}, {Zdziarski}, {Zech}, {Zefi}, {Ziegler}, {Zorn},
  {{\.Z}ywucka}, \& {H.~E.~S.~S. Collaboration}}]{2019MNRAS.482.3011A}
{Abdalla}, H., {Aharonian}, F., {Ait Benkhali}, F., {et~al.} 2019, \mnras, 482,
  3011, \dodoi{10.1093/mnras/sty2686}

\bibitem[{{Abdalla} {et~al.}(2020){Abdalla}, {Adam}, {Aharonian}, {Ait
  Benkhali}, {Ang{\"u}ner}, {Arakawa}, {Arcaro}, {Armand}, {Armstrong},
  {Ashkar}, {Backes}, {Baghmanyan}, {Barbosa Martins}, {Barnacka}, {Barnard},
  {Becherini}, {Berge}, {Bernl{\"o}hr}, {B{\"o}ttcher}, {Boisson}, {Bolmont},
  {Bonnefoy}, {Bregeon}, {Breuhaus}, {Brun}, {Brun}, {Bryan}, {B{\"u}chele},
  {Bulik}, {Bylund}, {Caroff}, {Carosi}, {Casanova}, {Chand}, {Chandra},
  {Chen}, {Cotter}, {Cury{\l}o}, {Davids}, {Davies}, {Deil}, {Devin}, {deWilt},
  {Dirson}, {Djannati-Ata{\"\i}}, {Dmytriiev}, {Donath}, {Doroshenko}, {Dyks},
  {Egberts}, {Eichhorn}, {Emery}, {Ernenwein}, {Feijen}, {Fegan}, {Fiasson},
  {Fontaine}, {Funk}, {F{\"u}{\ss}ling}, {Gabici}, {Gallant}, {Giavitto},
  {Giunti}, {Glawion}, {Glicenstein}, {Gottschall}, {Grondin}, {Hahn}, {Haupt},
  {Hermann}, {Hinton}, {Hofmann}, {Hoischen}, {Holch}, {Holler}, {H{\"o}rbe},
  {Horns}, {Huber}, {Iwasaki}, {Jamrozy}, {Jankowsky}, {Jankowsky},
  {Jardin-Blicq}, {Joshi}, {Jung-Richardt}, {Kastendieck}, {Katarzy{\'n}ski},
  {Katsuragawa}, {Katz}, {Khangulyan}, {Kh{\'e}lifi}, {Klepser},
  {Klu{\'z}niak}, {Komin}, {Konno}, {Kosack}, {Kostunin}, {Kreter}, {Lamanna},
  {Lemi{\`e}re}, {Lemoine-Goumard}, {Lenain}, {Leser}, {Levy}, {Lohse},
  {Lypova}, {Mackey}, {Majumdar}, {Malyshev}, {Malyshev}, {Marandon},
  {Marchegiani}, {Marcowith}, {Mares}, {Mart{\'\i}-Devesa}, {Marx}, {Maurin},
  {Meintjes}, {Moderski}, {Mohamed}, {Mohrmann}, {Moore}, {Morris}, {Moulin},
  {Muller}, {Murach}, {Nakashima}, {Nakashima}, {de Naurois}, {Ndiyavala},
  {Niederwanger}, {Niemiec}, {Oakes}, {O'Brien}, {Odaka}, {Ohm}, {de O{\~n}a
  Wilhelmi}, {Ostrowski}, {Panter}, {Parsons}, {Peyaud}, {Piel}, {Pita},
  {Poireau}, {Noel}, {Prokhorov}, {Prokoph}, {P{\"u}hlhofer}, {Punch},
  {Quirrenbach}, {Raab}, {Rauth}, {Reimer}, {Reimer}, {Remy}, {Renaud},
  {Rieger}, {Rinchiuso}, {Romoli}, {Rowell}, {Rudak}, {Ruiz-Velasco},
  {Sahakian}, {Sailer}, {Saito}, {Sanchez}, {Santangelo}, {Sasaki}, {Scalici},
  {Sch{\"u}ssler}, {Schutte}, {Schwanke}, {Schwemmer}, {Seglar-Arroyo},
  {Senniappan}, {Seyffert}, {Shafi}, {Shiningayamwe}, {Simoni}, {Sinha}, {Sol},
  {Specovius}, {Spencer}, {Spir-Jacob}, {Stawarz}, {Steenkamp}, {Stegmann},
  {Steppa}, {Takahashi}, {Tavernier}, {Taylor}, {Terrier}, {Tiziani},
  {Tluczykont}, {Tomankova}, {Trichard}, {Tsirou}, {Tsuji}, {Tuffs},
  {Uchiyama}, {van der Walt}, {van Eldik}, {van Rensburg}, {van Soelen},
  {Vasileiadis}, {Veh}, {Venter}, {Vincent}, {Vink}, {V{\"o}lk}, {Vuillaume},
  {Wadiasingh}, {Wagner}, {Watson}, {Werner}, {White}, {Wierzcholska}, {Yang},
  {Yoneda}, {Zacharias}, {Zanin}, {Zargaryan}, {Zdziarski}, {Zech}, {Zhu},
  {Zorn}, {{\.Z}ywucka}, \& {Cerruti}}]{2020MNRAS.494.5590A}
{Abdalla}, H., {Adam}, R., {Aharonian}, F., {et~al.} 2020, \mnras, 494, 5590,
  \dodoi{10.1093/mnras/staa999}

\bibitem[{{Abdo} {et~al.}(2010){Abdo}, {Ackermann}, {Agudo}, {Ajello}, {Aller},
  {Aller}, {Angelakis}, {Arkharov}, {Axelsson}, {Bach}, {Baldini}, {Ballet},
  {Barbiellini}, {Bastieri}, {Baughman}, {Bechtol}, {Bellazzini}, {Benitez},
  {Berdyugin}, {Berenji}, {Blandford}, {Bloom}, {Boettcher}, {Bonamente},
  {Borgland}, {Bregeon}, {Brez}, {Brigida}, {Bruel}, {Burnett}, {Burrows},
  {Buson}, {Caliandro}, {Calzoletti}, {Cameron}, {Capalbi}, {Caraveo},
  {Carosati}, {Casandjian}, {Cavazzuti}, {Cecchi}, {{\c{C}}elik}, {Charles},
  {Chaty}, {Chekhtman}, {Chen}, {Chiang}, {Chincarini}, {Ciprini}, {Claus},
  {Cohen-Tanugi}, {Colafrancesco}, {Cominsky}, {Conrad}, {Costamante},
  {Cutini}, {D'ammando}, {Deitrick}, {D'Elia}, {Dermer}, {de Angelis}, {de
  Palma}, {Digel}, {Donnarumma}, {Silva}, {Drell}, {Dubois}, {Dultzin},
  {Dumora}, {Falcone}, {Farnier}, {Favuzzi}, {Fegan}, {Focke}, {Forn{\'e}},
  {Fortin}, {Frailis}, {Fuhrmann}, {Fukazawa}, {Funk}, {Fusco}, {G{\'o}mez},
  {Gargano}, {Gasparrini}, {Gehrels}, {Germani}, {Giebels}, {Giglietto},
  {Giommi}, {Giordano}, {Giuliani}, {Glanzman}, {Godfrey}, {Grenier},
  {Gronwall}, {Grove}, {Guillemot}, {Guiriec}, {Gurwell}, {Hadasch},
  {Hanabata}, {Harding}, {Hayashida}, {Hays}, {Healey}, {Heidt}, {Hiriart},
  {Horan}, {Hoversten}, {Hughes}, {Itoh}, {Jackson}, {J{\'o}hannesson},
  {Johnson}, {Johnson}, {Jorstad}, {Kadler}, {Kamae}, {Katagiri}, {Kataoka},
  {Kawai}, {Kennea}, {Kerr}, {Kimeridze}, {Kn{\"o}dlseder}, {Kocian},
  {Kopatskaya}, {Koptelova}, {Konstantinova}, {Kovalev}, {Kovalev},
  {Kurtanidze}, {Kuss}, {Lande}, {Larionov}, {Latronico}, {Leto}, {Lindfors},
  {Longo}, {Loparco}, {Lott}, {Lovellette}, {Lubrano}, {Madejski}, {Makeev},
  {Marchegiani}, {Marscher}, {Marshall}, {Max-Moerbeck}, {Mazziotta},
  {McConville}, {McEnery}, {Meurer}, {Michelson}, {Mitthumsiri}, {Mizuno},
  {Moiseev}, {Monte}, {Monzani}, {Morselli}, {Moskalenko}, {Murgia},
  {Nestoras}, {Nilsson}, {Nizhelsky}, {Nolan}, {Norris}, {Nuss}, {Ohsugi},
  {Ojha}, {Omodei}, {Orlando}, {Ormes}, {Osborne}, {Ozaki}, {Pacciani},
  {Padovani}, {Pagani}, {Page}, {Paneque}, {Panetta}, {Parent}, {Pasanen},
  {Pavlidou}, {Pelassa}, {Pepe}, {Perri}, {Pesce-Rollins}, {Piranomonte},
  {Piron}, {Pittori}, {Porter}, {Puccetti}, {Rahoui}, {Rain{\`o}}, {Raiteri},
  {Rando}, {Razzano}, {Reimer}, {Reimer}, {Reposeur}, {Richards}, {Ritz},
  {Rochester}, {Rodriguez}, {Romani}, {Ros}, {Roth}, {Roustazadeh}, {Ryde},
  {Sadrozinski}, {Sadun}, {Sanchez}, {Sander}, {Saz Parkinson}, {Scargle},
  {Sellerholm}, {Sgr{\`o}}, {Shaw}, {Sigua}, {Siskind}, {Smith}, {Smith},
  {Spandre}, {Spinelli}, {Starck}, {Stevenson}, {Stratta}, {Strickman},
  {Suson}, {Tajima}, {Takahashi}, {Takahashi}, {Takalo}, {Tanaka}, {Thayer},
  {Thayer}, {Thompson}, {Tibaldo}, {Torres}, {Tosti}, {Tramacere}, {Uchiyama},
  {Usher}, {Vasileiou}, {Verrecchia}, {Vilchez}, {Villata}, {Vitale}, {Waite},
  {Wang}, {Winer}, {Wood}, {Ylinen}, {Zensus}, {Zhekanis}, \&
  {Ziegler}}]{2010ApJ...716...30A}
{Abdo}, A.~A., {Ackermann}, M., {Agudo}, I., {et~al.} 2010, \apj, 716, 30,
  \dodoi{10.1088/0004-637X/716/1/30}

\bibitem[{{Abdollahi} {et~al.}(2020){Abdollahi}, {Acero}, {Ackermann},
  {Ajello}, {Atwood}, {Axelsson}, {Baldini}, {Ballet}, {Barbiellini},
  {Bastieri}, {Becerra Gonzalez}, {Bellazzini}, {Berretta}, {Bissaldi},
  {Blandford}, {Bloom}, {Bonino}, {Bottacini}, {Brandt}, {Bregeon}, {Bruel},
  {Buehler}, {Burnett}, {Buson}, {Cameron}, {Caputo}, {Caraveo}, {Casandjian},
  {Castro}, {Cavazzuti}, {Charles}, {Chaty}, {Chen}, {Cheung}, {Chiaro},
  {Ciprini}, {Cohen-Tanugi}, {Cominsky}, {Coronado-Bl{\'a}zquez}, {Costantin},
  {Cuoco}, {Cutini}, {D'Ammando}, {DeKlotz}, {de la Torre Luque}, {de Palma},
  {Desai}, {Digel}, {Di Lalla}, {Di Mauro}, {Di Venere}, {Dom{\'\i}nguez},
  {Dumora}, {Fana Dirirsa}, {Fegan}, {Ferrara}, {Franckowiak}, {Fukazawa},
  {Funk}, {Fusco}, {Gargano}, {Gasparrini}, {Giglietto}, {Giommi}, {Giordano},
  {Giroletti}, {Glanzman}, {Green}, {Grenier}, {Griffin}, {Grondin}, {Grove},
  {Guiriec}, {Harding}, {Hayashi}, {Hays}, {Hewitt}, {Horan},
  {J{\'o}hannesson}, {Johnson}, {Kamae}, {Kerr}, {Kocevski}, {Kovac'evic'},
  {Kuss}, {Landriu}, {Larsson}, {Latronico}, {Lemoine-Goumard}, {Li},
  {Liodakis}, {Longo}, {Loparco}, {Lott}, {Lovellette}, {Lubrano}, {Madejski},
  {Maldera}, {Malyshev}, {Manfreda}, {Marchesini}, {Marcotulli},
  {Mart{\'\i}-Devesa}, {Martin}, {Massaro}, {Mazziotta}, {McEnery}, {Mereu},
  {Meyer}, {Michelson}, {Mirabal}, {Mizuno}, {Monzani}, {Morselli},
  {Moskalenko}, {Negro}, {Nuss}, {Ojha}, {Omodei}, {Orienti}, {Orlando},
  {Ormes}, {Palatiello}, {Paliya}, {Paneque}, {Pei}, {Pe{\~n}a-Herazo},
  {Perkins}, {Persic}, {Pesce-Rollins}, {Petrosian}, {Petrov}, {Piron}, {Poon},
  {Porter}, {Principe}, {Rain{\`o}}, {Rando}, {Razzano}, {Razzaque}, {Reimer},
  {Reimer}, {Remy}, {Reposeur}, {Romani}, {Saz Parkinson}, {Schinzel},
  {Serini}, {Sgr{\`o}}, {Siskind}, {Smith}, {Spandre}, {Spinelli}, {Strong},
  {Suson}, {Tajima}, {Takahashi}, {Tak}, {Thayer}, {Thompson}, {Tibaldo},
  {Torres}, {Torresi}, {Valverde}, {Van Klaveren}, {van Zyl}, {Wood},
  {Yassine}, \& {Zaharijas}}]{2020ApJS..247...33A}
{Abdollahi}, S., {Acero}, F., {Ackermann}, M., {et~al.} 2020, \apjs, 247, 33,
  \dodoi{10.3847/1538-4365/ab6bcb}

\bibitem[{{Abeysekara} {et~al.}(2015){Abeysekara}, {Archambault}, {Archer},
  {Aune}, {Barnacka}, {Benbow}, {Bird}, {Biteau}, {Buckley}, {Bugaev},
  {Cardenzana}, {Cerruti}, {Chen}, {Christiansen}, {Ciupik}, {Connolly},
  {Coppi}, {Cui}, {Dickinson}, {Dumm}, {Eisch}, {Errando}, {Falcone}, {Feng},
  {Finley}, {Fleischhack}, {Flinders}, {Fortin}, {Fortson}, {Furniss},
  {Gillanders}, {Griffin}, {Grube}, {Gyuk}, {H{\"u}tten}, {H{\r{a}}kansson},
  {Hanna}, {Holder}, {Humensky}, {Johnson}, {Kaaret}, {Kar}, {Kelley-Hoskins},
  {Khassen}, {Kieda}, {Krause}, {Krennrich}, {Kumar}, {Lang}, {Maier},
  {McArthur}, {McCann}, {Meagher}, {Moriarty}, {Mukherjee}, {Nieto},
  {O'Faol{\'a}in de Bhr{\'o}ithe}, {Ong}, {Otte}, {Park}, {Perkins},
  {Petrashyk}, {Pohl}, {Popkow}, {Pueschel}, {Quinn}, {Ragan}, {Ratliff},
  {Reynolds}, {Richards}, {Roache}, {Rousselle}, {Santander}, {Sembroski},
  {Shahinyan}, {Smith}, {Staszak}, {Telezhinsky}, {Todd}, {Tucci}, {Tyler},
  {Vassiliev}, {Vincent}, {Wakely}, {Weiner}, {Weinstein}, {Wilhelm},
  {Williams}, {Zitzer}, {VERITAS}, {Smith}, {SPOL}, {Holoien}, {Prieto},
  {Kochanek}, {Stanek}, {Shappee}, {ASAS-SN}, {Hovatta}, {Max-Moerbeck},
  {Pearson}, {Reeves}, {Richards}, {Readhead}, {OVRO}, {Madejski}, {NuSTAR},
  {Djorgovski}, {Drake}, {Graham}, {Mahabal}, \& {CRTS}}]{2015ApJ...815L..22A}
{Abeysekara}, A.~U., {Archambault}, S., {Archer}, A., {et~al.} 2015, \apjl,
  815, L22, \dodoi{10.1088/2041-8205/815/2/L22}

\bibitem[{{Abeysekara} {et~al.}(2016){Abeysekara}, {Archambault}, {Archer},
  {Benbow}, {Bird}, {Biteau}, {Buchovecky}, {Buckley}, {Bugaev}, {Byrum},
  {Cardenzana}, {Cerruti}, {Chen}, {Christiansen}, {Ciupik}, {Connolly}, {Cui},
  {Dickinson}, {Dumm}, {Eisch}, {Errando}, {Falcone}, {Feng}, {Finley},
  {Fleischhack}, {Flinders}, {Fortin}, {Fortson}, {Furniss}, {Gillanders},
  {Griffin}, {Grube}, {Gyuk}, {Huetten}, {Hanna}, {Holder}, {Humensky},
  {Johnson}, {Kaaret}, {Kar}, {Kelley-Hoskins}, {Kertzman}, {Kieda}, {Krause},
  {Krennrich}, {Lang}, {Maier}, {McArthur}, {McCann}, {Meagher}, {Moriarty},
  {Mukherjee}, {Nieto}, {O'Brien}, {O'Faol{\'a}in de Bhr{\'o}ithe}, {Ong},
  {Otte}, {Park}, {Pelassa}, {Petrashyk}, {Petry}, {Pohl}, {Popkow},
  {Pueschel}, {Quinn}, {Ragan}, {Ratliff}, {Reyes}, {Reynolds}, {Reynolds},
  {Richards}, {Roache}, {Rulten}, {Santander}, {Sembroski}, {Shahinyan},
  {Smith}, {Staszak}, {Telezhinsky}, {Tucci}, {Tyler}, {Vincent}, {Wakely},
  {Weiner}, {Weinstein}, {Wilhelm}, {Williams}, \&
  {Zitzer}}]{2016MNRAS.459.2550A}
---. 2016, \mnras, 459, 2550, \dodoi{10.1093/mnras/stw664}

\bibitem[{{Abeysekara} {et~al.}(2018){Abeysekara}, {Archer}, {Benbow}, {Bird},
  {Brill}, {Brose}, {Buckley}, {Christiansen}, {Chromey}, {Daniel}, {Falcone},
  {Feng}, {Finley}, {Fortson}, {Furniss}, {Gillanders}, {Gueta}, {Hanna},
  {Hervet}, {Holder}, {Hughes}, {Humensky}, {Johnson}, {Kaaret}, {Kar},
  {Kelley-Hoskins}, {Kertzman}, {Kieda}, {Krause}, {Krennrich}, {Lang},
  {Moriarty}, {Mukherjee}, {O'Brien}, {Ong}, {Otte}, {Park}, {Petrashyk},
  {Pohl}, {Pueschel}, {Quinn}, {Ragan}, {Reynolds}, {Richards}, {Roache},
  {Rulten}, {Sadeh}, {Santander}, {Scott}, {Sembroski}, {Shahinyan}, {Tyler},
  {Wakely}, {Weinstein}, {Wells}, {Wilcox}, {Wilhelm}, {Williams},
  {Williamson}, {Zitzer}, {VERITAS Collaboration}, \&
  {Kaur}}]{2018ApJ...861L..20A}
{Abeysekara}, A.~U., {Archer}, A., {Benbow}, W., {et~al.} 2018, \apjl, 861,
  L20, \dodoi{10.3847/2041-8213/aad053}

\bibitem[{{Acciari} {et~al.}(2008){Acciari}, {Aliu}, {Beilicke}, {Benbow},
  {B{\"o}ttcher}, {Bradbury}, {Buckley}, {Bugaev}, {Butt}, {Celik}, {Cesarini},
  {Ciupik}, {Chow}, {Cogan}, {Colin}, {Cui}, {Daniel}, {Ergin}, {Falcone},
  {Fegan}, {Finley}, {Finnegan}, {Fortin}, {Fortson}, {Furniss}, {Gall},
  {Gillanders}, {Grube}, {Guenette}, {Gyuk}, {Hanna}, {Hays}, {Holder},
  {Horan}, {Hui}, {Humensky}, {Imran}, {Kaaret}, {Karlsson}, {Kertzman},
  {Kieda}, {Konopelko}, {Krawczynski}, {Krennrich}, {Lang}, {LeBohec}, {Lee},
  {Maier}, {McCann}, {McCutcheon}, {Moriarty}, {Mukherjee}, {Nagai}, {Niemiec},
  {Ong}, {Pandel}, {Perkins}, {Petry}, {Pohl}, {Quinn}, {Ragan}, {Reyes},
  {Reynolds}, {Roache}, {Rose}, {Schroedter}, {Sembroski}, {Smith}, {Steele},
  {Swordy}, {Toner}, {Vassiliev}, {Wagner}, {Wakely}, {Ward}, {Weekes},
  {Weinstein}, {White}, {Williams}, {Wissel}, {Wood}, \&
  {Zitzer}}]{2008ApJ...684L..73A}
{Acciari}, V.~A., {Aliu}, E., {Beilicke}, M., {et~al.} 2008, \apjl, 684, L73,
  \dodoi{10.1086/592244}

\bibitem[{{Acciari} {et~al.}(2009{\natexlab{a}}){Acciari}, {Aliu}, {Arlen},
  {Beilicke}, {Benbow}, {Bradbury}, {Buckley}, {Bugaev}, {Butt}, {Byrum},
  {Celik}, {Cesarini}, {Ciupik}, {Chow}, {Cogan}, {Colin}, {Cui}, {Daniel},
  {Ergin}, {Falcone}, {Fegan}, {Finley}, {Fortin}, {Fortson}, {Furniss},
  {Gillanders}, {Grube}, {Guenette}, {Gyuk}, {Hanna}, {Hays}, {Holder},
  {Horan}, {Hui}, {Humensky}, {Imran}, {Kaaret}, {Karlsson}, {Kertzman},
  {Kieda}, {Kildea}, {Konopelko}, {Krawczynski}, {Krennrich}, {Lang},
  {LeBohec}, {Maier}, {McCann}, {McCutcheon}, {Moriarty}, {Mukherjee}, {Nagai},
  {Niemiec}, {Ong}, {Pandel}, {Perkins}, {Pohl}, {Quinn}, {Ragan}, {Reyes},
  {Reynolds}, {Rose}, {Schroedter}, {Sembroski}, {Smith}, {Steele}, {Swordy},
  {Toner}, {Valcarcel}, {Vassiliev}, {Wagner}, {Wakely}, {Ward}, {Weekes},
  {Weinstein}, {White}, {Williams}, {Wissel}, {Wood}, \&
  {Zitzer}}]{2009ApJ...695.1370A}
{Acciari}, V.~A., {Aliu}, E., {Arlen}, T., {et~al.} 2009{\natexlab{a}}, \apj,
  695, 1370, \dodoi{10.1088/0004-637X/695/2/1370}

\bibitem[{{Acciari} {et~al.}(2009{\natexlab{b}}){Acciari}, {Aliu}, {Arlen},
  {Beilicke}, {Benbow}, {B{\"o}ttcher}, {Bradbury}, {Buckley}, {Bugaev},
  {Butt}, {Byrum}, {Cannon}, {Celik}, {Cesarini}, {Chow}, {Ciupik}, {Cogan},
  {Cui}, {Daniel}, {Dickherber}, {Ergin}, {Falcone}, {Fegan}, {Finley},
  {Fortin}, {Fortson}, {Furniss}, {Gall}, {Gibbs}, {Gillanders}, {Godambe},
  {Grube}, {Guenette}, {Gyuk}, {Hanna}, {Hays}, {Holder}, {Horan}, {Hui},
  {Humensky}, {Imran}, {Kaaret}, {Karlsson}, {Kertzman}, {Kieda}, {Kildea},
  {Konopelko}, {Krawczynski}, {Krennrich}, {Lang}, {LeBohec}, {Maier},
  {McCann}, {McCutcheon}, {Millis}, {Moriarty}, {Mukherjee}, {Nagai}, {Ong},
  {Otte}, {Pandel}, {Perkins}, {Petry}, {Pizlo}, {Pohl}, {Quinn}, {Ragan},
  {Reyes}, {Reynolds}, {Roache}, {Rose}, {Schroedter}, {Sembroski}, {Smith},
  {Steele}, {Swordy}, {Theiling}, {Toner}, {Varlotta}, {Vassiliev}, {Wagner},
  {Wakely}, {Ward}, {Weekes}, {Weinstein}, {Williams}, {Wissel}, {Wood}, \&
  {Zitzer}}]{2009ApJ...693L.104A}
---. 2009{\natexlab{b}}, \apjl, 693, L104, \dodoi{10.1088/0004-637X/693/2/L104}

\bibitem[{{Acciari} {et~al.}(2010{\natexlab{a}}){Acciari}, {Aliu}, {Arlen},
  {Aune}, {Bautista}, {Beilicke}, {Benbow}, {B{\"o}ttcher}, {Boltuch},
  {Bradbury}, {Buckley}, {Bugaev}, {Byrum}, {Cannon}, {Cesarini}, {Ciupik},
  {Cui}, {Dickherber}, {Duke}, {Falcone}, {Finley}, {Finnegan}, {Fortson},
  {Furniss}, {Galante}, {Gall}, {Gibbs}, {Gillanders}, {Godambe}, {Grube},
  {Guenette}, {Gyuk}, {Hanna}, {Holder}, {Hui}, {Humensky}, {Imran}, {Kaaret},
  {Karlsson}, {Kertzman}, {Kieda}, {Konopelko}, {Krawczynski}, {Krennrich},
  {Lang}, {Lamerato}, {LeBohec}, {Maier}, {McArthur}, {McCann}, {McCutcheon},
  {Moriarty}, {Mukherjee}, {Ong}, {Otte}, {Pandel}, {Perkins}, {Petry},
  {Pichel}, {Pohl}, {Quinn}, {Ragan}, {Reyes}, {Reynolds}, {Roache}, {Rose},
  {Roustazadeh}, {Schroedter}, {Sembroski}, {Senturk}, {Smith}, {Steele},
  {Swordy}, {Te{\v{s}}i{\'c}}, {Theiling}, {Thibadeau}, {Varlotta},
  {Vassiliev}, {Vincent}, {Wagner}, {Wakely}, {Ward}, {Weekes}, {Weinstein},
  {Weisgarber}, {Williams}, {Wissel}, {Wood}, {Zitzer}, {Ackermann}, {Ajello},
  {Antolini}, {Baldini}, {Ballet}, {Barbiellini}, {Bastieri}, {Bechtol},
  {Bellazzini}, {Berenji}, {Blandford}, {Bloom}, {Bonamente}, {Borgland},
  {Bouvier}, {Bregeon}, {Brigida}, {Bruel}, {Buehler}, {Buson}, {Caliandro},
  {Cameron}, {Caraveo}, {Carrigan}, {Casandjian}, {Cavazzuti}, {Cecchi},
  {{\c{C}}elik}, {Charles}, {Chekhtman}, {Cheung}, {Chiang}, {Ciprini},
  {Claus}, {Cohen-Tanugi}, {Conrad}, {Dermer}, {de Palma}, {Silva}, {Drell},
  {Dubois}, {Dumora}, {Farnier}, {Favuzzi}, {Fegan}, {Fortin}, {Frailis},
  {Fukazawa}, {Funk}, {Fusco}, {Gargano}, {Gasparrini}, {Gehrels}, {Germani},
  {Giebels}, {Giglietto}, {Giordano}, {Giroletti}, {Glanzman}, {Godfrey},
  {Grenier}, {Grove}, {Guiriec}, {Hays}, {Horan}, {Hughes}, {J{\'o}hannesson},
  {Johnson}, {Johnson}, {Kamae}, {Katagiri}, {Kataoka}, {Kn{\"o}dlseder},
  {Kuss}, {Lande}, {Latronico}, {Lee}, {Llena Garde}, {Longo}, {Loparco},
  {Lott}, {Lovellette}, {Lubrano}, {Makeev}, {Mazziotta}, {Michelson},
  {Mitthumsiri}, {Mizuno}, {Moiseev}, {Monte}, {Monzani}, {Morselli},
  {Moskalenko}, {Murgia}, {Nolan}, {Norris}, {Nuss}, {Ohno}, {Ohsugi},
  {Omodei}, {Orlando}, {Ormes}, {Paneque}, {Panetta}, {Pelassa}, {Pepe},
  {Pesce-Rollins}, {Piron}, {Porter}, {Rain{\`o}}, {Rando}, {Razzano},
  {Reimer}, {Reimer}, {Ripken}, {Rodriguez}, {Roth}, {Sadrozinski}, {Sanchez},
  {Sander}, {Scargle}, {Sgr{\`o}}, {Siskind}, {Smith}, {Spandre}, {Spinelli},
  {Strickman}, {Suson}, {Takahashi}, {Tanaka}, {Thayer}, {Thayer}, {Thompson},
  {Tibaldo}, {Torres}, {Tosti}, {Tramacere}, {Usher}, {Vasileiou}, {Vilchez},
  {Vitale}, {Waite}, {Wang}, {Winer}, {Wood}, {Yang}, {Ylinen}, \&
  {Ziegler}}]{2010ApJ...715L..49A}
---. 2010{\natexlab{a}}, \apjl, 715, L49, \dodoi{10.1088/2041-8205/715/1/L49}

\bibitem[{{Acciari} {et~al.}(2010{\natexlab{b}}){Acciari}, {Aliu}, {Arlen},
  {Aune}, {Bautista}, {Beilicke}, {Benbow}, {B{\"o}ttcher}, {Boltuch},
  {Bradbury}, {Buckley}, {Bugaev}, {Byrum}, {Cannon}, {Cesarini}, {Chow},
  {Ciupik}, {Cogan}, {Cui}, {Duke}, {Falcone}, {Finley}, {Finnegan}, {Fortson},
  {Furniss}, {Galante}, {Gall}, {Gillanders}, {Godambe}, {Grube}, {Guenette},
  {Gyuk}, {Hanna}, {Holder}, {Hui}, {Humensky}, {Kaaret}, {Karlsson},
  {Kertzman}, {Kieda}, {Konopelko}, {Krawczynski}, {Krennrich}, {Lang},
  {LeBohec}, {Maier}, {McArthur}, {McCann}, {McCutcheon}, {Millis}, {Moriarty},
  {Nagai}, {Ong}, {Otte}, {Pandel}, {Perkins}, {Pichel}, {Pohl}, {Quinn},
  {Ragan}, {Reyes}, {Reynolds}, {Roache}, {Rose}, {Schroedter}, {Sembroski},
  {Senturk}, {Smith}, {Steele}, {Swordy}, {Theiling}, {Thibadeau}, {Varlotta},
  {Vassiliev}, {Vincent}, {Wagner}, {Wakely}, {Ward}, {Weekes}, {Weinstein},
  {Weisgarber}, {Williams}, {Wissel}, {Wood}, {Zitzer}, {VERITAS
  Collaboration}, {Abdo}, {Ackermann}, {Ajello}, {Baldini}, {Ballet},
  {Barbiellini}, {Bastieri}, {Baughman}, {Bechtol}, {Bellazzini}, {Berenji},
  {Blandford}, {Bloom}, {Bonamente}, {Borgland}, {Bregeon}, {Brez}, {Brigida},
  {Bruel}, {Burnett}, {Caliandro}, {Cameron}, {Caraveo}, {Casandjian},
  {Cavazzuti}, {Cecchi}, {{\c{C}}elik}, {Chekhtman}, {Cheung}, {Chiang},
  {Ciprini}, {Claus}, {Cohen-Tanugi}, {Conrad}, {Cutini}, {Dermer}, {de
  Angelis}, {de Palma}, {do Couto e Silva}, {Drell}, {Drlica-Wagner}, {Dubois},
  {Dumora}, {Farnier}, {Favuzzi}, {Fegan}, {Focke}, {Fortin}, {Frailis},
  {Fukazawa}, {Fusco}, {Gargano}, {Gasparrini}, {Gehrels}, {Germani},
  {Giebels}, {Giglietto}, {Giommi}, {Giordano}, {Glanzman}, {Godfrey},
  {Grenier}, {Grove}, {Guillemot}, {Guiriec}, {Hanabata}, {Hays}, {Hughes},
  {Jackson}, {J{\'o}hannesson}, {Johnson}, {Johnson}, {Kamae}, {Katagiri},
  {Kataoka}, {Kawai}, {Kerr}, {Kn{\"o}dlseder}, {Kocian}, {Kuss}, {Lande},
  {Latronico}, {Longo}, {Loparco}, {Lott}, {Lovellette}, {Lubrano}, {Madejski},
  {Makeev}, {Mazziotta}, {McEnery}, {Meurer}, {Michelson}, {Mitthumsiri},
  {Mizuno}, {Moiseev}, {Monte}, {Monzani}, {Morselli}, {Moskalenko}, {Murgia},
  {Nolan}, {Norris}, {Nuss}, {Ohsugi}, {Omodei}, {Orlando}, {Ormes}, {Paneque},
  {Parent}, {Pelassa}, {Pepe}, {Pesce-Rollins}, {Piron}, {Porter}, {Rain{\`o}},
  {Rando}, {Razzano}, {Reimer}, {Reimer}, {Reposeur}, {Rodriguez}, {Roth},
  {Ryde}, {Sadrozinski}, {Sanchez}, {Sander}, {Saz Parkinson}, {Scargle},
  {Sgr{\`o}}, {Shaw}, {Siskind}, {Smith}, {Spandre}, {Spinelli}, {Strickman},
  {Suson}, {Tajima}, {Takahashi}, {Tanaka}, {Thayer}, {Thayer}, {Thompson},
  {Tibaldo}, {Torres}, {Tosti}, {Tramacere}, {Uchiyama}, {Usher}, {Vasileiou},
  {Vilchez}, {Vitale}, {Waite}, {Wang}, {Winer}, {Wood}, {Ylinen}, {Ziegler},
  {Fermi LAT Collaboration}, {Barber}, \& {Terndrup}}]{2010ApJ...708L.100A}
---. 2010{\natexlab{b}}, \apjl, 708, L100, \dodoi{10.1088/2041-8205/708/2/L100}

\bibitem[{{Acciari} {et~al.}(2011){Acciari}, {Arlen}, {Aune}, {Beilicke},
  {Benbow}, {B{\"o}ttcher}, {Boltuch}, {Bradbury}, {Buckley}, {Bugaev},
  {Cannon}, {Cesarini}, {Ciupik}, {Cui}, {Dickherber}, {Duke}, {Errando},
  {Falcone}, {Finley}, {Finnegan}, {Fortson}, {Furniss}, {Galante}, {Gall},
  {Godambe}, {Grube}, {Guenette}, {Gyuk}, {Hanna}, {Holder}, {Huang}, {Hui},
  {Humensky}, {Imran}, {Kaaret}, {Karlsson}, {Kertzman}, {Kieda}, {Konopelko},
  {Krawczynski}, {Krennrich}, {Madhavan}, {Maier}, {McArthur}, {McCann},
  {Moriarty}, {Ong}, {Otte}, {Pandel}, {Perkins}, {Pichel}, {Pohl}, {Quinn},
  {Ragan}, {Reyes}, {Reynolds}, {Roache}, {Rose}, {Schroedter}, {Sembroski},
  {Steele}, {Swordy}, {Theiling}, {Thibadeau}, {Varlotta}, {Vassiliev},
  {Vincent}, {Wakely}, {Ward}, {Weekes}, {Weinstein}, {Weisgarber}, {Williams},
  {Wood}, {Zitzer}, {VERITAS Collaboration}, {Aleksi{\'c}}, {Antonelli},
  {Antoranz}, {Backes}, {Barrio}, {Bastieri}, {Becerra Gonz{\'a}lez},
  {Bednarek}, {Berdyugin}, {Berger}, {Bernardini}, {Biland}, {Blanch}, {Bock},
  {Boller}, {Bonnoli}, {Bordas}, {Borla Tridon}, {Bosch-Ramon}, {Bose},
  {Braun}, {Bretz}, {Camara}, {Carmona}, {Carosi}, {Colin}, {Colombo},
  {Contreras}, {Cortina}, {Covino}, {Dazzi}, {De Angelis}, {De Cea del Pozo},
  {De Lotto}, {De Maria}, {De Sabata}, {Delgado Mendez}, {Diago Ortega},
  {Doert}, {Dom{\'\i}nguez}, {Dominis Prester}, {Dorner}, {Doro}, {Elsaesser},
  {Errando}, {Ferenc}, {Fonseca}, {Font}, {Garc{\'\i}a L{\'o}pez},
  {Garczarczyk}, {Gaug}, {Giavitto}, {Godinovi{\'c}}, {Hadasch}, {Herrero},
  {Hildebrand}, {H{\"o}hne-M{\"o}nch}, {Hose}, {Hrupec}, {Jogler}, {Klepser},
  {Kr{\"a}henb{\"u}hl}, {Kranich}, {Krause}, {La Barbera}, {Leonardo},
  {Lindfors}, {Lombardi}, {Longo}, {L{\'o}pez}, {Lorenz}, {Majumdar},
  {Makariev}, {Maneva}, {Mankuzhiyil}, {Mannheim}, {Maraschi}, {Mariotti},
  {Mart{\'\i}nez}, {Mazin}, {Meucci}, {Miranda}, {Mirzoyan}, {Miyamoto},
  {Mold{\'o}n}, {Moralejo}, {Nieto}, {Nilsson}, {Orito}, {Oya}, {Paoletti},
  {Paredes}, {Partini}, {Pasanen}, {Pauss}, {Pegna}, {Perez-Torres}, {Persic},
  {Peruzzo}, {Pochon}, {Prada}, {Prada Moroni}, {Prandini}, {Puchades},
  {Puljak}, {Reichardt}, {Reinthal}, {Rhode}, {Rib{\'o}}, {Rico}, {Rissi},
  {R{\"u}gamer}, {Saggion}, {Saito}, {Saito}, {Salvati}, {S{\'a}nchez-Conde},
  {Satalecka}, {Scalzotto}, {Scapin}, {Schultz}, {Schweizer}, {Shayduk},
  {Shore}, {Sierpowska-Bartosik}, {Sillanp{\"a}{\"a}}, {Sitarek}, {Sobczynska},
  {Spanier}, {Spiro}, {Stamerra}, {Steinke}, {Storz}, {Strah}, {Struebig},
  {Suric}, {Takalo}, {Tavecchio}, {Temnikov}, {Terzi{\'c}}, {Tescaro},
  {Teshima}, {Torres}, {Vankov}, {Wagner}, {Weitzel}, {Zabalza}, {Zandanel},
  {Zanin}, {MAGIC Collaboration}, {Paneque}, \&
  {Hayashida}}]{2011ApJ...729....2A}
{Acciari}, V.~A., {Arlen}, T., {Aune}, T., {et~al.} 2011, \apj, 729, 2,
  \dodoi{10.1088/0004-637X/729/1/2}

\bibitem[{{Acciari} {et~al.}(2020){Acciari}, {Ansoldi}, {Antonelli}, {Engels},
  {Asano}, {Baack}, {Babi{\'c}}, {Banerjee}, {Barres de Almeida}, {Barrio},
  {Becerra Gonz{\'a}lez}, {Bednarek}, {Bellizzi}, {Bernardini}, {Berti},
  {Besenrieder}, {Bhattacharyya}, {Bigongiari}, {Biland}, {Blanch}, {Bonnoli},
  {Bo{\v{s}}njak}, {Busetto}, {Carosi}, {Ceribella}, {Cerruti}, {Chai},
  {Chilingaryan}, {Cikota}, {Colak}, {Colin}, {Colombo}, {Contreras},
  {Cortina}, {Covino}, {D'Elia}, {Da Vela}, {Dazzi}, {De Angelis}, {De Lotto},
  {Delfino}, {Delgado}, {Depaoli}, {Di Pierro}, {Di Venere}, {Do Souto
  Espi{\~n}eira}, {Dominis Prester}, {Donini}, {Dorner}, {Doro}, {Elsaesser},
  {Ramazani}, {Fattorini}, {Ferrara}, {Fidalgo}, {Foffano}, {Fonseca}, {Font},
  {Fruck}, {Fukami}, {Garc{\'\i}a L{\'o}pez}, {Garczarczyk}, {Gasparyan},
  {Gaug}, {Giglietto}, {Giordano}, {Godinovi{\'c}}, {Green}, {Guberman},
  {Hadasch}, {Hahn}, {Herrera}, {Hoang}, {Hrupec}, {H{\"u}tten}, {Inada},
  {Inoue}, {Ishio}, {Iwamura}, {Jouvin}, {Kerszberg}, {Kubo}, {Kushida},
  {Lamastra}, {Lelas}, {Leone}, {Lindfors}, {Lombardi}, {Longo}, {L{\'o}pez},
  {L{\'o}pez-Coto}, {L{\'o}pez-Oramas}, {Loporchio}, {Machado de Oliveira
  Fraga}, {Maggio}, {Majumdar}, {Makariev}, {Mallamaci}, {Maneva}, {Manganaro},
  {Mannheim}, {Maraschi}, {Mariotti}, {Mart{\'\i}nez}, {Mazin},
  {Mi{\'c}anovi{\'c}}, {Miceli}, {Minev}, {Miranda}, {Mirzoyan}, {Molina},
  {Moralejo}, {Morcuende}, {Moreno}, {Moretti}, {Munar-Adrover}, {Neustroev},
  {Nigro}, {Nilsson}, {Ninci}, {Nishijima}, {Noda}, {Nogu{\'e}s}, {Nozaki},
  {Paiano}, {Palatiello}, {Paneque}, {Paoletti}, {Paredes}, {Pe{\~n}il},
  {Peresano}, {Persic}, {Prada Moroni}, {Prandini}, {Puljak}, {Rhode},
  {Rib{\'o}}, {Rico}, {Righi}, {Rugliancich}, {Saha}, {Sahakyan}, {Saito},
  {Sakurai}, {Satalecka}, {Schmidt}, {Schweizer}, {Sitarek},
  {{\v{S}}nidari{\'c}}, {Sobczynska}, {Somero}, {Stamerra}, {Strom}, {Strzys},
  {Suda}, {Suri{\'c}}, {Takahashi}, {Tavecchio}, {Temnikov}, {Terzi{\'c}},
  {Teshima}, {Torres-Alb{\`a}}, {Tosti}, {Vagelli}, {van Scherpenberg},
  {Vanzo}, {Vazquez Acosta}, {Vigorito}, {Vitale}, {Vovk}, {Will}, {Zari{\'c}},
  {Arcaro}, {Carosi}, {D'Ammando}, {Tombesi}, \&
  {Lohfink}}]{2020ApJS..247...16A}
{Acciari}, V.~A., {Ansoldi}, S., {Antonelli}, L.~A., {et~al.} 2020, \apjs, 247,
  16, \dodoi{10.3847/1538-4365/ab5b98}

\bibitem[{{Acciari} {et~al.}(2021){Acciari}, {Ansoldi}, {Antonelli}, {Arbet
  Engels}, {Artero}, {Asano}, {Baack}, {Babi{\'c}}, {Baquero}, {Barres de
  Almeida}, {Barrio}, {Batkovi{\'c}}, {Becerra Gonz{\'a}lez}, {Bednarek},
  {Bellizzi}, {Bernardini}, {Bernardos}, {Berti}, {Besenrieder},
  {Bhattacharyya}, {Bigongiari}, {Biland}, {Blanch}, {Bo{\v{s}}njak},
  {Busetto}, {Carosi}, {Ceribella}, {Cerruti}, {Chai}, {Chilingarian},
  {Cikota}, {Colak}, {Colombo}, {Contreras}, {Cortina}, {Covino}, {D'Amico},
  {D'Elia}, {da Vela}, {Dazzi}, {de Angelis}, {de Lotto}, {Delfino}, {Delgado},
  {Delgado Mendez}, {Depaoli}, {Pierro}, {Venere}, {Do Souto Espi{\~n}eira},
  {Dominis Prester}, {Donini}, {Dorner}, {Doro}, {Elsaesser}, {Fallah
  Ramazani}, {Fattorini}, {Ferrara}, {Fonseca}, {Font}, {Fruck}, {Fukami},
  {Garc{\'\i}a L{\'o}pez}, {Garczarczyk}, {Gasparyan}, {Gaug}, {Giglietto},
  {Giordano}, {Gliwny}, {Godinovi{\'c}}, {Green}, {Green}, {Hadasch}, {Hahn},
  {Heckmann}, {Herrera}, {Hoang}, {Hrupec}, {H{\"u}tten}, {Inada}, {Inoue},
  {Ishio}, {Iwamura}, {Jim{\'e}nez}, {Jormanainen}, {Jouvin}, {Kajiwara},
  {Karjalainen}, {Kerszberg}, {Kobayashi}, {Kubo}, {Kushida}, {Lamastra},
  {Lelas}, {Leone}, {Lindfors}, {Lombardi}, {Longo}, {L{\'o}pez-Coto},
  {L{\'o}pez-Moya}, {L{\'o}pez-Oramas}, {Loporchio}, {Machado de Oliveira
  Fraga}, {Maggio}, {Majumdar}, {Makariev}, {Mallamaci}, {Maneva}, {Manganaro},
  {Mannheim}, {Maraschi}, {Mariotti}, {Mart{\'\i}nez}, {Mazin}, {Menchiari},
  {Mender}, {Mi{\'c}anovi{\'c}}, {Miceli}, {Miener}, {Minev}, {Miranda},
  {Mirzoyan}, {Molina}, {Moralejo}, {Morcuende}, {Moreno}, {Moretti},
  {Neustroev}, {Nigro}, {Nilsson}, {Nishijima}, {Noda}, {Nozaki}, {Ohtani},
  {Oka}, {Otero-Santos}, {Paiano}, {Palatiello}, {Paneque}, {Paoletti},
  {Paredes}, {Pavleti{\'c}}, {Pe{\~n}il}, {Perennes}, {Persic}, {Prada Moroni},
  {Prandini}, {Priyadarshi}, {Puljak}, {Rhode}, {Rib{\'o}}, {Rico}, {Righi},
  {Rugliancich}, {Saha}, {Sahakyan}, {Saito}, {Sakurai}, {Satalecka},
  {Saturni}, {Schleicher}, {Schmidt}, {Schweizer}, {Sitarek},
  {{\v{S}}nidari{\'c}}, {Sobczynska}, {Spolon}, {Stamerra}, {Strom}, {Strzys},
  {Suda}, {Suri{\'c}}, {Takahashi}, {Tavecchio}, {Temnikov}, {Terzi{\'c}},
  {Teshima}, {Tosti}, {Truzzi}, {Tutone}, {Ubach}, {van Scherpenberg}, {Vanzo},
  {Vazquez Acosta}, {Ventura}, {Verguilov}, {Vigorito}, {Vitale}, {Vovk},
  {Will}, {Wunderlich}, {Zari{\'c}}, {Bissaldi}, {Bonnoli}, {Cutini},
  {D'Ammando}, {Nabizadeh}, {Marchini}, {Orienti}, \& {MAGIC
  Collaboration}}]{2021MNRAS.507.1528A}
---. 2021, \mnras, 507, 1528, \dodoi{10.1093/mnras/stab1994}

\bibitem[{{Ackermann} {et~al.}(2011){Ackermann}, {Ajello}, {Allafort},
  {Antolini}, {Atwood}, {Axelsson}, {Baldini}, {Ballet}, {Barbiellini},
  {Bastieri}, {Bechtol}, {Bellazzini}, {Berenji}, {Blandford}, {Bloom},
  {Bonamente}, {Borgland}, {Bottacini}, {Bouvier}, {Bregeon}, {Brigida},
  {Bruel}, {Buehler}, {Burnett}, {Buson}, {Caliandro}, {Cameron}, {Caraveo},
  {Casandjian}, {Cavazzuti}, {Cecchi}, {Charles}, {Cheung}, {Chiang},
  {Ciprini}, {Claus}, {Cohen-Tanugi}, {Conrad}, {Costamante}, {Cutini}, {de
  Angelis}, {de Palma}, {Dermer}, {Digel}, {Silva}, {Drell}, {Dubois},
  {Escande}, {Favuzzi}, {Fegan}, {Ferrara}, {Finke}, {Focke}, {Fortin},
  {Frailis}, {Fukazawa}, {Funk}, {Fusco}, {Gargano}, {Gasparrini}, {Gehrels},
  {Germani}, {Giebels}, {Giglietto}, {Giommi}, {Giordano}, {Giroletti},
  {Glanzman}, {Godfrey}, {Grenier}, {Grove}, {Guiriec}, {Gustafsson},
  {Hadasch}, {Hayashida}, {Hays}, {Healey}, {Horan}, {Hou}, {Hughes},
  {Iafrate}, {J{\'o}hannesson}, {Johnson}, {Johnson}, {Kamae}, {Katagiri},
  {Kataoka}, {Kn{\"o}dlseder}, {Kuss}, {Lande}, {Larsson}, {Latronico},
  {Longo}, {Loparco}, {Lott}, {Lovellette}, {Lubrano}, {Madejski}, {Mazziotta},
  {McConville}, {McEnery}, {Michelson}, {Mitthumsiri}, {Mizuno}, {Moiseev},
  {Monte}, {Monzani}, {Moretti}, {Morselli}, {Moskalenko}, {Murgia},
  {Nakamori}, {Naumann-Godo}, {Nolan}, {Norris}, {Nuss}, {Ohno}, {Ohsugi},
  {Okumura}, {Omodei}, {Orienti}, {Orlando}, {Ormes}, {Ozaki}, {Paneque},
  {Parent}, {Pesce-Rollins}, {Pierbattista}, {Piranomonte}, {Piron}, {Pivato},
  {Porter}, {Rain{\`o}}, {Rando}, {Razzano}, {Razzaque}, {Reimer}, {Reimer},
  {Ritz}, {Rochester}, {Romani}, {Roth}, {Sanchez}, {Sbarra}, {Scargle},
  {Schalk}, {Sgr{\`o}}, {Shaw}, {Siskind}, {Spandre}, {Spinelli}, {Strong},
  {Suson}, {Tajima}, {Takahashi}, {Takahashi}, {Tanaka}, {Thayer}, {Thayer},
  {Thompson}, {Tibaldo}, {Tinivella}, {Torres}, {Tosti}, {Troja}, {Uchiyama},
  {Vandenbroucke}, {Vasileiou}, {Vianello}, {Vitale}, {Waite}, {Wallace},
  {Wang}, {Winer}, {Wood}, {Wood}, \& {Zimmer}}]{2011ApJ...743..171A}
{Ackermann}, M., {Ajello}, M., {Allafort}, A., {et~al.} 2011, \apj, 743, 171,
  \dodoi{10.1088/0004-637X/743/2/171}

\bibitem[{{Ackermann} {et~al.}(2012){Ackermann}, {Ajello}, {Allafort},
  {Schady}, {Baldini}, {Ballet}, {Barbiellini}, {Bastieri}, {Bellazzini},
  {Blandford}, {Bloom}, {Borgland}, {Bottacini}, {Bouvier}, {Bregeon},
  {Brigida}, {Bruel}, {Buehler}, {Buson}, {Caliandro}, {Cameron}, {Caraveo},
  {Cavazzuti}, {Cecchi}, {Charles}, {Chaves}, {Chekhtman}, {Cheung}, {Chiang},
  {Chiaro}, {Ciprini}, {Claus}, {Cohen-Tanugi}, {Conrad}, {Cutini},
  {D'Ammando}, {de Palma}, {Dermer}, {Digel}, {do Couto e Silva},
  {Dom{\'\i}nguez}, {Drell}, {Drlica-Wagner}, {Favuzzi}, {Fegan}, {Focke},
  {Franckowiak}, {Fukazawa}, {Funk}, {Fusco}, {Gargano}, {Gasparrini},
  {Gehrels}, {Germani}, {Giglietto}, {Giordano}, {Giroletti}, {Glanzman},
  {Godfrey}, {Grenier}, {Grove}, {Guiriec}, {Gustafsson}, {Hadasch},
  {Hayashida}, {Hays}, {Jackson}, {Jogler}, {Kataoka}, {Kn{\"o}dlseder},
  {Kuss}, {Lande}, {Larsson}, {Latronico}, {Longo}, {Loparco}, {Lovellette},
  {Lubrano}, {Mazziotta}, {McEnery}, {Mehault}, {Michelson}, {Mizuno}, {Monte},
  {Monzani}, {Morselli}, {Moskalenko}, {Murgia}, {Tramacere}, {Nuss},
  {Greiner}, {Ohno}, {Ohsugi}, {Omodei}, {Orienti}, {Orlando}, {Ormes},
  {Paneque}, {Perkins}, {Pesce-Rollins}, {Piron}, {Pivato}, {Porter},
  {Rain{\`o}}, {Rando}, {Razzano}, {Razzaque}, {Reimer}, {Reimer}, {Reyes},
  {Ritz}, {Rau}, {Romoli}, {Roth}, {S{\'a}nchez-Conde}, {Sanchez}, {Scargle},
  {Sgr{\`o}}, {Siskind}, {Spandre}, {Spinelli}, {Stawarz}, {Suson},
  {Takahashi}, {Tanaka}, {Thayer}, {Thompson}, {Tibaldo}, {Tinivella},
  {Torres}, {Tosti}, {Troja}, {Usher}, {Vandenbroucke}, {Vasileiou},
  {Vianello}, {Vitale}, {Waite}, {Winer}, {Wood}, \&
  {Wood}}]{2012Sci...338.1190A}
---. 2012, Science, 338, 1190, \dodoi{10.1126/science.1227160}

\bibitem[{{Aharonian} {et~al.}(2006){Aharonian}, {Akhperjanian}, {Bazer-Bachi},
  {Beilicke}, {Benbow}, {Berge}, {Bernl{\"o}hr}, {Boisson}, {Bolz}, {Borrel},
  {Braun}, {Breitling}, {Brown}, {B{\"u}hler}, {B{\"u}sching}, {Carrigan},
  {Chadwick}, {Chounet}, {Cornils}, {Costamante}, {Degrange}, {Dickinson},
  {Djannati-Ata{\"\i}}, {O'C. Drury}, {Dubus}, {Egberts}, {Emmanoulopoulos},
  {Espigat}, {Feinstein}, {Ferrero}, {Fontaine}, {Funk}, {Funk}, {Gallant},
  {Giebels}, {Glicenstein}, {Goret}, {Hadjichristidis}, {Hauser}, {Hauser},
  {Heinzelmann}, {Henri}, {Hermann}, {Hinton}, {Hofmann}, {Holleran}, {Horns},
  {Jacholkowska}, {de Jager}, {Kh{\'e}lifi}, {Komin}, {Konopelko}, {Latham},
  {Le Gallou}, {Lemi{\`e}re}, {Lemoine-Goumard}, {Lohse}, {Martin},
  {Martineau-Huynh}, {Marcowith}, {Masterson}, {McComb}, {de Naurois},
  {Nedbal}, {Nolan}, {Noutsos}, {Orford}, {Osborne}, {Ouchrif}, {Panter},
  {Pelletier}, {Pita}, {P{\"u}hlhofer}, {Punch}, {Raubenheimer}, {Raue},
  {Rayner}, {Reimer}, {Reimer}, {Ripken}, {Rob}, {Rolland}, {Rowell},
  {Sahakian}, {Saug{\'e}}, {Schlenker}, {Schlickeiser}, {Schwanke}, {Sol},
  {Spangler}, {Spanier}, {Steenkamp}, {Stegmann}, {Superina}, {Tavernet},
  {Terrier}, {Th{\'e}oret}, {Tluczykont}, {van Eldik}, {Vasileiadis}, {Venter},
  {Vincent}, {V{\"o}lk}, {Wagner}, \& {Ward}}]{2006AA...455..461A}
{Aharonian}, F., {Akhperjanian}, A.~G., {Bazer-Bachi}, A.~R., {et~al.} 2006,
  \aap, 455, 461, \dodoi{10.1051/0004-6361:20054732}

\bibitem[{{Aharonian} {et~al.}(2007{\natexlab{a}}){Aharonian}, {Akhperjanian},
  {Bazer-Bachi}, {Beilicke}, {Benbow}, {Berge}, {Bernl{\"o}hr}, {Boisson},
  {Bolz}, {Borrel}, {Braun}, {Brion}, {Brown}, {B{\"u}hler}, {B{\"u}sching},
  {Boutelier}, {Carrigan}, {Chadwick}, {Chounet}, {Coignet}, {Cornils},
  {Costamante}, {Degrange}, {Dickinson}, {Djannati-Ata{\"\i}}, {O'C. Drury},
  {Dubus}, {Egberts}, {Emmanoulopoulos}, {Espigat}, {Farnier}, {Feinstein},
  {Ferrero}, {Fiasson}, {Fontaine}, {Funk}, {Funk}, {F{\"u}{\ss}ling},
  {Gallant}, {Giebels}, {Glicenstein}, {Gl{\"u}ck}, {Goret}, {Hadjichristidis},
  {Hauser}, {Hauser}, {Heinzelmann}, {Henri}, {Hermann}, {Hinton}, {Hoffmann},
  {Hofmann}, {Holleran}, {Hoppe}, {Horns}, {Jacholkowska}, {de Jager},
  {Kendziorra}, {Kerschhaggl}, {Kh{\'e}lifi}, {Komin}, {Kosack}, {Lamanna},
  {Latham}, {Le Gallou}, {Lemi{\`e}re}, {Lemoine-Goumard}, {Lohse}, {Martin},
  {Martineau-Huynh}, {Marcowith}, {Masterson}, {Maurin}, {McComb}, {Moulin},
  {de Naurois}, {Nedbal}, {Nolan}, {Noutsos}, {Olive}, {Orford}, {Osborne},
  {Panter}, {Pelletier}, {Petrucci}, {Pita}, {P{\"u}hlhofer}, {Punch},
  {Ranchon}, {Raubenheimer}, {Raue}, {Rayner}, {Ripken}, {Rob}, {Rolland},
  {Rosier-Lees}, {Rowell}, {Sahakian}, {Santangelo}, {Saug{\'e}}, {Schlenker},
  {Schlickeiser}, {Schr{\"o}der}, {Schwanke}, {Schwarzburg}, {Schwemmer},
  {Shalchi}, {Sol}, {Spangler}, {Spanier}, {Steenkamp}, {Stegmann}, {Superina},
  {Tam}, {Tavernet}, {Terrier}, {Tluczykont}, {van Eldik}, {Vasileiadis},
  {Venter}, {Vialle}, {Vincent}, {V{\"o}lk}, {Wagner}, \&
  {Ward}}]{2007AA...470..475A}
---. 2007{\natexlab{a}}, \aap, 470, 475, \dodoi{10.1051/0004-6361:20077057}

\bibitem[{{Aharonian} {et~al.}(2007{\natexlab{b}}){Aharonian}, {Akhperjanian},
  {Barres de Almeida}, {Bazer-Bachi}, {Behera}, {Beilicke}, {Benbow},
  {Bernl{\"o}hr}, {Boisson}, {Bolz}, {Borrel}, {Braun}, {Brion}, {Brown},
  {B{\"u}hler}, {Bulik}, {B{\"u}sching}, {Boutelier}, {Carrigan}, {Chadwick},
  {Chounet}, {Clapson}, {Coignet}, {Cornils}, {Costamante}, {Dalton},
  {Degrange}, {Dickinson}, {Djannati-Ata{\"\i}}, {Domainko}, {O'C. Drury},
  {Dubois}, {Dubus}, {Dyks}, {Egberts}, {Emmanoulopoulos}, {Espigat},
  {Farnier}, {Feinstein}, {Fiasson}, {F{\"o}rster}, {Fontaine}, {Funk},
  {F{\"u}{\ss}ling}, {Gallant}, {Giebels}, {Glicenstein}, {Gl{\"u}ck}, {Goret},
  {Hadjichristidis}, {Hauser}, {Hauser}, {Heinzelmann}, {Henri}, {Hermann},
  {Hinton}, {Hoffmann}, {Hofmann}, {Holleran}, {Hoppe}, {Horns},
  {Jacholkowska}, {de Jager}, {Jung}, {Katarzy{\'n}ski}, {Kendziorra},
  {Kerschhaggl}, {Kh{\'e}lifi}, {Keogh}, {Komin}, {Kosack}, {Lamanna},
  {Latham}, {Lemi{\`e}re}, {Lemoine-Goumard}, {Lenain}, {Lohse}, {Martin},
  {Martineau-Huynh}, {Marcowith}, {Masterson}, {Maurin}, {Maurin}, {McComb},
  {Moderski}, {Moulin}, {de Naurois}, {Nedbal}, {Nolan}, {Ohm}, {Olive}, {de
  O{\~n}a Wilhelmi}, {Orford}, {Osborne}, {Ostrowski}, {Panter}, {Pedaletti},
  {Pelletier}, {Petrucci}, {Pita}, {P{\"u}hlhofer}, {Punch}, {Ranchon},
  {Raubenheimer}, {Raue}, {Rayner}, {Renaud}, {Ripken}, {Rob}, {Rolland},
  {Rosier-Lees}, {Rowell}, {Rudak}, {Ruppel}, {Sahakian}, {Santangelo},
  {Schlickeiser}, {Sch{\"o}ck}, {Schr{\"o}der}, {Schwanke}, {Schwarzburg},
  {Schwemmer}, {Shalchi}, {Sol}, {Spangler}, {Stawarz}, {Steenkamp},
  {Stegmann}, {Superina}, {Tam}, {Tavernet}, {Terrier}, {van Eldik},
  {Vasileiadis}, {Venter}, {Vialle}, {Vincent}, {Vivier}, {V{\"o}lk}, {Volpe},
  {Wagner}, {Ward}, {Zdziarski}, \& {Zech}}]{2007AA...475L...9A}
{Aharonian}, F., {Akhperjanian}, A.~G., {Barres de Almeida}, U., {et~al.}
  2007{\natexlab{b}}, \aap, 475, L9, \dodoi{10.1051/0004-6361:20078462}

\bibitem[{{Aharonian} {et~al.}(2007{\natexlab{c}}){Aharonian}, {Akhperjanian},
  {Barres de Almeida}, {Bazer-Bachi}, {Behera}, {Beilicke}, {Benbow},
  {Bernl{\"o}hr}, {Boisson}, {Bolz}, {Borrel}, {Braun}, {Brion}, {Brown},
  {B{\"u}hler}, {Bulik}, {B{\"u}sching}, {Boutelier}, {Carrigan}, {Chadwick},
  {Chounet}, {Clapson}, {Coignet}, {Cornils}, {Costamante}, {Dalton},
  {Degrange}, {Dickinson}, {Djannati-Ata{\"\i}}, {Domainko}, {O'C. Drury},
  {Dubois}, {Dubus}, {Dyks}, {Egberts}, {Emmanoulopoulos}, {Espigat},
  {Farnier}, {Feinstein}, {Fiasson}, {F{\"o}rster}, {Fontaine}, {Funk},
  {F{\"u}{\ss}ling}, {Gallant}, {Giebels}, {Glicenstein}, {Gl{\"u}ck}, {Goret},
  {Hadjichristidis}, {Hauser}, {Hauser}, {Heinzelmann}, {Henri}, {Hermann},
  {Hinton}, {Hoffmann}, {Hofmann}, {Holleran}, {Hoppe}, {Horns},
  {Jacholkowska}, {de Jager}, {Jung}, {Katarzy{\'n}ski}, {Kendziorra},
  {Kerschhaggl}, {Kh{\'e}lifi}, {Keogh}, {Komin}, {Kosack}, {Lamanna},
  {Latham}, {Lemi{\`e}re}, {Lemoine-Goumard}, {Lenain}, {Lohse}, {Martin},
  {Martineau-Huynh}, {Marcowith}, {Masterson}, {Maurin}, {Maurin}, {McComb},
  {Moderski}, {Moulin}, {de Naurois}, {Nedbal}, {Nolan}, {Ohm}, {Olive}, {de
  O{\~n}a Wilhelmi}, {Orford}, {Osborne}, {Ostrowski}, {Panter}, {Pedaletti},
  {Pelletier}, {Petrucci}, {Pita}, {P{\"u}hlhofer}, {Punch}, {Ranchon},
  {Raubenheimer}, {Raue}, {Rayner}, {Renaud}, {Ripken}, {Rob}, {Rolland},
  {Rosier-Lees}, {Rowell}, {Rudak}, {Ruppel}, {Sahakian}, {Santangelo},
  {Schlickeiser}, {Sch{\"o}ck}, {Schr{\"o}der}, {Schwanke}, {Schwarzburg},
  {Schwemmer}, {Shalchi}, {Sol}, {Spangler}, {Stawarz}, {Steenkamp},
  {Stegmann}, {Superina}, {Tam}, {Tavernet}, {Terrier}, {van Eldik},
  {Vasileiadis}, {Venter}, {Vialle}, {Vincent}, {Vivier}, {V{\"o}lk}, {Volpe},
  {Wagner}, {Ward}, {Zdziarski}, \& {Zech}}]{2007AA...473L..25A}
---. 2007{\natexlab{c}}, \aap, 473, L25, \dodoi{10.1051/0004-6361:20078412}

\bibitem[{{Aharonian} {et~al.}(2008{\natexlab{a}}){Aharonian}, {Akhperjanian},
  {Barres de Almeida}, {Bazer-Bachi}, {Behera}, {Beilicke}, {Benbow},
  {Bernl{\"o}hr}, {Boisson}, {Bolz}, {Borrel}, {Braun}, {Brion}, {Brown},
  {B{\"u}hler}, {Bulik}, {B{\"u}sching}, {Boutelier}, {Carrigan}, {Chadwick},
  {Chounet}, {Clapson}, {Coignet}, {Cornils}, {Costamante}, {Dalton},
  {Degrange}, {Dickinson}, {Djannati-Ata{\"\i}}, {Domainko}, {O'C. Drury},
  {Dubois}, {Dubus}, {Dyks}, {Egberts}, {Emmanoulopoulos}, {Espigat},
  {Farnier}, {Feinstein}, {Fiasson}, {F{\"o}rster}, {Fontaine}, {Funk},
  {F{\"u}{\ss}ling}, {Gallant}, {Giebels}, {Glicenstein}, {Gl{\"u}ck}, {Goret},
  {Hadjichristidis}, {Hauser}, {Hauser}, {Heinzelmann}, {Henri}, {Hermann},
  {Hinton}, {Hoffmann}, {Hofmann}, {Holleran}, {Hoppe}, {Horns},
  {Jacholkowska}, {de Jager}, {Jung}, {Katarzy{\'n}ski}, {Kendziorra},
  {Kerschhaggl}, {Kh{\'e}lifi}, {Keogh}, {Komin}, {Kosack}, {Lamanna},
  {Latham}, {Lemi{\`e}re}, {Lemoine-Goumard}, {Lenain}, {Lohse}, {Martin},
  {Martineau-Huynh}, {Marcowith}, {Masterson}, {Maurin}, {Maurin}, {McComb},
  {Moderski}, {Moulin}, {de Naurois}, {Nedbal}, {Nolan}, {Ohm}, {Olive}, {de
  O{\~n}a Wilhelmi}, {Orford}, {Osborne}, {Ostrowski}, {Panter}, {Pedaletti},
  {Pelletier}, {Petrucci}, {Pita}, {P{\"u}hlhofer}, {Punch}, {Ranchon},
  {Raubenheimer}, {Raue}, {Rayner}, {Renaud}, {Ripken}, {Rob}, {Rolland},
  {Rosier-Lees}, {Rowell}, {Rudak}, {Ruppel}, {Sahakian}, {Santangelo},
  {Schlickeiser}, {Sch{\"o}ck}, {Schr{\"o}der}, {Schwanke}, {Schwarzburg},
  {Schwemmer}, {Shalchi}, {Sol}, {Spangler}, {Stawarz}, {Steenkamp},
  {Stegmann}, {Superina}, {Tam}, {Tavernet}, {Terrier}, {van Eldik},
  {Vasileiadis}, {Venter}, {Vialle}, {Vincent}, {Vivier}, {V{\"o}lk}, {Volpe},
  {Wagner}, {Ward}, {Zdziarski}, \& {Zech}}]{2008AA...477..481A}
---. 2008{\natexlab{a}}, \aap, 477, 481, \dodoi{10.1051/0004-6361:20078603}

\bibitem[{{Aharonian} {et~al.}(2008{\natexlab{b}}){Aharonian}, {Akhperjanian},
  {Barres de Almeida}, {Bazer-Bachi}, {Behera}, {Beilicke}, {Benbow},
  {Bernl{\"o}hr}, {Boisson}, {Borrel}, {Braun}, {Brion}, {Brucker},
  {B{\"u}hler}, {Bulik}, {B{\"u}sching}, {Boutelier}, {Carrigan}, {Chadwick},
  {Chaves}, {Chounet}, {Clapson}, {Coignet}, {Cornils}, {Costamante}, {Dalton},
  {Degrange}, {Dickinson}, {Djannati-Ata{\"\i}}, {Domainko}, {O'C. Drury},
  {Dubois}, {Dubus}, {Dyks}, {Egberts}, {Emmanoulopoulos}, {Espigat},
  {Farnier}, {Feinstein}, {Fiasson}, {F{\"o}rster}, {Fontaine},
  {F{\"u}{\ss}ling}, {Gabici}, {Gallant}, {Giebels}, {Glicenstein},
  {Gl{\"u}ck}, {Goret}, {Hadjichristidis}, {Hauser}, {Hauser}, {Heinzelmann},
  {Henri}, {Hermann}, {Hinton}, {Hoffmann}, {Hofmann}, {Holleran}, {Hoppe},
  {Horns}, {Jacholkowska}, {de Jager}, {Jung}, {Katarzy{\'n}ski}, {Kaufmann},
  {Kendziorra}, {Kerschhaggl}, {Khangulyan}, {Kh{\'e}lifi}, {Keogh}, {Komin},
  {Kosack}, {Lamanna}, {Latham}, {Lenain}, {Lohse}, {Martin},
  {Martineau-Huynh}, {Marcowith}, {Masterson}, {Maurin}, {McComb}, {Moderski},
  {Moulin}, {Naumann-Godo}, {de Naurois}, {Nedbal}, {Nekrassov}, {Nolan},
  {Ohm}, {Olive}, {de O{\~n}a Wilhelmi}, {Orford}, {Osborne}, {Ostrowski},
  {Panter}, {Pedaletti}, {Pelletier}, {Petrucci}, {Pita}, {P{\"u}hlhofer},
  {Punch}, {Quirrenbach}, {Raubenheimer}, {Raue}, {Rayner}, {Renaud}, {Rieger},
  {Ripken}, {Rob}, {Rosier-Lees}, {Rowell}, {Rudak}, {Ruppel}, {Sahakian},
  {Santangelo}, {Schlickeiser}, {Sch{\"o}ck}, {Schr{\"o}der}, {Schwanke},
  {Schwarzburg}, {Schwemmer}, {Shalchi}, {Sol}, {Spangler}, {Stawarz},
  {Steenkamp}, {Stegmann}, {Superina}, {Tam}, {Tavernet}, {Terrier}, {van
  Eldik}, {Vasileiadis}, {Venter}, {Vialle}, {Vincent}, {Vivier}, {V{\"o}lk},
  {Volpe}, {Wagner}, {Ward}, {Zdziarski}, \& {Zech}}]{2008AA...481L.103A}
---. 2008{\natexlab{b}}, \aap, 481, L103, \dodoi{10.1051/0004-6361:200809603}

\bibitem[{{Aharonian} {et~al.}(2010){Aharonian}, {Akhperjanian}, {Anton},
  {Barres de Almeida}, {Bazer-Bachi}, {Becherini}, {Behera}, {Benbow},
  {Bernl{\"o}hr}, {Bochow}, {Boisson}, {Bolmont}, {Borrel}, {Brucker}, {Brun},
  {Brun}, {B{\"u}hler}, {Bulik}, {B{\"u}sching}, {Boutelier}, {Chadwick},
  {Charbonnier}, {Chaves}, {Cheesebrough}, {Chounet}, {Clapson}, {Coignet},
  {Dalton}, {Daniel}, {Davids}, {Degrange}, {Deil}, {Dickinson},
  {Djannati-Ata{\"\i}}, {Domainko}, {O'C. Drury}, {Dubois}, {Dubus}, {Dyks},
  {Dyrda}, {Egberts}, {Emmanoulopoulos}, {Espigat}, {Farnier}, {Feinstein},
  {Fiasson}, {F{\"o}rster}, {Fontaine}, {F{\"u}{\ss}ling}, {Gabici}, {Gallant},
  {G{\'e}rard}, {Gerbig}, {Giebels}, {Glicenstein}, {Gl{\"u}ck}, {Goret},
  {G{\"o}ring}, {Hauser}, {Hauser}, {Heinz}, {Heinzelmann}, {Henri}, {Hermann},
  {Hinton}, {Hoffmann}, {Hofmann}, {Holleran}, {Hoppe}, {Horns},
  {Jacholkowska}, {de Jager}, {Jahn}, {Jung}, {Katarzy{\'n}ski}, {Katz},
  {Kaufmann}, {Kendziorra}, {Kerschhaggl}, {Khangulyan}, {Kh{\'e}lifi},
  {Keogh}, {Klu{\'z}niak}, {Kneiske}, {Komin}, {Kosack}, {Lamanna}, {Lenain},
  {Lohse}, {Marandon}, {Martin}, {Martineau-Huynh}, {Marcowith}, {Masbou},
  {Maurin}, {McComb}, {Medina}, {Moderski}, {Moulin}, {Naumann-Godo}, {de
  Naurois}, {Nedbal}, {Nekrassov}, {Nicholas}, {Niemiec}, {Nolan}, {Ohm},
  {Olive}, {de O{\~n}a Wilhelmi}, {Orford}, {Ostrowski}, {Panter}, {Paz
  Arribas}, {Pedaletti}, {Pelletier}, {Petrucci}, {Pita}, {P{\"u}hlhofer},
  {Punch}, {Quirrenbach}, {Raubenheimer}, {Raue}, {Rayner}, {Renaud}, {Rieger},
  {Ripken}, {Rob}, {Rosier-Lees}, {Rowell}, {Rudak}, {Rulten}, {Ruppel},
  {Sahakian}, {Santangelo}, {Schlickeiser}, {Sch{\"o}ck}, {Schr{\"o}der},
  {Schwanke}, {Schwarzburg}, {Schwemmer}, {Shalchi}, {Sikora}, {Skilton},
  {Sol}, {Spangler}, {Stawarz}, {Steenkamp}, {Stegmann}, {Stinzing},
  {Superina}, {Szostek}, {Tam}, {Tavernet}, {Terrier}, {Tibolla}, {Tluczykont},
  {van Eldik}, {Vasileiadis}, {Venter}, {Venter}, {Vialle}, {Vincent},
  {Vivier}, {V{\"o}lk}, {Volpe}, {Wagner}, {Ward}, {Zdziarski}, \&
  {Zech}}]{2010AA...521A..69A}
{Aharonian}, F., {Akhperjanian}, A.~G., {Anton}, G., {et~al.} 2010, \aap, 521,
  A69, \dodoi{10.1051/0004-6361/200912363}

\bibitem[{{Aharonian}(2004)}]{2004vhec.book.....A}
{Aharonian}, F.~A. 2004, {Very high energy cosmic gamma radiation : a crucial
  window on the extreme Universe}, \dodoi{10.1142/4657}

\bibitem[{{Aharonian} {et~al.}(2008{\natexlab{c}}){Aharonian}, {Khangulyan}, \&
  {Costamante}}]{2008MNRAS.387.1206A}
{Aharonian}, F.~A., {Khangulyan}, D., \& {Costamante}, L. 2008{\natexlab{c}},
  \mnras, 387, 1206, \dodoi{10.1111/j.1365-2966.2008.13315.x}

\bibitem[{{Ahnen} {et~al.}(2016){Ahnen}, {Ansoldi}, {Antonelli}, {Antoranz},
  {Arcaro}, {Babic}, {Banerjee}, {Bangale}, {Barres de Almeida}, {Barrio},
  {Becerra Gonz{\'a}lez}, {Bednarek}, {Bernardini}, {Berti}, {Biasuzzi},
  {Biland}, {Blanch}, {Bonnefoy}, {Bonnoli}, {Borracci}, {Bretz}, {Buson},
  {Carosi}, {Chatterjee}, {Clavero}, {Colin}, {Colombo}, {Contreras},
  {Cortina}, {Covino}, {Da Vela}, {Dazzi}, {De Angelis}, {De Lotto}, {de
  O{\~n}a Wilhelmi}, {Di Pierro}, {Doert}, {Dom{\'\i}nguez}, {Dominis Prester},
  {Dorner}, {Doro}, {Einecke}, {Eisenacher Glawion}, {Elsaesser},
  {Engelkemeier}, {Fallah Ramazani}, {Fern{\'a}ndez-Barral}, {Fidalgo},
  {Fonseca}, {Font}, {Frantzen}, {Fruck}, {Galindo}, {Garc{\'\i}a L{\'o}pez},
  {Garczarczyk}, {Garrido Terrats}, {Gaug}, {Giammaria}, {Godinovi{\'c}},
  {Gora}, {Guberman}, {Hadasch}, {Hahn}, {Hayashida}, {Herrera}, {Hose},
  {Hrupec}, {Hughes}, {Idec}, {Kodani}, {Konno}, {Kubo}, {Kushida}, {La
  Barbera}, {Lelas}, {Lindfors}, {Lombardi}, {Longo}, {L{\'o}pez},
  {L{\'o}pez-Coto}, {Majumdar}, {Makariev}, {Mallot}, {Maneva}, {Manganaro},
  {Mannheim}, {Maraschi}, {Marcote}, {Mariotti}, {Mart{\'\i}nez}, {Mazin},
  {Menzel}, {Miranda}, {Mirzoyan}, {Moralejo}, {Moretti}, {Nakajima},
  {Neustroev}, {Niedzwiecki}, {Nievas Rosillo}, {Nilsson}, {Nishijima}, {Noda},
  {Nogu{\'e}s}, {Paiano}, {Palacio}, {Palatiello}, {Paneque}, {Paoletti},
  {Paredes}, {Paredes-Fortuny}, {Pedaletti}, {Peresano}, {Perri}, {Persic},
  {Poutanen}, {Prada Moroni}, {Prandini}, {Puljak}, {Garcia}, {Reichardt},
  {Rhode}, {Rib{\'o}}, {Rico}, {Saito}, {Satalecka}, {Schroeder}, {Schweizer},
  {Shore}, {Sillanp{\"a}{\"a}}, {Sitarek}, {Snidaric}, {Sobczynska},
  {Stamerra}, {Strzys}, {Suri{\'c}}, {Takalo}, {Tavecchio}, {Temnikov},
  {Terzi{\'c}}, {Tescaro}, {Teshima}, {Torres}, {Toyama}, {Treves}, {Vanzo},
  {Verguilov}, {Vovk}, {Ward}, {Will}, {Wu}, {Zanin}, \&
  {Desiante}}]{2016AA...595A..98A}
{Ahnen}, M.~L., {Ansoldi}, S., {Antonelli}, L.~A., {et~al.} 2016, \aap, 595,
  A98, \dodoi{10.1051/0004-6361/201629461}

\bibitem[{{Albert} {et~al.}(2006{\natexlab{a}}){Albert}, {Aliu}, {Anderhub},
  {Antoranz}, {Armada}, {Asensio}, {Baixeras}, {Barrio}, {Bartko}, {Bastieri},
  {Becker}, {Bednarek}, {Berger}, {Bigongiari}, {Biland}, {Bisesi}, {Bock},
  {Bordas}, {Bosch-Ramon}, {Bretz}, {Britvitch}, {Camara}, {Carmona},
  {Chilingarian}, {Ciprini}, {Coarasa}, {Commichau}, {Contreras}, {Cortina},
  {Curtef}, {Danielyan}, {Dazzi}, {De Angelis}, {de los Reyes}, {De Lotto},
  {Domingo-Santamar{\'\i}a}, {Dorner}, {Doro}, {Errando}, {Fagiolini},
  {Ferenc}, {Fern{\'a}ndez}, {Firpo}, {Flix}, {Fonseca}, {Font}, {Fuchs},
  {Galante}, {Garczarczyk}, {Gaug}, {Giller}, {Goebel}, {Hakobyan},
  {Hayashida}, {Hengstebeck}, {H{\"o}hne}, {Hose}, {Hsu}, {Jacon}, {Kalekin},
  {Kosyra}, {Kranich}, {Laatiaoui}, {Laille}, {Lenisa}, {Liebing}, {Lindfors},
  {Lombardi}, {Longo}, {L{\'o}pez}, {L{\'o}pez}, {Lorenz}, {Majumdar},
  {Maneva}, {Mannheim}, {Mansutti}, {Mariotti}, {Mart{\'\i}nez}, {Mazin},
  {Merck}, {Meucci}, {Meyer}, {Miranda}, {Mirzoyan}, {Mizobuchi}, {Moralejo},
  {Nilsson}, {Ninkovic}, {O{\~n}a-Wilhelmi}, {Ordu{\~n}a}, {Otte}, {Oya},
  {Paneque}, {Paoletti}, {Paredes}, {Pasanen}, {Pascoli}, {Pauss}, {Pegna},
  {Persic}, {Peruzzo}, {Piccioli}, {Poller}, {Prandini}, {Raymers}, {Rhode},
  {Rib{\'o}}, {Rico}, {Riegel}, {Rissi}, {Robert}, {R{\"u}gamer}, {Saggion},
  {S{\'a}nchez}, {Sartori}, {Scalzotto}, {Scapin}, {Schmitt}, {Schweizer},
  {Shayduk}, {Shinozaki}, {Shore}, {Sidro}, {Sillanp{\"a}{\"a}}, {Sobczynska},
  {Stamerra}, {Stark}, {Takalo}, {Temnikov}, {Tescaro}, {Teshima}, {Tonello},
  {Torres}, {Torres}, {Turini}, {Vankov}, {Vitale}, {Wagner}, {Wibig},
  {Wittek}, {Zanin}, \& {Zapatero}}]{2006ApJ...648L.105A}
{Albert}, J., {Aliu}, E., {Anderhub}, H., {et~al.} 2006{\natexlab{a}}, \apjl,
  648, L105, \dodoi{10.1086/508020}

\bibitem[{{Albert} {et~al.}(2006{\natexlab{b}}){Albert}, {Aliu}, {Anderhub},
  {Antoranz}, {Armada}, {Asensio}, {Baixeras}, {Barrio}, {Bartko}, {Bastieri},
  {Bednarek}, {Berger}, {Bigongiari}, {Biland}, {Bisesi}, {Bock}, {Bretz},
  {Britvitch}, {Camara}, {Chilingarian}, {Ciprini}, {Coarasa}, {Commichau},
  {Contreras}, {Cortina}, {Danielyan}, {Dazzi}, {De Angelis}, {de los Reyes},
  {De Lotto}, {Domingo-Santamar{\'\i}a}, {Dorner}, {Doro}, {Errando}, {Ferenc},
  {Fern{\'a}ndez}, {Firpo}, {Flix}, {Fonseca}, {Font}, {Galante},
  {Garczarczyk}, {Gaug}, {Gebauer}, {Giannitrapani}, {Giller}, {Goebel},
  {Hakobyan}, {Hayashida}, {Hengstebeck}, {H{\"o}hne}, {Hose}, {Jacon},
  {Kalekin}, {Kranich}, {Laille}, {Lenisa}, {Liebing}, {Lindfors}, {Longo},
  {L{\'o}pez}, {L{\'o}pez}, {Lorenz}, {Lucarelli}, {Majumdar}, {Maneva},
  {Mannheim}, {Mariotti}, {Mart{\'\i}nez}, {Mase}, {Mazin}, {Merck}, {Merck},
  {Meucci}, {Meyer}, {Miranda}, {Mirzoyan}, {Mizobuchi}, {Moralejo}, {Nilsson},
  {O{\~n}a-Wilhelmi}, {Ordu{\~n}a}, {Otte}, {Oya}, {Paneque}, {Paoletti},
  {Pasanen}, {Pascoli}, {Pauss}, {Pavel}, {Pegna}, {Peruzzo}, {Piccioli},
  {Pin}, {Prandini}, {Rico}, {Rhode}, {Riegel}, {Rissi}, {Robert}, {Rossato},
  {R{\"u}gamer}, {Saggion}, {S{\'a}nchez}, {Sartori}, {Scalzotto}, {Schmitt},
  {Schweizer}, {Shayduk}, {Shinozaki}, {Sidro}, {Sillanp{\"a}{\"a}},
  {Sobczynska}, {Stamerra}, {Stark}, {Takalo}, {Temnikov}, {Tescaro},
  {Teshima}, {Tonello}, {Torres}, {Torres}, {Turini}, {Vankov}, {Vitale},
  {Wagner}, {Wibig}, {Wittek}, \& {Zapatero}}]{2006ApJ...639..761A}
---. 2006{\natexlab{b}}, \apj, 639, 761, \dodoi{10.1086/499421}

\bibitem[{{Albert} {et~al.}(2007{\natexlab{a}}){Albert}, {Aliu}, {Anderhub},
  {Antoranz}, {Armada}, {Asensio}, {Baixeras}, {Barrio}, {Bartko}, {Bastieri},
  {Becker}, {Bednarek}, {Berger}, {Bigongiari}, {Biland}, {Bock}, {Bordas},
  {Bosch-Ramon}, {Bretz}, {Britvitch}, {Camara}, {Carmona}, {Chilingarian},
  {Ciprini}, {Coarasa}, {Commichau}, {Contreras}, {Cortina}, {Curtef},
  {Danielyan}, {Dazzi}, {De Angelis}, {de los Reyes}, {De Lotto},
  {Domingo-Santamar{\'\i}a}, {Dorner}, {Doro}, {Errando}, {Fagiolini},
  {Ferenc}, {Fern{\'a}ndez}, {Firpo}, {Flix}, {Fonseca}, {Font}, {Fuchs},
  {Galante}, {Garczarczyk}, {Gaug}, {Giller}, {Goebel}, {Hakobyan},
  {Hayashida}, {Hengstebeck}, {H{\"o}hne}, {Hose}, {Hsu}, {Jacon}, {Jogler},
  {Kalekin}, {Kosyra}, {Kranich}, {Kritzer}, {Laatiaoui}, {Laille}, {Liebing},
  {Lindfors}, {Lombardi}, {Longo}, {L{\'o}pez}, {L{\'o}pez}, {Lorenz},
  {Majumdar}, {Maneva}, {Mannheim}, {Mansutti}, {Mariotti}, {Mart{\'\i}nez},
  {Mazin}, {Merck}, {Meucci}, {Meyer}, {Miranda}, {Mirzoyan}, {Mizobuchi},
  {Moralejo}, {Nilsson}, {Ninkovic}, {O{\~n}a-Wilhelmi}, {Ordu{\~n}a}, {Otte},
  {Oya}, {Paneque}, {Paoletti}, {Paredes}, {Pasanen}, {Pascoli}, {Pauss},
  {Pegna}, {Persic}, {Peruzzo}, {Piccioli}, {Poller}, {Prandini}, {Raymers},
  {Rhode}, {Rib{\'o}}, {Rico}, {Rissi}, {Robert}, {R{\"u}gamer}, {Saggion},
  {S{\'a}nchez}, {Sartori}, {Scalzotto}, {Scapin}, {Schmitt}, {Schweizer},
  {Shayduk}, {Shinozaki}, {Shore}, {Sidro}, {Sillanp{\"a}{\"a}}, {Sobczynska},
  {Stamerra}, {Stark}, {Takalo}, {Temnikov}, {Tescaro}, {Teshima}, {Tonello},
  {Torres}, {Torres}, {Turini}, {Vankov}, {Vitale}, {Wagner}, {Wibig},
  {Wittek}, {Zanin}, \& {Zapatero}}]{2007ApJ...663..125A}
---. 2007{\natexlab{a}}, \apj, 663, 125, \dodoi{10.1086/518221}

\bibitem[{{Albert} {et~al.}(2007{\natexlab{b}}){Albert}, {Aliu}, {Anderhub},
  {Antoranz}, {Armada}, {Baixeras}, {Barrio}, {Bartko}, {Bastieri}, {Becker},
  {Bednarek}, {Berger}, {Bigongiari}, {Biland}, {Bock}, {Bordas},
  {Bosch-Ramon}, {Bretz}, {Britvitch}, {Camara}, {Carmona}, {Chilingarian},
  {Coarasa}, {Commichau}, {Contreras}, {Cortina}, {Costado}, {Curtef},
  {Danielyan}, {Dazzi}, {De Angelis}, {Delgado}, {de los Reyes}, {De Lotto},
  {Domingo-Santamar{\'\i}a}, {Dorner}, {Doro}, {Errando}, {Fagiolini},
  {Ferenc}, {Fern{\'a}ndez}, {Firpo}, {Flix}, {Fonseca}, {Font}, {Fuchs},
  {Galante}, {Garc{\'\i}a-L{\'o}pez}, {Garczarczyk}, {Gaug}, {Giller},
  {Goebel}, {Hakobyan}, {Hayashida}, {Hengstebeck}, {Herrero}, {H{\"o}hne},
  {Hose}, {Hsu}, {Jacon}, {Jogler}, {Kosyra}, {Kranich}, {Kritzer}, {Laille},
  {Lindfors}, {Lombardi}, {Longo}, {L{\'o}pez}, {L{\'o}pez}, {Lorenz},
  {Majumdar}, {Maneva}, {Mannheim}, {Mansutti}, {Mariotti}, {Mart{\'\i}nez},
  {Mazin}, {Merck}, {Meucci}, {Meyer}, {Miranda}, {Mirzoyan}, {Mizobuchi},
  {Moralejo}, {Nilsson}, {Ninkovic}, {O{\~n}a-Wilhelmi}, {Otte}, {Oya},
  {Paneque}, {Panniello}, {Paoletti}, {Paredes}, {Pasanen}, {Pascoli}, {Pauss},
  {Pegna}, {Persic}, {Peruzzo}, {Piccioli}, {Poller}, {Prandini}, {Puchades},
  {Raymers}, {Rhode}, {Rib{\'o}}, {Rico}, {Rissi}, {Robert}, {R{\"u}gamer},
  {Saggion}, {S{\'a}nchez}, {Sartori}, {Scalzotto}, {Scapin}, {Schmitt},
  {Schweizer}, {Shayduk}, {Shinozaki}, {Shore}, {Sidro}, {Sillanp{\"a}{\"a}},
  {Sobczynska}, {Stamerra}, {Stark}, {Takalo}, {Temnikov}, {Tescaro},
  {Teshima}, {Tonello}, {Torres}, {Turini}, {Vankov}, {Vitale}, {Wagner},
  {Wibig}, {Wittek}, {Zandanel}, {Zanin}, \& {Zapatero}}]{2007ApJ...666L..17A}
---. 2007{\natexlab{b}}, \apjl, 666, L17, \dodoi{10.1086/521550}

\bibitem[{{Aleksi{\'c}} {et~al.}(2011){Aleksi{\'c}}, {Antonelli}, {Antoranz},
  {Backes}, {Barrio}, {Bastieri}, {Becerra Gonz{\'a}lez}, {Bednarek},
  {Berdyugin}, {Berger}, {Bernardini}, {Biland}, {Blanch}, {Bock}, {Boller},
  {Bonnoli}, {Borla Tridon}, {Braun}, {Bretz}, {Ca{\~n}ellas}, {Carmona},
  {Carosi}, {Colin}, {Colombo}, {Contreras}, {Cortina}, {Cossio}, {Covino},
  {Dazzi}, {De Angelis}, {De Cea del Pozo}, {De Lotto}, {Delgado Mendez},
  {Diago Ortega}, {Doert}, {Dom{\'\i}nguez}, {Dominis Prester}, {Dorner},
  {Doro}, {Elsaesser}, {Ferenc}, {Fonseca}, {Font}, {Fruck}, {Garc{\'\i}a
  L{\'o}pez}, {Garczarczyk}, {Garrido}, {Giavitto}, {Godinovi{\'c}}, {Hadasch},
  {H{\"a}fner}, {Herrero}, {Hildebrand}, {H{\"o}hne-M{\"o}nch}, {Hose},
  {Hrupec}, {Huber}, {Jogler}, {Klepser}, {Kr{\"a}henb{\"u}hl}, {Krause}, {La
  Barbera}, {Lelas}, {Leonardo}, {Lindfors}, {Lombardi}, {L{\'o}pez}, {Lorenz},
  {Makariev}, {Maneva}, {Mankuzhiyil}, {Mannheim}, {Maraschi}, {Mariotti},
  {Mart{\'\i}nez}, {Mazin}, {Meucci}, {Miranda}, {Mirzoyan}, {Miyamoto},
  {Mold{\'o}n}, {Moralejo}, {Nieto}, {Nilsson}, {Orito}, {Oya}, {Paneque},
  {Paoletti}, {Pardo}, {Paredes}, {Partini}, {Pasanen}, {Pauss},
  {Perez-Torres}, {Persic}, {Peruzzo}, {Pilia}, {Pochon}, {Prada}, {Prada
  Moroni}, {Prandini}, {Puljak}, {Reichardt}, {Reinthal}, {Rhode}, {Rib{\'o}},
  {Rico}, {R{\"u}gamer}, {Saggion}, {Saito}, {Saito}, {Salvati}, {Satalecka},
  {Scalzotto}, {Scapin}, {Schultz}, {Schweizer}, {Shayduk}, {Shore},
  {Sillanp{\"a}{\"a}}, {Sitarek}, {Sobczynska}, {Spanier}, {Spiro}, {Stamerra},
  {Steinke}, {Storz}, {Strah}, {Suri{\'c}}, {Takalo}, {Tavecchio}, {Temnikov},
  {Terzi{\'c}}, {Tescaro}, {Teshima}, {Thom}, {Tibolla}, {Torres}, {Treves},
  {Vankov}, {Vogler}, {Wagner}, {Weitzel}, {Zabalza}, {Zandanel}, {Zanin},
  {MAGIC Collaboration}, {Tanaka}, {Wood}, \& {Buson}}]{2011ApJ...730L...8A}
{Aleksi{\'c}}, J., {Antonelli}, L.~A., {Antoranz}, P., {et~al.} 2011, \apjl,
  730, L8, \dodoi{10.1088/2041-8205/730/1/L8}

\bibitem[{{Aleksi{\'c}} {et~al.}(2012{\natexlab{a}}){Aleksi{\'c}}, {Alvarez},
  {Antonelli}, {Antoranz}, {Asensio}, {Backes}, {Barrio}, {Bastieri}, {Becerra
  Gonz{\'a}lez}, {Bednarek}, {Berdyugin}, {Berger}, {Bernardini}, {Biland},
  {Blanch}, {Bock}, {Boller}, {Bonnoli}, {Borla Tridon}, {Braun}, {Bretz},
  {Ca{\~n}ellas}, {Carmona}, {Carosi}, {Colin}, {Colombo}, {Contreras},
  {Cortina}, {Cossio}, {Covino}, {Dazzi}, {de Angelis}, {de Caneva}, {de Cea
  Del Pozo}, {de Lotto}, {Delgado Mendez}, {Diago Ortega}, {Doert},
  {Dom{\'\i}nguez}, {Dominis Prester}, {Dorner}, {Doro}, {Elsaesser}, {Ferenc},
  {Fonseca}, {Font}, {Fruck}, {Garc{\'\i}a L{\'o}pez}, {Garczarczyk},
  {Garrido}, {Giavitto}, {Godinovi{\'c}}, {Hadasch}, {H{\"a}fner}, {Herrero},
  {Hildebrand}, {H{\"o}hne-M{\"o}nch}, {Hose}, {Hrupec}, {Huber}, {Jogler},
  {Kellermann}, {Klepser}, {Kr{\"a}henb{\"u}hl}, {Krause}, {La Barbera},
  {Lelas}, {Leonardo}, {Lindfors}, {Lombardi}, {L{\'o}pez}, {L{\'o}pez-Oramas},
  {Lorenz}, {Makariev}, {Maneva}, {Mankuzhiyil}, {Mannheim}, {Maraschi},
  {Mariotti}, {Mart{\'\i}nez}, {Mazin}, {Meucci}, {Miranda}, {Mirzoyan},
  {Miyamoto}, {Mold{\'o}n}, {Moralejo}, {Munar-Adrover}, {Nieto}, {Nilsson},
  {Orito}, {Oya}, {Paneque}, {Paoletti}, {Pardo}, {Paredes}, {Partini},
  {Pasanen}, {Pauss}, {Perez-Torres}, {Persic}, {Peruzzo}, {Pilia}, {Pochon},
  {Prada}, {Prada Moroni}, {Prandini}, {Puljak}, {Reichardt}, {Reinthal},
  {Rhode}, {Rib{\'o}}, {Rico}, {R{\"u}gamer}, {Saggion}, {Saito}, {Saito},
  {Salvati}, {Satalecka}, {Scalzotto}, {Scapin}, {Schultz}, {Schweizer},
  {Shayduk}, {Shore}, {Sillanp{\"a}{\"a}}, {Sitarek}, {Snidaric}, {Sobczynska},
  {Spanier}, {Spiro}, {Stamatescu}, {Stamerra}, {Steinke}, {Storz}, {Strah},
  {Suri{\'c}}, {Takalo}, {Takami}, {Tavecchio}, {Temnikov}, {Terzi{\'c}},
  {Tescaro}, {Teshima}, {Tibolla}, {Torres}, {Treves}, {Uellenbeck}, {Vankov},
  {Vogler}, {Wagner}, {Weitzel}, {Zabalza}, {Zandanel}, {Zanin}, {Kadenius},
  {Weidinger}, \& {Buson}}]{2012AA...539A.118A}
{Aleksi{\'c}}, J., {Alvarez}, E.~A., {Antonelli}, L.~A., {et~al.}
  2012{\natexlab{a}}, \aap, 539, A118, \dodoi{10.1051/0004-6361/201117967}

\bibitem[{{Aleksi{\'c}} {et~al.}(2012{\natexlab{b}}){Aleksi{\'c}}, {Alvarez},
  {Antonelli}, {Antoranz}, {Ansoldi}, {Asensio}, {Backes}, {Barres de Almeida},
  {Barrio}, {Bastieri}, {Becerra Gonz{\'a}lez}, {Bednarek}, {Berger},
  {Bernardini}, {Biland}, {Blanch}, {Bock}, {Boller}, {Bonnoli}, {Borla
  Tridon}, {Bretz}, {Ca{\~n}ellas}, {Carmona}, {Carosi}, {Colin}, {Colombo},
  {Contreras}, {Cortina}, {Cossio}, {Covino}, {Da Vela}, {Dazzi}, {De Angelis},
  {De Caneva}, {De Cea del Pozo}, {De Lotto}, {Delgado Mendez}, {Diago Ortega},
  {Doert}, {Dom{\'\i}nguez}, {Dominis Prester}, {Dorner}, {Doro}, {Eisenacher},
  {Elsaesser}, {Ferenc}, {Fonseca}, {Font}, {Fruck}, {Garc{\'\i}a L{\'o}pez},
  {Garczarczyk}, {Garrido Terrats}, {Gaug}, {Giavitto}, {Godinovi{\'c}},
  {Gonz{\'a}lez Mu{\~n}oz}, {Gozzini}, {Hadasch}, {H{\"a}fner}, {Herrero},
  {Hildebrand}, {Hose}, {Hrupec}, {Huber}, {Jankowski}, {Jogler}, {Kadenius},
  {Kellermann}, {Klepser}, {Kr{\"a}henb{\"u}hl}, {Krause}, {La Barbera},
  {Lelas}, {Leonardo}, {Lewandowska}, {Lindfors}, {Lombardi}, {L{\'o}pez},
  {L{\'o}pez-Coto}, {L{\'o}pez-Oramas}, {Lorenz}, {Makariev}, {Maneva},
  {Mankuzhiyil}, {Mannheim}, {Maraschi}, {Mariotti}, {Mart{\'\i}nez}, {Mazin},
  {Meucci}, {Miranda}, {Mirzoyan}, {Mold{\'o}n}, {Moralejo}, {Munar-Adrover},
  {Niedzwiecki}, {Nieto}, {Nilsson}, {Nowak}, {Orito}, {Paiano}, {Paneque},
  {Paoletti}, {Pardo}, {Paredes}, {Partini}, {Perez-Torres}, {Persic}, {Pilia},
  {Pochon}, {Prada}, {Prada Moroni}, {Prandini}, {Puerto Gimenez}, {Puljak},
  {Reichardt}, {Reinthal}, {Rhode}, {Rib{\'o}}, {Rico}, {R{\"u}gamer},
  {Saggion}, {Saito}, {Saito}, {Salvati}, {Satalecka}, {Scalzotto}, {Scapin},
  {Schultz}, {Schweizer}, {Shore}, {Sillanp{\"a}{\"a}}, {Sitarek}, {Snidaric},
  {Sobczynska}, {Spanier}, {Spiro}, {Stamatescu}, {Stamerra}, {Steinke},
  {Storz}, {Strah}, {Sun}, {Suri{\'c}}, {Takalo}, {Takami}, {Tavecchio},
  {Temnikov}, {Terzi{\'c}}, {Tescaro}, {Teshima}, {Tibolla}, {Torres},
  {Treves}, {Uellenbeck}, {Vogler}, {Wagner}, {Weitzel}, {Zabalza}, {Zandanel},
  {Zanin}, {Berdyugin}, {Buson}, {J{\"a}rvel{\"a}}, {Larsson},
  {L{\"a}hteenm{\"a}ki}, \& {Tammi}}]{2012AA...544A.142A}
---. 2012{\natexlab{b}}, \aap, 544, A142, \dodoi{10.1051/0004-6361/201219133}

\bibitem[{{Aleksi{\'c}} {et~al.}(2014{\natexlab{a}}){Aleksi{\'c}}, {Ansoldi},
  {Antonelli}, {Antoranz}, {Babic}, {Bangale}, {Barres de Almeida}, {Barrio},
  {Becerra Gonz{\'a}lez}, {Bednarek}, {Bernardini}, {Biland}, {Blanch},
  {Bonnefoy}, {Bonnoli}, {Borracci}, {Bretz}, {Carmona}, {Carosi}, {Carreto
  Fidalgo}, {Colin}, {Colombo}, {Contreras}, {Cortina}, {Covino}, {da Vela},
  {Dazzi}, {de Angelis}, {de Caneva}, {de Lotto}, {Delgado Mendez}, {Doert},
  {Dom{\'\i}nguez}, {Dominis Prester}, {Dorner}, {Doro}, {Einecke},
  {Eisenacher}, {Elsaesser}, {Farina}, {Ferenc}, {Fonseca}, {Font}, {Frantzen},
  {Fruck}, {Garc{\'\i}a L{\'o}pez}, {Garczarczyk}, {Garrido Terrats}, {Gaug},
  {Godinovi{\'c}}, {Gonz{\'a}lez Mu{\~n}oz}, {Gozzini}, {Hadasch}, {Hayashida},
  {Herrera}, {Herrero}, {Hildebrand}, {Hose}, {Hrupec}, {Idec}, {Kadenius},
  {Kellermann}, {Kodani}, {Konno}, {Krause}, {Kubo}, {Kushida}, {La Barbera},
  {Lelas}, {Lewandowska}, {Lindfors}, {Lombardi}, {L{\'o}pez},
  {L{\'o}pez-Coto}, {L{\'o}pez-Oramas}, {Lorenz}, {Lozano}, {Makariev},
  {Mallot}, {Maneva}, {Mankuzhiyil}, {Mannheim}, {Maraschi}, {Marcote},
  {Mariotti}, {Mart{\'\i}nez}, {Mazin}, {Menzel}, {Meucci}, {Miranda},
  {Mirzoyan}, {Moralejo}, {Munar-Adrover}, {Nakajima}, {Niedzwiecki},
  {Nilsson}, {Nishijima}, {Noda}, {Nowak}, {Orito}, {Overkemping}, {Paiano},
  {Palatiello}, {Paneque}, {Paoletti}, {Paredes}, {Paredes-Fortuny}, {Partini},
  {Persic}, {Prada}, {Prada Moroni}, {Prandini}, {Preziuso}, {Puljak},
  {Reinthal}, {Rhode}, {Rib{\'o}}, {Rico}, {Rodriguez Garcia}, {R{\"u}gamer},
  {Saggion}, {Saito}, {Saito}, {Satalecka}, {Scalzotto}, {Scapin}, {Schultz},
  {Schweizer}, {Shore}, {Sillanp{\"a}{\"a}}, {Sitarek}, {Snidaric},
  {Sobczynska}, {Spanier}, {Stamatescu}, {Stamerra}, {Steinbring}, {Storz},
  {Strzys}, {Sun}, {Suri{\'c}}, {Takalo}, {Takami}, {Tavecchio}, {Temnikov},
  {Terzi{\'c}}, {Tescaro}, {Teshima}, {Thaele}, {Tibolla}, {Torres}, {Toyama},
  {Treves}, {Uellenbeck}, {Vogler}, {Wagner}, {Zandanel}, {Zanin}, \& {MAGIC
  Collaboration}}]{2014AA...572A.121A}
{Aleksi{\'c}}, J., {Ansoldi}, S., {Antonelli}, L.~A., {et~al.}
  2014{\natexlab{a}}, \aap, 572, A121, \dodoi{10.1051/0004-6361/201424254}

\bibitem[{{Aleksi{\'c}} {et~al.}(2014{\natexlab{b}}){Aleksi{\'c}}, {Antonelli},
  {Antoranz}, {Babic}, {Barres de Almeida}, {Barrio}, {Becerra Gonz{\'a}lez},
  {Bednarek}, {Berger}, {Bernardini}, {Biland}, {Blanch}, {Bock}, {Boller},
  {Bonnefoy}, {Bonnoli}, {Borla Tridon}, {Borracci}, {Bretz}, {Carmona},
  {Carosi}, {Carreto Fidalgo}, {Colin}, {Colombo}, {Contreras}, {Cortina},
  {Cossio}, {Covino}, {Da Vela}, {Dazzi}, {De Angelis}, {De Caneva}, {Delgado
  Mendez}, {De Lotto}, {Doert}, {Dom{\'\i}nguez}, {Dominis Prester}, {Dorner},
  {Doro}, {Eisenacher}, {Elsaesser}, {Farina}, {Ferenc}, {Fonseca}, {Font},
  {Fruck}, {Garc{\'\i}a L{\'o}pez}, {Garczarczyk}, {Garrido Terrats}, {Gaug},
  {Giavitto}, {Godinovi{\'c}}, {Gonz{\'a}lez Mu{\~n}oz}, {Gozzini}, {Hadamek},
  {Hadasch}, {H{\"a}fner}, {Herrero}, {Hose}, {Hrupec}, {Idec}, {Kadenius},
  {Knoetig}, {Kr{\"a}henb{\"u}hl}, {Krause}, {Kushida}, {La Barbera}, {Lelas},
  {Lewandowska}, {Lindfors}, {Lombardi}, {L{\'o}pez-Coto}, {L{\'o}pez},
  {L{\'o}pez-Oramas}, {Lorenz}, {Lozano}, {Makariev}, {Mallot}, {Maneva},
  {Mankuzhiyil}, {Mannheim}, {Maraschi}, {Marcote}, {Mariotti},
  {Mart{\'\i}nez}, {Masbou}, {Mazin}, {Meucci}, {Miranda}, {Mirzoyan},
  {Mold{\'o}n}, {Moralejo}, {Munar-Adrover}, {Nakajima}, {Niedzwiecki},
  {Nilsson}, {Nowak}, {Orito}, {Overkemping}, {Paiano}, {Palatiello},
  {Paneque}, {Paoletti}, {Paredes}, {Partini}, {Persic}, {Prada}, {Prada
  Moroni}, {Prandini}, {Preziuso}, {Puljak}, {Reichardt}, {Reinthal}, {Rhode},
  {Rib{\'o}}, {Rico}, {R{\"u}gamer}, {Saggion}, {Saito}, {Saito}, {Salvati},
  {Satalecka}, {Scalzotto}, {Scapin}, {Schultz}, {Schweizer}, {Shore},
  {Sillanp{\"a}{\"a}}, {Sitarek}, {Snidaric}, {Sobczynska}, {Spanier}, {Spiro},
  {Stamatescu}, {Stamerra}, {Steinke}, {Storz}, {Sun}, {Suri{\'c}}, {Takalo},
  {Takami}, {Tavecchio}, {Temnikov}, {Terzi{\'c}}, {Tescaro}, {Teshima},
  {Thaele}, {Tibolla}, {Torres}, {Toyama}, {Treves}, {Uellenbeck}, {Vogler},
  {Wagner}, {Weitzel}, {Zandanel}, {Zanin}, \& {MAGIC
  Collaboration}}]{2014AA...563A..91A}
{Aleksi{\'c}}, J., {Antonelli}, L.~A., {Antoranz}, P., {et~al.}
  2014{\natexlab{b}}, \aap, 563, A91, \dodoi{10.1051/0004-6361/201321938}

\bibitem[{{Aleksi{\'c}} {et~al.}(2015{\natexlab{a}}){Aleksi{\'c}}, {Ansoldi},
  {Antonelli}, {Antoranz}, {Babic}, {Bangale}, {Barres de Almeida}, {Barrio},
  {Becerra Gonz{\'a}lez}, {Bednarek}, {Berger}, {Bernardini}, {Biland},
  {Blanch}, {Bonnefoy}, {Bonnoli}, {Borracci}, {Bretz}, {Carmona}, {Carosi},
  {Carreto Fidalgo}, {Colin}, {Colombo}, {Contreras}, {Cortina}, {Covino}, {da
  Vela}, {Dazzi}, {de Angelis}, {de Caneva}, {de Lotto}, {Delgado Mendez},
  {Doert}, {Dom{\'\i}nguez}, {Dominis Prester}, {Dorner}, {Doro}, {Einecke},
  {Eisenacher}, {Elsaesser}, {Farina}, {Ferenc}, {Fonseca}, {Font}, {Frantzen},
  {Fruck}, {Garc{\'\i}a L{\'o}pez}, {Garczarczyk}, {Garrido Terrats}, {Gaug},
  {Godinovi{\'c}}, {Gonz{\'a}lez Mu{\~n}oz}, {Gozzini}, {Hadasch}, {Hayashida},
  {Herrera}, {Herrero}, {Hildebrand}, {Hose}, {Hrupec}, {Idec}, {Kadenius},
  {Kellermann}, {Kodani}, {Konno}, {Krause}, {Kubo}, {Kushida}, {La Barbera},
  {Lelas}, {Lewandowska}, {Lindfors}, {Lombardi}, {L{\'o}pez},
  {L{\'o}pez-Coto}, {L{\'o}pez-Oramas}, {Lorenz}, {Lozano}, {Makariev},
  {Mallot}, {Maneva}, {Mankuzhiyil}, {Mannheim}, {Maraschi}, {Marcote},
  {Mariotti}, {Mart{\'\i}nez}, {Mazin}, {Menzel}, {Meucci}, {Miranda},
  {Mirzoyan}, {Moralejo}, {Munar-Adrover}, {Nakajima}, {Niedzwiecki},
  {Nilsson}, {Nishijima}, {Noda}, {Nowak}, {Orito}, {Overkemping}, {Paiano},
  {Palatiello}, {Paneque}, {Paoletti}, {Paredes}, {Paredes-Fortuny}, {Partini},
  {Persic}, {Prada}, {Moroni}, {Prandini}, {Preziuso}, {Puljak}, {Reinthal},
  {Rhode}, {Rib{\'o}}, {Rico}, {Rodriguez Garcia}, {R{\"u}gamer}, {Saggion},
  {Saito}, {Saito}, {Satalecka}, {Scalzotto}, {Scapin}, {Schultz}, {Schweizer},
  {Shore}, {Sillanp{\"a}{\"a}}, {Sitarek}, {Snidaric}, {Sobczynska}, {Spanier},
  {Stamatescu}, {Stamerra}, {Steinbring}, {Storz}, {Sun}, {Suri{\'c}},
  {Takalo}, {Takami}, {Tavecchio}, {Temnikov}, {Terzi{\'c}}, {Tescaro},
  {Teshima}, {Thaele}, {Tibolla}, {Torres}, {Toyama}, {Treves}, {Uellenbeck},
  {Vogler}, {Wagner}, {Zandanel}, {Zanin}, {MAGIC Collaboration}, {Tronconi},
  {Buson}, \& {Borghese}}]{2015MNRAS.446..217A}
{Aleksi{\'c}}, J., {Ansoldi}, S., {Antonelli}, L.~A., {et~al.}
  2015{\natexlab{a}}, \mnras, 446, 217, \dodoi{10.1093/mnras/stu2024}

\bibitem[{{Aleksi{\'c}} {et~al.}(2015{\natexlab{b}}){Aleksi{\'c}}, {Ansoldi},
  {Antonelli}, {Antoranz}, {Babic}, {Bangale}, {Barrio}, {Becerra
  Gonz{\'a}lez}, {Bednarek}, {Bernardini}, {Biasuzzi}, {Biland}, {Blanch},
  {Bonnefoy}, {Bonnoli}, {Borracci}, {Bretz}, {Carmona}, {Carosi}, {Colin},
  {Colombo}, {Contreras}, {Cortina}, {Covino}, {Da Vela}, {Dazzi}, {De
  Angelis}, {De Caneva}, {De Lotto}, {de O{\~n}a Wilhelmi}, {Delgado Mendez},
  {Di Pierro}, {Dominis Prester}, {Dorner}, {Doro}, {Einecke}, {Eisenacher},
  {Elsaesser}, {Fern{\'a}ndez-Barral}, {Fidalgo}, {Fonseca}, {Font},
  {Frantzen}, {Fruck}, {Galindo}, {Garc{\'\i}a L{\'o}pez}, {Garczarczyk},
  {Garrido Terrats}, {Gaug}, {Godinovi{\'c}}, {Gonz{\'a}lez Mu{\~n}oz},
  {Gozzini}, {Hadasch}, {Hanabata}, {Hayashida}, {Herrera}, {Hose}, {Hrupec},
  {Idec}, {Kadenius}, {Kellermann}, {Knoetig}, {Kodani}, {Konno}, {Krause},
  {Kubo}, {Kushida}, {La Barbera}, {Lelas}, {Lewandowska}, {Lindfors},
  {Lombardi}, {Longo}, {L{\'o}pez}, {L{\'o}pez-Coto}, {L{\'o}pez-Oramas},
  {Lorenz}, {Lozano}, {Makariev}, {Mallot}, {Maneva}, {Mannheim}, {Maraschi},
  {Marcote}, {Mariotti}, {Mart{\'\i}nez}, {Mazin}, {Menzel}, {Miranda},
  {Mirzoyan}, {Moralejo}, {Munar-Adrover}, {Nakajima}, {Neustroev},
  {Niedzwiecki}, {Nievas Rosillo}, {Nilsson}, {Nishijima}, {Noda}, {Orito},
  {Overkemping}, {Paiano}, {Palatiello}, {Paneque}, {Paoletti}, {Paredes},
  {Paredes-Fortuny}, {Persic}, {Poutanen}, {Prada Moroni}, {Prandini},
  {Puljak}, {Reinthal}, {Rhode}, {Rib{\'o}}, {Rico}, {Rodriguez Garcia},
  {Saito}, {Saito}, {Satalecka}, {Scalzotto}, {Scapin}, {Schultz}, {Schweizer},
  {Shore}, {Sillanp{\"a}{\"a}}, {Sitarek}, {Snidaric}, {Sobczynska},
  {Stamerra}, {Steinbring}, {Strzys}, {Takalo}, {Takami}, {Tavecchio},
  {Temnikov}, {Terzi{\'c}}, {Tescaro}, {Teshima}, {Thaele}, {Torres}, {Toyama},
  {Treves}, {Vogler}, {Will}, {Zanin}, {Berger}, {Buson}, {D'Ammando},
  {Gasparrini}, {Hovatta}, {Max-Moerbeck}, {Readhead}, \&
  {Richards}}]{2015MNRAS.451..739A}
---. 2015{\natexlab{b}}, \mnras, 451, 739, \dodoi{10.1093/mnras/stv895}

\bibitem[{{Aleksi{\'c}} {et~al.}(2016){Aleksi{\'c}}, {Ansoldi}, {Antonelli},
  {Antoranz}, {Arcaro}, {Babic}, {Bangale}, {Barres de Almeida}, {Barrio},
  {Becerra Gonz{\'a}lez}, {Bednarek}, {Bernardini}, {Biasuzzi}, {Biland},
  {Blanch}, {Bonnefoy}, {Bonnoli}, {Borracci}, {Bretz}, {Carmona}, {Carosi},
  {Colin}, {Colombo}, {Contreras}, {Cortina}, {Covino}, {Da Vela}, {Dazzi}, {De
  Angelis}, {De Caneva}, {De Lotto}, {de O{\~n}a Wilhelmi}, {Delgado Mendez},
  {Di Pierro}, {Dominis Prester}, {Dorner}, {Doro}, {Einecke}, {Eisenacher},
  {Elsaesser}, {Fern{\'a}ndez-Barral}, {Fidalgo}, {Fonseca}, {Font},
  {Frantzen}, {Fruck}, {Galindo}, {Garc{\'\i}a L{\'o}pez}, {Garczarczyk},
  {Garrido Terrats}, {Gaug}, {Godinovi{\'c}}, {Gonz{\'a}lez Mu{\~n}oz},
  {Gozzini}, {Hadasch}, {Hanabata}, {Hayashida}, {Herrera}, {Hose}, {Hrupec},
  {Idec}, {Kadenius}, {Kellermann}, {Knoetig}, {Kodani}, {Konno}, {Krause},
  {Kubo}, {Kushida}, {La Barbera}, {Lelas}, {Lewandowska}, {Lindfors},
  {Lombardi}, {Longo}, {L{\'o}pez}, {L{\'o}pez-Coto}, {L{\'o}pez-Oramas},
  {Lorenz}, {Lozano}, {Makariev}, {Mallot}, {Maneva}, {Mannheim}, {Maraschi},
  {Marcote}, {Mariotti}, {Mart{\'\i}nez}, {Mazin}, {Menzel}, {Miranda},
  {Mirzoyan}, {Moralejo}, {Munar-Adrover}, {Nakajima}, {Neustroev},
  {Niedzwiecki}, {Nievas Rosillo}, {Nilsson}, {Nishijima}, {Noda}, {Orito},
  {Overkemping}, {Paiano}, {Palatiello}, {Paneque}, {Paoletti}, {Paredes},
  {Paredes-Fortuny}, {Persic}, {Poutanen}, {Prada Moroni}, {Prandini},
  {Puljak}, {Reinthal}, {Rhode}, {Rib{\'o}}, {Rico}, {Rodriguez Garcia},
  {Saito}, {Saito}, {Satalecka}, {Scalzotto}, {Scapin}, {Schweizer}, {Shore},
  {Sillanp{\"a}{\"a}}, {Sitarek}, {Snidaric}, {Sobczynska}, {Stamerra},
  {Steinbring}, {Strzys}, {Takalo}, {Takami}, {Tavecchio}, {Temnikov},
  {Terzi{\'c}}, {Tescaro}, {Teshima}, {Thaele}, {Torres}, {Toyama}, {Treves},
  {Vogler}, {Will}, {Zanin}, {Buson}, {D'Ammando}, {L{\"a}hteenm{\"a}ki},
  {Hovatta}, {Kovalev}, {Lister}, {Max-Moerbeck}, {Mundell}, {Pushkarev},
  {Rastorgueva-Foi}, {Readhead}, {Richards}, {Tammi}, {Sanchez}, {Tornikoski},
  {Savolainen}, \& {Steele}}]{2016AA...591A..10A}
---. 2016, \aap, 591, A10, \dodoi{10.1051/0004-6361/201527176}

\bibitem[{{Aliu} {et~al.}(2011){Aliu}, {Aune}, {Beilicke}, {Benbow},
  {B{\"o}ttcher}, {Bouvier}, {Bradbury}, {Buckley}, {Bugaev}, {Cannon},
  {Cesarini}, {Ciupik}, {Connolly}, {Cui}, {Decerprit}, {Dickherber}, {Duke},
  {Errando}, {Falcone}, {Feng}, {Finnegan}, {Fortson}, {Furniss}, {Galante},
  {Gall}, {Gillanders}, {Godambe}, {Griffin}, {Grube}, {Gyuk}, {Hanna},
  {Hivick}, {Holder}, {Huan}, {Hughes}, {Hui}, {Humensky}, {Kaaret},
  {Karlsson}, {Kertzman}, {Kieda}, {Krawczynski}, {Krennrich}, {Maier},
  {Majumdar}, {McArthur}, {McCann}, {Moriarty}, {Mukherjee}, {Nelson}, {Ong},
  {Orr}, {Otte}, {Park}, {Perkins}, {Pichel}, {Pohl}, {Prokoph}, {Quinn},
  {Ragan}, {Reyes}, {Reynolds}, {Roache}, {Rose}, {Ruppel}, {Saxon},
  {Sembroski}, {Skole}, {Smith}, {Staszak}, {Te{\v{s}}i{\'c}}, {Theiling},
  {Thibadeau}, {Tsurusaki}, {Tyler}, {Varlotta}, {Vassiliev}, {Wakely},
  {Weekes}, {Weinstein}, {Williams}, {Zitzer}, {VERITAS Collaboration},
  {Ciprini}, {Fumagalli}, {Kaplan}, {Paneque}, \&
  {Prochaska}}]{2011ApJ...742..127A}
{Aliu}, E., {Aune}, T., {Beilicke}, M., {et~al.} 2011, \apj, 742, 127,
  \dodoi{10.1088/0004-637X/742/2/127}

\bibitem[{{Aliu} {et~al.}(2012){Aliu}, {Archambault}, {Arlen}, {Aune},
  {Beilicke}, {Benbow}, {B{\"o}ttcher}, {Bouvier}, {Bradbury}, {Buckley},
  {Bugaev}, {Byrum}, {Cannon}, {Cesarini}, {Ciupik}, {Collins-Hughes},
  {Connolly}, {Coppi}, {Cui}, {Decerprit}, {Dickherber}, {Dumm}, {Errando},
  {Falcone}, {Feng}, {Finley}, {Finnegan}, {Fortson}, {Furniss}, {Galante},
  {Gall}, {Godambe}, {Griffin}, {Grube}, {Gyuk}, {Hanna}, {Hawkins}, {Holder},
  {Huan}, {Hughes}, {Humensky}, {Kaaret}, {Karlsson}, {Kertzman}, {Khassen},
  {Kieda}, {Krawczynski}, {Krennrich}, {Lang}, {Lee}, {Madhavan}, {Maier},
  {Majumdar}, {McArthur}, {McCann}, {Moriarty}, {Mukherjee}, {Ong}, {Orr},
  {Otte}, {Palma}, {Park}, {Perkins}, {Pichel}, {Pohl}, {Prokoph}, {Quinn},
  {Ragan}, {Reyes}, {Reynolds}, {Roache}, {Rose}, {Ruppel}, {Saxon},
  {Schroedter}, {Sembroski}, {{\c{S}}ent{\"u}rk}, {Smith}, {Staszak},
  {Telezhinsky}, {Te{\v{s}}i{\'c}}, {Theiling}, {Thibadeau}, {Tsurusaki},
  {Varlotta}, {Vivier}, {Wakely}, {Ward}, {Weekes}, {Weinstein}, {Weisgarber},
  {Williams}, {Zitzer}, {Fortin}, \& {Horan}}]{2012ApJ...750...94A}
{Aliu}, E., {Archambault}, S., {Arlen}, T., {et~al.} 2012, \apj, 750, 94,
  \dodoi{10.1088/0004-637X/750/2/94}

\bibitem[{{Allen} {et~al.}(2017){Allen}, {Archambault}, {Archer}, {Benbow},
  {Bird}, {Bourbeau}, {Brose}, {Buchovecky}, {Buckley}, {Bugaev}, {Cardenzana},
  {Cerruti}, {Chen}, {Christiansen}, {Connolly}, {Cui}, {Daniel}, {Eisch},
  {Falcone}, {Feng}, {Fernandez-Alonso}, {Finley}, {Fleischhack}, {Flinders},
  {Fortson}, {Furniss}, {Gillanders}, {Griffin}, {Grube}, {H{\"u}tten},
  {H{\r{a}}kansson}, {Hanna}, {Hervet}, {Holder}, {Hughes}, {Humensky},
  {Johnson}, {Kaaret}, {Kar}, {Kelley-Hoskins}, {Kertzman}, {Kieda}, {Krause},
  {Krennrich}, {Kumar}, {Lang}, {Maier}, {McArthur}, {McCann}, {Meagher},
  {Moriarty}, {Mukherjee}, {Nguyen}, {Nieto}, {O'Brien}, {de Bhr{\'o}ithe},
  {Ong}, {Otte}, {Park}, {Petrashyk}, {Pichel}, {Pohl}, {Popkow}, {Pueschel},
  {Quinn}, {Ragan}, {Reynolds}, {Richards}, {Roache}, {Rovero}, {Rulten},
  {Sadeh}, {Santander}, {Sembroski}, {Shahinyan}, {Telezhinsky}, {Tucci},
  {Tyler}, {Wakely}, {Weinstein}, {Wilhelm}, \&
  {Williams}}]{2017MNRAS.471.2117A}
{Allen}, C., {Archambault}, S., {Archer}, A., {et~al.} 2017, \mnras, 471, 2117,
  \dodoi{10.1093/mnras/stx1756}

\bibitem[{{Anastassopoulos} {et~al.}(2017){Anastassopoulos}, {Aune}, {Barth},
  {Belov}, {Br{\"a}uninger}, {Cantatore}, {Carmona}, {Castel}, {Cetin},
  {Christensen}, {Collar}, {Dafni}, {Davenport}, {Decker}, {Dermenev}, {Desch},
  {Eleftheriadis}, {Fanourakis}, {Ferrer-Ribas}, {Fischer}, {Garc{\'\i}a},
  {Gardikiotis}, {Garza}, {Gazis}, {Geralis}, {Giomataris}, {Gninenko},
  {Hailey}, {Hasinoff}, {Hoffmann}, {Iguaz}, {Irastorza}, {Jakobsen}, {Jacoby},
  {Jakov{\v{c}}i{\'c}}, {Kaminski}, {Karuza}, {Kralj}, {Kr{\v{c}}mar},
  {Kostoglou}, {Krieger}, {Laki{\'c}}, {Laurent}, {Liolios},
  {Ljubi{\v{c}}i{\'c}}, {Luz{\'o}n}, {Maroudas}, {Miceli}, {Neff}, {Ortega},
  {Papaevangelou}, {Paraschou}, {Pivovaroff}, {Raffelt}, {Rosu}, {Ruz},
  {Ch{\'o}liz}, {Savvidis}, {Schmidt}, {Semertzidis}, {Solanki}, {Stewart},
  {Vafeiadis}, {Vogel}, {Yildiz}, \& {Zioutas}}]{2017NatPh..13..584A}
{Anastassopoulos}, V., {Aune}, S., {Barth}, K., {et~al.} 2017, Nature Physics,
  13, 584, \dodoi{10.1038/nphys4109}

\bibitem[{{Anderhub} {et~al.}(2009){Anderhub}, {Antonelli}, {Antoranz},
  {Backes}, {Baixeras}, {Balestra}, {Barrio}, {Bastieri}, {Becerra
  Gonz{\'a}lez}, {Becker}, {Bednarek}, {Berdyugin}, {Berger}, {Bernardini},
  {Biland}, {Bock}, {Bonnoli}, {Bordas}, {Borla Tridon}, {Bosch-Ramon}, {Bose},
  {Braun}, {Bretz}, {Britzger}, {Camara}, {Carmona}, {Carosi}, {Colin},
  {Commichau}, {Contreras}, {Cortina}, {Costado}, {Covino}, {Dazzi}, {De
  Angelis}, {de Cea del Pozo}, {De los Reyes}, {De Lotto}, {De Maria}, {De
  Sabata}, {Delgado Mendez}, {Dom{\'\i}nguez}, {Dominis Prester}, {Dorner},
  {Doro}, {Elsaesser}, {Errando}, {Ferenc}, {Fern{\'a}ndez}, {Firpo},
  {Fonseca}, {Font}, {Galante}, {Garc{\'\i}a L{\'o}pez}, {Garczarczyk}, {Gaug},
  {Godinovic}, {Goebel}, {Hadasch}, {Herrero}, {Hildebrand},
  {H{\"o}hne-M{\"o}nch}, {Hose}, {Hrupec}, {Hsu}, {Jogler}, {Klepser},
  {Kranich}, {La Barbera}, {Laille}, {Leonardo}, {Lindfors}, {Lombardi},
  {Longo}, {L{\'o}pez}, {Lorenz}, {Majumdar}, {Maneva}, {Mankuzhiyil},
  {Mannheim}, {Maraschi}, {Mariotti}, {Mart{\'\i}nez}, {Mazin}, {Meucci},
  {Miranda}, {Mirzoyan}, {Miyamoto}, {Mold{\'o}n}, {Moles}, {Moralejo},
  {Nieto}, {Nilsson}, {Ninkovic}, {Orito}, {Oya}, {Paoletti}, {Paredes},
  {Pasanen}, {Pascoli}, {Pauss}, {Pegna}, {Perez-Torres}, {Persic}, {Peruzzo},
  {Prada}, {Prandini}, {Puchades}, {Puljak}, {Reichardt}, {Rhode}, {Rib{\'o}},
  {Rico}, {Rissi}, {Robert}, {R{\"u}gamer}, {Saggion}, {Sainio}, {Saito},
  {Salvati}, {S{\'a}nchez-Conde}, {Satalecka}, {Scalzotto}, {Scapin},
  {Schweizer}, {Shayduk}, {Shore}, {Sierpowska-Bartosik}, {Sillanp{\"a}{\"a}},
  {Sitarek}, {Sobczynska}, {Spanier}, {Spiro}, {Stamerra}, {Stark}, {Suric},
  {Takalo}, {Tavecchio}, {Temnikov}, {Tescaro}, {Teshima}, {Torres}, {Turini},
  {Vankov}, {Wagner}, {Villforth}, {Zabalza}, {Zandanel}, {Zanin}, \&
  {Zapatero}}]{2009ApJ...704L.129A}
{Anderhub}, H., {Antonelli}, L.~A., {Antoranz}, P., {et~al.} 2009, \apjl, 704,
  L129, \dodoi{10.1088/0004-637X/704/2/L129}

\bibitem[{Angelis {et~al.}(2007)Angelis, Persic, \&
  Roncadelli}]{2007Constraints}
Angelis, A.~D., Persic, M., \& Roncadelli, M. 2007

\bibitem[{{Archambault} {et~al.}(2013){Archambault}, {Arlen}, {Aune}, {Behera},
  {Beilicke}, {Benbow}, {Bird}, {Bouvier}, {Buckley}, {Bugaev}, {Byrum},
  {Cesarini}, {Ciupik}, {Connolly}, {Cui}, {Errando}, {Falcone}, {Federici},
  {Feng}, {Finley}, {Fortson}, {Furniss}, {Galante}, {Gall}, {Gillanders},
  {Griffin}, {Grube}, {Gyuk}, {Hanna}, {Holder}, {Hughes}, {Humensky},
  {Kaaret}, {Kertzman}, {Khassen}, {Kieda}, {Krawczynski}, {Krennrich},
  {Kumar}, {Lang}, {Madhavan}, {Maier}, {Majumdar}, {McArthur}, {McCann},
  {Millis}, {Moriarty}, {Mukherjee}, {O'Faol{\'a}in de Bhr{\'o}ithe}, {Ong},
  {Otte}, {Park}, {Perkins}, {Pohl}, {Popkow}, {Prokoph}, {Quinn}, {Ragan},
  {Reyes}, {Reynolds}, {Richards}, {Roache}, {Saxon}, {Sembroski}, {Smith},
  {Staszak}, {Telezhinsky}, {Theiling}, {Varlotta}, {Vassiliev}, {Vincent},
  {Wakely}, {Weekes}, {Weinstein}, {Welsing}, {Williams}, {Zitzer}, {VERITAS
  Collaboration}, {B{\"o}ttcher}, {Fegan}, {Fortin}, {Halpern}, {Kovalev},
  {Lister}, {Liu}, {Pushkarev}, \& {Smith}}]{2013ApJ...776...69A}
{Archambault}, S., {Arlen}, T., {Aune}, T., {et~al.} 2013, \apj, 776, 69,
  \dodoi{10.1088/0004-637X/776/2/69}

\bibitem[{{Archambault} {et~al.}(2015){Archambault}, {Archer}, {Beilicke},
  {Benbow}, {Bird}, {Biteau}, {Bouvier}, {Bugaev}, {Cardenzana}, {Cerruti},
  {Chen}, {Ciupik}, {Connolly}, {Cui}, {Dickinson}, {Dumm}, {Eisch}, {Errando},
  {Falcone}, {Feng}, {Finley}, {Fleischhack}, {Fortin}, {Fortson}, {Furniss},
  {Gillanders}, {Griffin}, {Griffiths}, {Grube}, {Gyuk}, {H{\r{a}}kansson},
  {Hanna}, {Holder}, {Humensky}, {Johnson}, {Kaaret}, {Kar}, {Kertzman},
  {Khassen}, {Kieda}, {Krause}, {Krennrich}, {Kumar}, {Lang}, {Maier},
  {McArthur}, {McCann}, {Meagher}, {Millis}, {Moriarty}, {Mukherjee}, {Nieto},
  {O'Faol{\'a}in de Bhr{\'o}ithe}, {Ong}, {Otte}, {Park}, {Pohl}, {Popkow},
  {Prokoph}, {Pueschel}, {Quinn}, {Ragan}, {Reyes}, {Reynolds}, {Richards},
  {Roache}, {Santander}, {Sembroski}, {Shahinyan}, {Smith}, {Staszak},
  {Telezhinsky}, {Tucci}, {Tyler}, {Varlotta}, {Vincent}, {Wakely},
  {Weinstein}, {Welsing}, {Wilhelm}, {Williams}, {Zitzer}, {Veritas
  Collaboration}, \& {Hughes}}]{2015ApJ...808..110A}
{Archambault}, S., {Archer}, A., {Beilicke}, M., {et~al.} 2015, \apj, 808, 110,
  \dodoi{10.1088/0004-637X/808/2/110}

\bibitem[{{Archambault} {et~al.}(2016){Archambault}, {Archer}, {Barnacka},
  {Behera}, {Beilicke}, {Benbow}, {Berger}, {Bird}, {B{\"o}ttcher}, {Buckley},
  {Bugaev}, {Cardenzana}, {Cerruti}, {Chen}, {Christiansen}, {Ciupik},
  {Collins-Hughes}, {Connolly}, {Cui}, {Dickinson}, {Dumm}, {Eisch}, {Errando},
  {Falcone}, {Federici}, {Feng}, {Finley}, {Fleischhack}, {Fortson}, {Furniss},
  {Gillanders}, {Godambe}, {Griffin}, {Griffiths}, {Grube}, {Gyuk},
  {H{\r{a}}kansson}, {Hanna}, {Holder}, {Hughes}, {Johnson}, {Kaaret}, {Kar},
  {Kertzman}, {Khassen}, {Kieda}, {Krawczynski}, {Kumar}, {Lang}, {Madhavan},
  {Maier}, {McArthur}, {McCann}, {Meagher}, {Millis}, {Moriarty}, {Nelson},
  {Nieto}, {de Bhr{\'o}ithe}, {Ong}, {Otte}, {Park}, {Perkins}, {Pohl},
  {Popkow}, {Prokoph}, {Pueschel}, {Quinn}, {Ragan}, {Rajotte}, {Reyes},
  {Reynolds}, {Richards}, {Roache}, {Sembroski}, {Shahinyan}, {Smith},
  {Staszak}, {Sweeney}, {Telezhinsky}, {Tucci}, {Tyler}, {Varlotta},
  {Vassiliev}, {Wakely}, {Welsing}, {Wilhelm}, {Williams}, \&
  {Zitzer}}]{2016MNRAS.461..202A}
{Archambault}, S., {Archer}, A., {Barnacka}, A., {et~al.} 2016, \mnras, 461,
  202, \dodoi{10.1093/mnras/stw1319}

\bibitem[{{Arias} {et~al.}(2010){Arias}, {Jaeckel}, {Redondo}, \&
  {Ringwald}}]{2010PhRvD..82k5018A}
{Arias}, P., {Jaeckel}, J., {Redondo}, J., \& {Ringwald}, A. 2010, \prd, 82,
  115018, \dodoi{10.1103/PhysRevD.82.115018}

\bibitem[{{Astropy Collaboration} {et~al.}(2013){Astropy Collaboration},
  {Robitaille}, {Tollerud}, {Greenfield}, {Droettboom}, {Bray}, {Aldcroft},
  {Davis}, {Ginsburg}, {Price-Whelan}, {Kerzendorf}, {Conley}, {Crighton},
  {Barbary}, {Muna}, {Ferguson}, {Grollier}, {Parikh}, {Nair}, {Unther},
  {Deil}, {Woillez}, {Conseil}, {Kramer}, {Turner}, {Singer}, {Fox}, {Weaver},
  {Zabalza}, {Edwards}, {Azalee Bostroem}, {Burke}, {Casey}, {Crawford},
  {Dencheva}, {Ely}, {Jenness}, {Labrie}, {Lim}, {Pierfederici}, {Pontzen},
  {Ptak}, {Refsdal}, {Servillat}, \& {Streicher}}]{2013A&A...558A..33A}
{Astropy Collaboration}, {Robitaille}, T.~P., {Tollerud}, E.~J., {et~al.} 2013,
  \aap, 558, A33, \dodoi{10.1051/0004-6361/201322068}

\bibitem[{{Baumgartner} {et~al.}(2013){Baumgartner}, {Tueller}, {Markwardt},
  {Skinner}, {Barthelmy}, {Mushotzky}, {Evans}, \&
  {Gehrels}}]{2013ApJS..207...19B}
{Baumgartner}, W.~H., {Tueller}, J., {Markwardt}, C.~B., {et~al.} 2013, \apjs,
  207, 19, \dodoi{10.1088/0067-0049/207/2/19}

\bibitem[{{Benbow}(2011)}]{2011ICRC....8...51B}
{Benbow}, W. 2011, in International Cosmic Ray Conference, Vol.~8,
  International Cosmic Ray Conference, 51, \dodoi{10.7529/ICRC2011/V08/0747}

\bibitem[{{Boller} {et~al.}(2016){Boller}, {Freyberg}, {Tr{\"u}mper}, {Haberl},
  {Voges}, \& {Nandra}}]{2016A&A...588A.103B}
{Boller}, T., {Freyberg}, M.~J., {Tr{\"u}mper}, J., {et~al.} 2016, \aap, 588,
  A103, \dodoi{10.1051/0004-6361/201525648}

\bibitem[{{Bony (de)} {et~al.}(2022){Bony (de)}, {Bylund}, {Meyer}, {Noel},
  {Sanchez}, {Abdalla}, {Aharonian}, {Ait-Benkhali}, {Anguener}, {Arcaro},
  {Armand}, {Armstrong}, {Ashkar}, {Backes}, {Baghmanyan}, {Barbosa Martins},
  {Barnacka}, {Barnard}, {Batzofin}, {Becherini}, {Berge}, {Bernloehr}, {Bi},
  {B{\"o}ttcher}, {Boisson}, {Bolmont}, {Breuhaus}, {Brose}, {Brun}, {Bulik},
  {Cangemi}, {Caroff}, {Casanova}, {Catalano}, {Chambery}, {Chand}, {Chen},
  {Cotter}, {Curlo}, {Dalgleish}, {Damascene Mbarubucyeye}, {Davids}, {Davies},
  {Devin}, {Djannati-Ata{\"\i}}, {Dmytriiev}, {Donath}, {Doroshenko}, {Dreyer},
  {Du Plessis}, {Duffy}, {Egberts}, {Einecke}, {Ernenwein}, {Fegan}, {Feijen},
  {Fiasson}, {Fichet de Clairfontaine}, {Fontaine}, {Frans}, {Fuessling},
  {Funk}, {Gabici}, {Gallant}, {Giavitto}, {Giunti}, {Glawion}, {Glicenstein},
  {Grondin}, {Hattingh}, {Haupt}, {Hermann}, {Hinton}, {Hofmann}, {Hoischen},
  {Holch}, {Holler}, {Horns}, {Huang}, {Huber}, {H{\"o}rbe}, {Jamrozy},
  {Jankowsky}, {Joshi}, {Jung}, {Kasai}, {Katarzynski}, {Katz}, {Khangulyan},
  {Khelifi}, {Klepser}, {Kluzniak}, {Komin}, {Konno}, {Kosack}, {Kostunin},
  {Kreter}, {Kukec Mezek}, {Kundu}, {Lamanna}, {Le Stum}, {Lemiere},
  {Lemoine-Goumard}, {Lenain}, {Leuschner}, {Levy}, {Lohse}, {Luashvili},
  {Lypova}, {Mackey}, {Majumdar}, {Malyshev}, {Malyshev}, {Marandon},
  {Marchegiani}, {Marcowith}, {Mares}, {Marti'i-Devesa}, {Marx}, {Maurin},
  {Meintjes}, {Mitchell}, {Moderski}, {Mohrmann}, {Montanari}, {Moore},
  {Morris}, {Moulin}, {Muller}, {Murach}, {Nakashima}, {Naurois (de)},
  {Nayerhoda}, {Davids}, {Niemiec}, {O'Brien}, {Oberholzer}, {Ohm},
  {Olivera-Nieto}, {Ona-Wilhelmi (de)}, {Ostrowski}, {Panny}, {Panter},
  {Parsons}, {Peron}, {Pita}, {Poireau}, {Prokhorov}, {Prokoph}, {Puehlhofer},
  {Punch}, {Quirrenbach}, {Reichherzer}, {Reimer}, {Reimer}, {Remy}, {Renaud},
  {Reville}, {Rieger}, {Romoli}, {Rowell}, {Rudak}, {Rueda Ricarte}, {Ruiz
  Velasco}, {Sahakian}, {Sailer}, {Salzmann}, {Santangelo}, {Sasaki},
  {Schaefer}, {Schutte}, {Schwanke}, {Sch{\"u}ssler}, {Senniappan}, {Seyffert},
  {Shapopi}, {Shiningayamwe}, {Simoni}, {Sinha}, {Sol}, {Spackman},
  {Specovius}, {Spencer}, {Spir-Jacob}, {Stawarz}, {Steenkamp}, {Stegmann},
  {Steinmassl}, {Steppa}, {Sun}, {Takahashi}, {Tanaka}, {Tavernier}, {Taylor},
  {Terrier}, {Thiersen}, {Thorpe-Morgan}, {Tluczykont}, {Tomankova}, {Tsirou},
  {Tsuji}, {Tuffs}, {Uchiyama}, {van der Walt}, {van Eldik}, {van Rensburg},
  {van Soelen}, {Vasileiadis}, {Veh}, {Venter}, {Vincent}, {Vink}, {V{\"o}lk},
  {Wagner}, {Watson}, {Werner}, {White}, {Wierzcholska}, {Wong}, {Yassin},
  {Yusafzai}, {Zacharias}, {Zanin}, {Zargaryan}, {Zdziarski}, {Zech}, {Zhu},
  {Zmija}, {Zouari}, \& {{\.Z}ywucka}}]{2022icrc.confE.823B}
{Bony (de)}, M., {Bylund}, T., {Meyer}, M., {et~al.} 2022, in 37th
  International Cosmic Ray Conference. 12-23 July 2021. Berlin, 823.
\newblock \doarXiv{2108.02232}

\bibitem[{{Breit} \& {Wheeler}(1934)}]{1934PhRv...46.1087B}
{Breit}, G., \& {Wheeler}, J.~A. 1934, Physical Review, 46, 1087,
  \dodoi{10.1103/PhysRev.46.1087}

\bibitem[{{Cenedese} {et~al.}(2022){Cenedese}, {Franceschini}, \&
  {Galanti}}]{2022MNRAS.516..216C}
{Cenedese}, F., {Franceschini}, A., \& {Galanti}, G. 2022, \mnras, 516, 216,
  \dodoi{10.1093/mnras/stac2123}

\bibitem[{{Cerruti}(2011)}]{2011ICRC....8..109C}
{Cerruti}, M. 2011, in International Cosmic Ray Conference, Vol.~8,
  International Cosmic Ray Conference, 109, \dodoi{10.7529/ICRC2011/V08/0913}

\bibitem[{{Costamante}(2013)}]{2013IJMPD..2230025C}
{Costamante}, L. 2013, International Journal of Modern Physics D, 22, 1330025,
  \dodoi{10.1142/S0218271813300255}

\bibitem[{{Couti{\~n}o de Leon} {et~al.}(2019){Couti{\~n}o de Leon}, {Alonso},
  {Rosa-Gonzalez}, \& {Longinotti}}]{2019ICRC...36..654C}
{Couti{\~n}o de Leon}, S., {Alonso}, A.~C., {Rosa-Gonzalez}, D., \&
  {Longinotti}, A.~L. 2019, in International Cosmic Ray Conference, Vol.~36,
  36th International Cosmic Ray Conference (ICRC2019), 654,
  \dodoi{10.22323/1.358.0654}

\bibitem[{{de Angelis} {et~al.}(2011){de Angelis}, {Galanti}, \&
  {Roncadelli}}]{2011PhRvD..84j5030D}
{de Angelis}, A., {Galanti}, G., \& {Roncadelli}, M. 2011, \prd, 84, 105030,
  \dodoi{10.1103/PhysRevD.84.105030}

\bibitem[{{De Angelis} {et~al.}(2013){De Angelis}, {Galanti}, \&
  {Roncadelli}}]{2013MNRAS.432.3245D}
{De Angelis}, A., {Galanti}, G., \& {Roncadelli}, M. 2013, \mnras, 432, 3245,
  \dodoi{10.1093/mnras/stt684}

\bibitem[{{de Angelis} {et~al.}(2009){de Angelis}, {Mansutti}, {Persic}, \&
  {Roncadelli}}]{2009MNRAS.394L..21D}
{de Angelis}, A., {Mansutti}, O., {Persic}, M., \& {Roncadelli}, M. 2009,
  \mnras, 394, L21, \dodoi{10.1111/j.1745-3933.2008.00602.x}

\bibitem[{{de Angelis} {et~al.}(2007){de Angelis}, {Roncadelli}, \&
  {Mansutti}}]{2007PhRvD..76l1301D}
{de Angelis}, A., {Roncadelli}, M., \& {Mansutti}, O. 2007, \prd, 76, 121301,
  \dodoi{10.1103/PhysRevD.76.121301}

\bibitem[{{Dessert} {et~al.}(2022){Dessert}, {Dunsky}, \&
  {Safdi}}]{2022PhRvD.105j3034D}
{Dessert}, C., {Dunsky}, D., \& {Safdi}, B.~R. 2022, \prd, 105, 103034,
  \dodoi{10.1103/PhysRevD.105.103034}

\bibitem[{{Dom{\'\i}nguez} {et~al.}(2011){Dom{\'\i}nguez}, {Primack},
  {Rosario}, {Prada}, {Gilmore}, {Faber}, {Koo}, {Somerville},
  {P{\'e}rez-Torres}, {P{\'e}rez-Gonz{\'a}lez}, {Huang}, {Davis},
  {Guhathakurta}, {Barmby}, {Conselice}, {Lozano}, {Newman}, \&
  {Cooper}}]{2011MNRAS.410.2556D}
{Dom{\'\i}nguez}, A., {Primack}, J.~R., {Rosario}, D.~J., {et~al.} 2011,
  \mnras, 410, 2556, \dodoi{10.1111/j.1365-2966.2010.17631.x}

\bibitem[{{Donato} {et~al.}(2005){Donato}, {Sambruna}, \&
  {Gliozzi}}]{2005A&A...433.1163D}
{Donato}, D., {Sambruna}, R.~M., \& {Gliozzi}, M. 2005, \aap, 433, 1163,
  \dodoi{10.1051/0004-6361:20034555}

\bibitem[{{Dwek} \& {Krennrich}(2013)}]{2013APh....43..112D}
{Dwek}, E., \& {Krennrich}, F. 2013, Astroparticle Physics, 43, 112,
  \dodoi{10.1016/j.astropartphys.2012.09.003}

\bibitem[{{Ehret}(2008)}]{2008arXiv0812.3495E}
{Ehret}, K. 2008, arXiv e-prints, arXiv:0812.3495.
\newblock \doarXiv{0812.3495}

\bibitem[{{Essey} \& {Kusenko}(2012)}]{2012ApJ...751L..11E}
{Essey}, W., \& {Kusenko}, A. 2012, \apjl, 751, L11,
  \dodoi{10.1088/2041-8205/751/1/L11}

\bibitem[{{Foffano} {et~al.}(2019){Foffano}, {Prandini}, {Franceschini}, \&
  {Paiano}}]{2019ICRC...36..676F}
{Foffano}, L., {Prandini}, E., {Franceschini}, A., \& {Paiano}, S. 2019, in
  International Cosmic Ray Conference, Vol.~36, 36th International Cosmic Ray
  Conference (ICRC2019), 676.
\newblock \doarXiv{1907.13076}

\bibitem[{{Fortin}(2008)}]{2008AIPC.1085..565F}
{Fortin}, P. 2008, in American Institute of Physics Conference Series, Vol.
  1085, American Institute of Physics Conference Series, ed. F.~A. {Aharonian},
  W.~{Hofmann}, \& F.~{Rieger}, 565--568, \dodoi{10.1063/1.3076735}

\bibitem[{{Franceschini} \& {Rodighiero}(2017)}]{2017A&A...603A..34F}
{Franceschini}, A., \& {Rodighiero}, G. 2017, \aap, 603, A34,
  \dodoi{10.1051/0004-6361/201629684}

\bibitem[{{Franceschini} \& {Rodighiero}(2018)}]{2018A&A...614C...1F}
---. 2018, \aap, 614, C1, \dodoi{10.1051/0004-6361/201629684e}

\bibitem[{Furlanetto \& Loeb(2001)}]{Furlanetto_2001}
Furlanetto, S.~R., \& Loeb, A. 2001, The Astrophysical Journal, 556, 619,
  \dodoi{10.1086/321630}

\bibitem[{{Galanti} \& {Roncadelli}(2018)}]{2018PhRvD..98d3018G}
{Galanti}, G., \& {Roncadelli}, M. 2018, \prd, 98, 043018,
  \dodoi{10.1103/PhysRevD.98.043018}

\bibitem[{Galanti {et~al.}(2015)Galanti, Roncadelli, Angelis, \&
  Bignami}]{2015Advantages}
Galanti, G., Roncadelli, M., Angelis, A.~D., \& Bignami, G.~F. 2015, Physics

\bibitem[{{Galanti} {et~al.}(2020){Galanti}, {Roncadelli}, {De Angelis}, \&
  {Bignami}}]{2020MNRAS.493.1553G}
{Galanti}, G., {Roncadelli}, M., {De Angelis}, A., \& {Bignami}, G.~F. 2020,
  \mnras, 493, 1553, \dodoi{10.1093/mnras/stz3410}

\bibitem[{{Gat{\'e}} {et~al.}(2017){Gat{\'e}}, {H.~E.~S.~S. Collaboration}, \&
  {Fitoussi}}]{2017ICRC...35..645G}
{Gat{\'e}}, F., {H.~E.~S.~S. Collaboration}, \& {Fitoussi}, T. 2017, in
  International Cosmic Ray Conference, Vol. 301, 35th International Cosmic Ray
  Conference (ICRC2017), 645.
\newblock \doarXiv{1708.09612}

\bibitem[{{Gilmore} {et~al.}(2012){Gilmore}, {Somerville}, {Primack}, \&
  {Dom{\'\i}nguez}}]{2012MNRAS.422.3189G}
{Gilmore}, R.~C., {Somerville}, R.~S., {Primack}, J.~R., \& {Dom{\'\i}nguez},
  A. 2012, \mnras, 422, 3189, \dodoi{10.1111/j.1365-2966.2012.20841.x}

\bibitem[{{Grasso} \& {Rubinstein}(2001)}]{2001PhR...348..163G}
{Grasso}, D., \& {Rubinstein}, H.~R. 2001, \physrep, 348, 163,
  \dodoi{10.1016/S0370-1573(00)00110-1}

\bibitem[{{H.~E.~S.~S. Collaboration} {et~al.}(2010{\natexlab{a}}){H.~E.~S.~S.
  Collaboration}, {Abramowski}, {Acero}, {Aharonian}, {Akhperjanian}, {Anton},
  {Barres de Almeida}, {Bazer-Bachi}, {Becherini}, {Benbow}, {Bernl{\"o}hr},
  {Bochow}, {Boisson}, {Bolmont}, {Borrel}, {Brucker}, {Brun}, {Brun},
  {B{\"u}hler}, {Bulik}, {B{\"u}sching}, {Boutelier}, {Chadwick},
  {Charbonnier}, {Chaves}, {Cheesebrough}, {Chounet}, {Clapson}, {Coignet},
  {Conrad}, {Costamante}, {Dalton}, {Daniel}, {Davids}, {Degrange}, {Deil},
  {Dickinson}, {Djannati-Ata{\"\i}}, {Domainko}, {O'C. Drury}, {Dubois},
  {Dubus}, {Dyks}, {Dyrda}, {Egberts}, {Eger}, {Espigat}, {Fallon}, {Farnier},
  {Fegan}, {Feinstein}, {Fernandes}, {Fiasson}, {F{\"o}rster}, {Fontaine},
  {F{\"u}{\ss}ling}, {Gabici}, {Gallant}, {G{\'e}rard}, {Gerbig}, {Giebels},
  {Glicenstein}, {Gl{\"u}ck}, {Goret}, {G{\"o}ring}, {Hampf}, {Hauser},
  {Heinz}, {Heinzelmann}, {Henri}, {Hermann}, {Hinton}, {Hoffmann}, {Hofmann},
  {Hofverberg}, {Holleran}, {Hoppe}, {Horns}, {Jacholkowska}, {de Jager},
  {Jahn}, {Jung}, {Katarzy{\'n}ski}, {Katz}, {Kaufmann}, {Kerschhaggl},
  {Khangulyan}, {Kh{\'e}lifi}, {Keogh}, {Klochkov}, {Klu{\'z}niak}, {Kneiske},
  {Komin}, {Kosack}, {Kossakowski}, {Lamanna}, {Lenain}, {Lohse}, {Lu},
  {Marandon}, {Marcowith}, {Masbou}, {Maurin}, {McComb}, {Medina},
  {M{\'e}hault}, {Moderski}, {Moulin}, {Naumann-Godo}, {de Naurois}, {Nedbal},
  {Nekrassov}, {Nguyen}, {Nicholas}, {Niemiec}, {Nolan}, {Ohm}, {Olive}, {de
  O{\~n}a Wilhelmi}, {Opitz}, {Orford}, {Ostrowski}, {Panter}, {Paz Arribas},
  {Pedaletti}, {Pelletier}, {Petrucci}, {Pita}, {P{\"u}hlhofer}, {Punch},
  {Quirrenbach}, {Raubenheimer}, {Raue}, {Rayner}, {Reimer}, {Renaud}, {de los
  Reyes}, {Rieger}, {Ripken}, {Rob}, {Rosier-Lees}, {Rowell}, {Rudak},
  {Rulten}, {Ruppel}, {Ryde}, {Sahakian}, {Santangelo}, {Schlickeiser},
  {Sch{\"o}ck}, {Sch{\"o}nwald}, {Schwanke}, {Schwarzburg}, {Schwemmer},
  {Shalchi}, {Sushch}, {Sikora}, {Skilton}, {Sol}, {Stawarz}, {Steenkamp},
  {Stegmann}, {Stinzing}, {Superina}, {Szostek}, {Tam}, {Tavernet}, {Terrier},
  {Tibolla}, {Tluczykont}, {Valerius}, {van Eldik}, {Vasileiadis}, {Venter},
  {Venter}, {Vialle}, {Viana}, {Vincent}, {Vivier}, {V{\"o}lk}, {Volpe},
  {Vorobiov}, {Wagner}, {Ward}, {Zdziarski}, {Zech}, \&
  {Zechlin}}]{2010AA...520A..83H}
{H.~E.~S.~S. Collaboration}, {Abramowski}, A., {Acero}, F., {et~al.}
  2010{\natexlab{a}}, \aap, 520, A83, \dodoi{10.1051/0004-6361/201014484}

\bibitem[{{H.~E.~S.~S. Collaboration} {et~al.}(2010{\natexlab{b}}){H.~E.~S.~S.
  Collaboration}, {Acero}, {Aharonian}, {Akhperjanian}, {Anton}, {Barres de
  Almeida}, {Bazer-Bachi}, {Becherini}, {Behera}, {Benbow}, {Bernl{\"o}hr},
  {Bochow}, {Boisson}, {Bolmont}, {Borrel}, {Brucker}, {Brun}, {Brun},
  {B{\"u}hler}, {Bulik}, {B{\"u}sching}, {Boutelier}, {Chadwick},
  {Charbonnier}, {Chaves}, {Cheesebrough}, {Chounet}, {Clapson}, {Coignet},
  {Costamante}, {Dalton}, {Daniel}, {Davids}, {Degrange}, {Deil}, {Dickinson},
  {Djannati-Ata{\"\i}}, {Domainko}, {O'C. Drury}, {Dubois}, {Dubus}, {Dyks},
  {Dyrda}, {Egberts}, {Eger}, {Espigat}, {Fallon}, {Farnier}, {Fegan},
  {Feinstein}, {Fiasson}, {F{\"o}rster}, {Fontaine}, {F{\"u}{\ss}ling},
  {Gabici}, {Gallant}, {G{\'e}rard}, {Gerbig}, {Giebels}, {Glicenstein},
  {Gl{\"u}ck}, {Goret}, {G{\"o}ring}, {Hauser}, {Heinz}, {Heinzelmann},
  {Henri}, {Hermann}, {Hinton}, {Hoffmann}, {Hofmann}, {Hofverberg},
  {Holleran}, {Hoppe}, {Horns}, {Jacholkowska}, {de Jager}, {Jahn}, {Jung},
  {Katarzy{\'n}ski}, {Katz}, {Kaufmann}, {Kerschhaggl}, {Khangulyan},
  {Kh{\'e}lifi}, {Keogh}, {Klochkov}, {Klu{\'z}niak}, {Kneiske}, {Komin},
  {Kosack}, {Kossakowski}, {Lamanna}, {Lenain}, {Lohse}, {Marandon},
  {Martineau-Huynh}, {Marcowith}, {Masbou}, {Maurin}, {McComb}, {Medina},
  {M{\'e}hault}, {Moderski}, {Moulin}, {Naumann-Godo}, {de Naurois}, {Nedbal},
  {Nekrassov}, {Nicholas}, {Niemiec}, {Nolan}, {Ohm}, {Olive}, {de O{\~n}a
  Wilhelmi}, {Orford}, {Ostrowski}, {Panter}, {Paz Arribas}, {Pedaletti},
  {Pelletier}, {Petrucci}, {Pita}, {P{\"u}hlhofer}, {Punch}, {Quirrenbach},
  {Raubenheimer}, {Raue}, {Rayner}, {Renaud}, {Rieger}, {Ripken}, {Rob},
  {Rosier-Lees}, {Rowell}, {Rudak}, {Rulten}, {Ruppel}, {Sahakian},
  {Santangelo}, {Schlickeiser}, {Sch{\"o}ck}, {Schwanke}, {Schwarzburg},
  {Schwemmer}, {Shalchi}, {Sikora}, {Skilton}, {Sol}, {Stawarz}, {Steenkamp},
  {Stegmann}, {Stinzing}, {Superina}, {Szostek}, {Tam}, {Tavernet}, {Terrier},
  {Tibolla}, {Tluczykont}, {van Eldik}, {Vasileiadis}, {Venter}, {Venter},
  {Vialle}, {Vincent}, {Vivier}, {V{\"o}lk}, {Volpe}, {Wagner}, {Ward},
  {Zdziarski}, \& {Zech}}]{2010AA...511A..52H}
{H.~E.~S.~S. Collaboration}, {Acero}, F., {Aharonian}, F., {et~al.}
  2010{\natexlab{b}}, \aap, 511, A52, \dodoi{10.1051/0004-6361/200913073}

\bibitem[{{H.~E.~S.~S. Collaboration} {et~al.}(2012{\natexlab{a}}){H.~E.~S.~S.
  Collaboration}, {Abramowski}, {Acero}, {Aharonian}, {Akhperjanian}, {Anton},
  {Balzer}, {Barnacka}, {Becherini}, {Becker}, {Bernl{\"o}hr}, {Birsin},
  {Biteau}, {Bochow}, {Boisson}, {Bolmont}, {Bordas}, {Brucker}, {Brun},
  {Brun}, {Bulik}, {B{\"u}sching}, {Carrigan}, {Casanova}, {Cerruti},
  {Chadwick}, {Charbonnier}, {Chaves}, {Cheesebrough}, {Cologna}, {Conrad},
  {Dalton}, {Daniel}, {Davids}, {Degrange}, {Deil}, {Dickinson},
  {Djannati-Ata{\"\i}}, {Domainko}, {Drury}, {Dubus}, {Dutson}, {Dyks},
  {Dyrda}, {Egberts}, {Eger}, {Espigat}, {Fallon}, {Fegan}, {Feinstein},
  {Fernandes}, {Fiasson}, {Fontaine}, {F{\"o}rster}, {F{\"u}{\ss}ling},
  {Gallant}, {Gast}, {G{\'e}rard}, {Gerbig}, {Giebels}, {Glicenstein},
  {Gl{\"u}ck}, {G{\"o}ring}, {H{\"a}ffner}, {Hague}, {Hahn}, {Hampf}, {Harris},
  {Hauser}, {Heinz}, {Heinzelmann}, {Henri}, {Hermann}, {Hillert}, {Hinton},
  {Hofmann}, {Hofverberg}, {Holler}, {Horns}, {Jacholkowska}, {de Jager},
  {Jahn}, {Jamrozy}, {Jung}, {Kastendieck}, {Katarzy{\'n}ski}, {Katz},
  {Kaufmann}, {Keogh}, {Kh{\'e}lifi}, {Klochkov}, {Klu{\'z}niak}, {Kneiske},
  {Komin}, {Kosack}, {Kossakowski}, {Krayzel}, {Laffon}, {Lamanna}, {Lenain},
  {Lennarz}, {Lohse}, {Lopatin}, {Lu}, {Marandon}, {Marcowith}, {Masbou},
  {Maxted}, {Mayer}, {McComb}, {Medina}, {M{\'e}hault}, {Moderski}, {Mohamed},
  {Moulin}, {Naumann}, {Naumann-Godo}, {de Naurois}, {Nedbal}, {Nekrassov},
  {Nguyen}, {Nicholas}, {Niemiec}, {Nolan}, {Ohm}, {de O{\~n}a Wilhelmi},
  {Opitz}, {Ostrowski}, {Oya}, {Panter}, {Paz Arribas}, {Pekeur}, {Pelletier},
  {Perez}, {Petrucci}, {Peyaud}, {Pita}, {P{\"u}hlhofer}, {Punch},
  {Quirrenbach}, {Raue}, {Rayner}, {Reimer}, {Reimer}, {Renaud}, {de los
  Reyes}, {Rieger}, {Ripken}, {Rob}, {Rosier-Lees}, {Rowell}, {Rudak},
  {Rulten}, {Sahakian}, {Sanchez}, {Santangelo}, {Schlickeiser}, {Schulz},
  {Schwanke}, {Schwarzburg}, {Schwemmer}, {Sheidaei}, {Skilton}, {Sol},
  {Spengler}, {Stawarz}, {Steenkamp}, {Stegmann}, {Stinzing}, {Stycz},
  {Sushch}, {Szostek}, {Tavernet}, {Terrier}, {Tluczykont}, {Valerius}, {van
  Eldik}, {Vasileiadis}, {Venter}, {Viana}, {Vincent}, {V{\"o}lk}, {Volpe},
  {Vorobiov}, {Vorster}, {Wagner}, {Ward}, {White}, {Wierzcholska},
  {Zacharias}, {Zajczyk}, {Zdziarski}, {Zech}, \&
  {Zechlin}}]{2012AA...542A..94H}
{H.~E.~S.~S. Collaboration}, {Abramowski}, A., {Acero}, F., {et~al.}
  2012{\natexlab{a}}, \aap, 542, A94, \dodoi{10.1051/0004-6361/201218910}

\bibitem[{{H.~E.~S.~S. Collaboration} {et~al.}(2012{\natexlab{b}}){H.~E.~S.~S.
  Collaboration}, {Abramowski}, {Acero}, {Aharonian}, {Akhperjanian}, {Anton},
  {Balzer}, {Barnacka}, {Barres de Almeida}, {Becherini}, {Becker}, {Behera},
  {Bernloehr}, {Birsin}, {Biteau}, {Bochow}, {Boisson}, {Bolmont}, {Bordas},
  {Brucker}, {Brun}, {Brun}, {Bulik}, {Buesching}, {Carrigan}, {Casanova},
  {Cerruti}, {Chadwick}, {Charbonnier}, {Chaves}, {Cheesebrough}, {Chounet},
  {Clapson}, {Coignet}, {Cologna}, {Conrad}, {Dalton}, {Daniel}, {Davids},
  {Degrange}, {Deil}, {Dickinson}, {Djannati-Ataie}, {Domainko}, {Drury},
  {Dubois}, {Dubus}, {Dutson}, {Dyks}, {Dyrda}, {Egberts}, {Eger}, {Espigat},
  {Fallon}, {Farnier}, {Feinstein}, {Fernandes}, {Fiasson}, {Fontaine},
  {Foerster}, {Fuesling}, {Gallant}, {Gast}, {Gerard}, {Gerbig}, {Giebels},
  {Glicenstein}, {Glueck}, {Goret}, {Goering}, {Haeffner}, {Hague}, {Hampf},
  {Hauser}, {Heinz}, {Heinzelmann}, {Henri}, {Hermann}, {Hinton}, {Hoffmann},
  {Hofmann}, {Hofverberg}, {Holler}, {Horns}, {Jacholkowska}, {de Jager},
  {Jahn}, {Jamrozy}, {Jung}, {Kastendieck}, {Katarzynski}, {Katz}, {Kaufmann},
  {Keogh}, {Khangulyan}, {Khelifi}, {Klochkov}, {Kluzniak}, {Kneiske}, {Komin},
  {Kosack}, {Kossakowski}, {Laffon}, {Lamanna}, {Lennarz}, {Lohse}, {Lopatin},
  {Lu}, {Marandon}, {Marcowith}, {Masbou}, {Maurin}, {Maxted}, {Mayer},
  {McComb}, {Medina}, {Mehault}, {Moderski}, {Moulin}, {Naumann},
  {Naumann-Godo}, {de Naurois}, {Nedbal}, {Nekrassov}, {Nguyen}, {Nicholas},
  {Niemiec}, {Nolan}, {Ohm}, {de Ona Wilhelmi}, {Opitz}, {Ostrowski}, {Oya},
  {Panter}, {Paz Arribas}, {Pedaletti}, {Pelletier}, {Petrucci}, {Pita},
  {Puehlhofer}, {Punch}, {Quirrenbach}, {Raue}, {Rayner}, {Reimer}, {Reimer},
  {Renaud}, {de Los Reyes}, {Rieger}, {Ripken}, {Rob}, {Rosier-Lees}, {Rowell},
  {Rudak}, {Rulten}, {Ruppel}, {Sahakian}, {Sanchez}, {Santangelo},
  {Schlickeiser}, {Schoeck}, {Schulz}, {Schwanke}, {Schwarzburg}, {Schwemmer},
  {Sheidaei}, {Sikora}, {Skilton}, {Sol}, {Spengler}, {Stawarz}, {Steenkamp},
  {Stegmann}, {Stinzing}, {Stycz}, {Sushch}, {Szostek}, {Tavernet}, {Terrier},
  {Tluczykont}, {Valerius}, {van Eldik}, {Vasileiadis}, {Venter}, {Vialle},
  {Viana}, {Vincent}, {Voelk}, {Volpe}, {Vorobiov}, {Vorster}, {Wagner},
  {Ward}, {White}, {Wierzcholska}, {Zacharias}, {Zajczyk}, {Zdziarski}, {Zech},
  {Zechlin}, {Costamante}, {Fegan}, \& {Ajello}}]{2012AA...538A.103H}
---. 2012{\natexlab{b}}, \aap, 538, A103, \dodoi{10.1051/0004-6361/201118406}

\bibitem[{{H.~E.~S.~S. Collaboration} {et~al.}(2013{\natexlab{a}}){H.~E.~S.~S.
  Collaboration}, {Abramowski}, {Acero}, {Aharonian}, {Akhperjanian},
  {Ang{\"u}ner}, {Anton}, {Balenderan}, {Balzer}, {Barnacka}, {Becherini},
  {Becker Tjus}, {Bernl{\"o}hr}, {Birsin}, {Bissaldi}, {Biteau}, {Boisson},
  {Bolmont}, {Bordas}, {Brucker}, {Brun}, {Brun}, {Bulik}, {Carrigan},
  {Casanova}, {Cerruti}, {Chadwick}, {Chalme-Calvet}, {Chaves}, {Cheesebrough},
  {Chr{\'e}tien}, {Colafrancesco}, {Cologna}, {Conrad}, {Couturier}, {Dalton},
  {Daniel}, {Davids}, {Degrange}, {Deil}, {deWilt}, {Dickinson},
  {Djannati-Ata{\"\i}}, {Domainko}, {O'C. Drury}, {Dubus}, {Dutson}, {Dyks},
  {Dyrda}, {Edwards}, {Egberts}, {Eger}, {Espigat}, {Farnier}, {Fegan},
  {Feinstein}, {Fernandes}, {Fernandez}, {Fiasson}, {Fontaine}, {F{\"o}rster},
  {F{\"u}{\ss}ling}, {Gajdus}, {Gallant}, {Garrigoux}, {Gast}, {Giebels},
  {Glicenstein}, {G{\"o}ring}, {Grondin}, {Grudzi{\'n}ska}, {H{\"a}ffner},
  {Hague}, {Hahn}, {Harris}, {Heinzelmann}, {Henri}, {Hermann}, {Hervet},
  {Hillert}, {Hinton}, {Hofmann}, {Hofverberg}, {Holler}, {Horns},
  {Jacholkowska}, {Jahn}, {Jamrozy}, {Janiak}, {Jankowsky}, {Jung},
  {Kastendieck}, {Katarzy{\'n}ski}, {Katz}, {Kaufmann}, {Kh{\'e}lifi},
  {Kieffer}, {Klepser}, {Klochkov}, {Klu{\'z}niak}, {Kneiske}, {Kolitzus},
  {Komin}, {Kosack}, {Krakau}, {Krayzel}, {Kr{\"u}ger}, {Laffon}, {Lamanna},
  {Lefaucheur}, {Lemoine-Goumard}, {Lenain}, {Lennarz}, {Lohse}, {Lopatin},
  {Lu}, {Marandon}, {Marcowith}, {Maurin}, {Maxted}, {Mayer}, {McComb},
  {Medina}, {M{\'e}hault}, {Menzler}, {Meyer}, {Moderski}, {Mohamed}, {Moulin},
  {Murach}, {Naumann}, {de Naurois}, {Nedbal}, {Niemiec}, {Nolan}, {Oakes},
  {Ohm}, {de O{\~n}a Wilhelmi}, {Opitz}, {Ostrowski}, {Oya}, {Panter},
  {Parsons}, {Paz Arribas}, {Pekeur}, {Pelletier}, {Perez}, {Petrucci},
  {Peyaud}, {Pita}, {Poon}, {P{\"u}hlhofer}, {Punch}, {Quirrenbach}, {Raab},
  {Raue}, {Reimer}, {Reimer}, {Renaud}, {de los Reyes}, {Rieger}, {Rob},
  {Rosier-Lees}, {Rowell}, {Rudak}, {Rulten}, {Sahakian}, {Sanchez},
  {Santangelo}, {Schlickeiser}, {Sch{\"u}ssler}, {Schulz}, {Schwanke},
  {Schwarzburg}, {Schwemmer}, {Sol}, {Spengler}, {Spie{\ss}}, {Stawarz},
  {Steenkamp}, {Stegmann}, {Stinzing}, {Stycz}, {Sushch}, {Szostek},
  {Tavernet}, {Terrier}, {Tluczykont}, {Trichard}, {Valerius}, {van Eldik},
  {Vasileiadis}, {Venter}, {Viana}, {Vincent}, {V{\"o}lk}, {Volpe}, {Vorster},
  {Wagner}, {Wagner}, {Ward}, {Weidinger}, {White}, {Wierzcholska}, {Willmann},
  {W{\"o}rnlein}, {Wouters}, {Zacharias}, {Zajczyk}, {Zdziarski}, {Zech}, \&
  {Zechlin}}]{2013AA...554A..72H}
---. 2013{\natexlab{a}}, \aap, 554, A72, \dodoi{10.1051/0004-6361/201220996}

\bibitem[{{H.~E.~S.~S. Collaboration} {et~al.}(2013{\natexlab{b}}){H.~E.~S.~S.
  Collaboration}, {Abramowski}, {Acero}, {Aharonian}, {Ait Benkhali},
  {Akhperjanian}, {Ang{\"u}ner}, {Anton}, {Balenderan}, {Balzer}, {Barnacka},
  {Becherini}, {Becker Tjus}, {Bernl{\"o}hr}, {Birsin}, {Bissaldi}, {Biteau},
  {B{\"o}ttcher}, {Boisson}, {Bolmont}, {Bordas}, {Brucker}, {Brun}, {Brun},
  {Bulik}, {Carrigan}, {Casanova}, {Cerruti}, {Chadwick}, {Chalme-Calvet},
  {Chaves}, {Cheesebrough}, {Chr{\'e}tien}, {Colafrancesco}, {Cologna},
  {Conrad}, {Couturier}, {Dalton}, {Daniel}, {Davids}, {Degrange}, {Deil},
  {deWilt}, {Dickinson}, {Djannati-Ata{\"\i}}, {Domainko}, {Drury}, {Dubus},
  {Dutson}, {Dyks}, {Dyrda}, {Edwards}, {Egberts}, {Eger}, {Espigat},
  {Farnier}, {Fegan}, {Feinstein}, {Fernandes}, {Fernandez}, {Fiasson},
  {Fontaine}, {F{\"o}rster}, {F{\"u}{\ss}ling}, {Gajdus}, {Gallant},
  {Garrigoux}, {Giebels}, {Glicenstein}, {Grondin}, {Grudzi{\'n}ska},
  {H{\"a}ffner}, {Hague}, {Hahn}, {Harris}, {Heinzelmann}, {Henri}, {Hermann},
  {Hervet}, {Hillert}, {Hinton}, {Hofmann}, {Hofverberg}, {Holler}, {Horns},
  {Jacholkowska}, {Jahn}, {Jamrozy}, {Janiak}, {Jankowsky}, {Jung},
  {Kastendieck}, {Katarzy{\'n}ski}, {Katz}, {Kaufmann}, {Kh{\'e}lifi},
  {Kieffer}, {Klepser}, {Klochkov}, {Klu{\'z}niak}, {Kneiske}, {Kolitzus},
  {Komin}, {Kosack}, {Krakau}, {Krayzel}, {Kr{\"u}ger}, {Laffon}, {Lamanna},
  {Lefaucheur}, {Lemoine-Goumard}, {Lenain}, {Lennarz}, {Lohse}, {Lopatin},
  {Lu}, {Marandon}, {Marcowith}, {Marx}, {Maurin}, {Maxted}, {Mayer}, {McComb},
  {Medina}, {M{\'e}hault}, {Menzler}, {Meyer}, {Moderski}, {Mohamed}, {Moulin},
  {Murach}, {Naumann}, {de Naurois}, {Nedbal}, {Niemiec}, {Nolan}, {Oakes},
  {Ohm}, {de O{\~n}a Wilhelmi}, {Opitz}, {Ostrowski}, {Oya}, {Panter},
  {Parsons}, {Paz Arribas}, {Pekeur}, {Pelletier}, {Perez}, {Petrucci},
  {Peyaud}, {Pita}, {Poon}, {P{\"u}hlhofer}, {Punch}, {Quirrenbach}, {Raab},
  {Raue}, {Reimer}, {Reimer}, {Renaud}, {de los Reyes}, {Rieger}, {Rob},
  {Rosier-Lees}, {Rowell}, {Rudak}, {Rulten}, {Sahakian}, {Sanchez},
  {Santangelo}, {Schlickeiser}, {Sch{\"u}ssler}, {Schulz}, {Schwanke},
  {Schwarzburg}, {Schwemmer}, {Sol}, {Spengler}, {Spies}, {Stawarz},
  {Steenkamp}, {Stegmann}, {Stinzing}, {Stycz}, {Sushch}, {Szostek},
  {Tavernet}, {Terrier}, {Tluczykont}, {Trichard}, {Valerius}, {van Eldik},
  {Vasileiadis}, {Venter}, {Viana}, {Vincent}, {V{\"o}lk}, {Volpe}, {Vorster},
  {Wagner}, {Wagner}, {Ward}, {Weidinger}, {Weitzel}, {White}, {Wierzcholska},
  {Willmann}, {W{\"o}rnlein}, {Wouters}, {Zacharias}, {Zajczyk}, {Zdziarski},
  {Zech}, \& {Zechlin}}]{2013AA...559A.136H}
---. 2013{\natexlab{b}}, \aap, 559, A136, \dodoi{10.1051/0004-6361/201321639}

\bibitem[{{H.~E.~S.~S. Collaboration} {et~al.}(2013{\natexlab{c}}){H.~E.~S.~S.
  Collaboration}, {Abramowski}, {Acero}, {Akhperjanian}, {Anton}, {Balenderan},
  {Balzer}, {Barnacka}, {Becherini}, {Becker Tjus}, {Behera}, {Bernl{\"o}hr},
  {Birsin}, {Biteau}, {Bochow}, {Boisson}, {Bolmont}, {Bordas}, {Brucker},
  {Brun}, {Brun}, {Bulik}, {Carrigan}, {Casanova}, {Cerruti}, {Chadwick},
  {Chaves}, {Cheesebrough}, {Colafrancesco}, {Cologna}, {Conrad}, {Couturier},
  {Dalton}, {Daniel}, {Davids}, {Degrange}, {Deil}, {deWilt}, {Dickinson},
  {Djannati-Ata{\"\i}}, {Domainko}, {O'C. Drury}, {Dubus}, {Dutson}, {Dyks},
  {Dyrda}, {Egberts}, {Eger}, {Espigat}, {Fallon}, {Farnier}, {Fegan},
  {Feinstein}, {Fernandes}, {Fernandez}, {Fiasson}, {Fontaine}, {F{\"o}rster},
  {F{\"u}{\ss}ling}, {Gajdus}, {Gallant}, {Garrigoux}, {Gast}, {Giebels},
  {Glicenstein}, {Gl{\"u}ck}, {G{\"o}ring}, {Grondin}, {Grudzi{\'n}ska},
  {H{\"a}ffner}, {Hague}, {Hahn}, {Hampf}, {Harris}, {Heinz}, {Heinzelmann},
  {Henri}, {Hermann}, {Hillert}, {Hinton}, {Hofmann}, {Hofverberg}, {Holler},
  {Horns}, {Jacholkowska}, {Jahn}, {Jamrozy}, {Jung}, {Kastendieck},
  {Katarzy{\'n}ski}, {Katz}, {Kaufmann}, {Kh{\'e}lifi}, {Klepser}, {Klochkov},
  {Klu{\'z}niak}, {Kneiske}, {Kolitzus}, {Komin}, {Kosack}, {Kossakowski},
  {Krayzel}, {Kr{\"u}ger}, {Laffon}, {Lamanna}, {Lefaucheur},
  {Lemoine-Goumard}, {Lenain}, {Lennarz}, {Lohse}, {Lopatin}, {Lu}, {Marandon},
  {Marcowith}, {Masbou}, {Maurin}, {Maxted}, {Mayer}, {McComb}, {Medina},
  {M{\'e}hault}, {Menzler}, {Moderski}, {Mohamed}, {Moulin}, {Naumann},
  {Naumann-Godo}, {de Naurois}, {Nedbal}, {Nguyen}, {Niemiec}, {Nolan}, {Ohm},
  {de O{\~n}a Wilhelmi}, {Opitz}, {Ostrowski}, {Oya}, {Panter}, {Parsons}, {Paz
  Arribas}, {Pekeur}, {Pelletier}, {Perez}, {Petrucci}, {Peyaud}, {Pita},
  {P{\"u}hlhofer}, {Punch}, {Quirrenbach}, {Raab}, {Raue}, {Reimer}, {Reimer},
  {Renaud}, {de los Reyes}, {Rieger}, {Ripken}, {Rob}, {Rosier-Lees}, {Rowell},
  {Rudak}, {Rulten}, {Sahakian}, {Sanchez}, {Santangelo}, {Schlickeiser},
  {Schulz}, {Schwanke}, {Schwarzburg}, {Schwemmer}, {Sheidaei}, {Skilton},
  {Sol}, {Spengler}, {Stawarz}, {Steenkamp}, {Stegmann}, {Stinzing}, {Stycz},
  {Sushch}, {Szostek}, {Tavernet}, {Terrier}, {Tluczykont}, {Trichard},
  {Valerius}, {van Eldik}, {Vasileiadis}, {Venter}, {Viana}, {Vincent},
  {V{\"o}lk}, {Volpe}, {Vorobiov}, {Vorster}, {Wagner}, {Ward}, {White},
  {Wierzcholska}, {Wouters}, {Zacharias}, {Zajczyk}, {Zdziarski}, {Zech},
  {Zechlin}, \& {Pelat}}]{2013AA...552A.118H}
---. 2013{\natexlab{c}}, \aap, 552, A118, \dodoi{10.1051/0004-6361/201321108}

\bibitem[{{H.~E.~S.~S. Collaboration} {et~al.}(2020){H.~E.~S.~S.
  Collaboration}, {Abdalla}, {Adam}, {Aharonian}, {Ait Benkhali},
  {Ang{\"u}ner}, {Arakawa}, {Arcaro}, {Armand}, {Ashkar}, {Backes}, {Barbosa
  Martins}, {Barnard}, {Becherini}, {Berge}, {Bernl{\"o}hr}, {Blackwell},
  {B{\"o}ttcher}, {Boisson}, {Bolmont}, {Bonnefoy}, {Bregeon}, {Breuhaus},
  {Brun}, {Brun}, {Bryan}, {B{\"u}chele}, {Bulik}, {Bylund}, {Capasso},
  {Caroff}, {Carosi}, {Casanova}, {Cerruti}, {Chand}, {Chandra}, {Chen},
  {Colafrancesco}, {Cury{\l}o}, {Davids}, {Deil}, {Devin}, {deWilt}, {Dirson},
  {Djannati-Ata{\"\i}}, {Dmytriiev}, {Donath}, {Doroshenko}, {Drury}, {Dyks},
  {Egberts}, {Emery}, {Ernenwein}, {Eschbach}, {Feijen}, {Fegan}, {Fiasson},
  {Fontaine}, {Funk}, {F{\"u}{\ss}ling}, {Gabici}, {Gallant}, {Gat{\'e}},
  {Giavitto}, {Glawion}, {Glicenstein}, {Gottschall}, {Grondin}, {Hahn},
  {Haupt}, {Heinzelmann}, {Henri}, {Hermann}, {Hinton}, {Hofmann}, {Hoischen},
  {Holch}, {Holler}, {Horns}, {Huber}, {Iwasaki}, {Jamrozy}, {Jankowsky},
  {Jankowsky}, {Jardin-Blicq}, {Jung-Richardt}, {Kastendieck},
  {Katarzy{\'n}ski}, {Katsuragawa}, {Katz}, {Khangulyan}, {Kh{\'e}lifi},
  {King}, {Klepser}, {Klu{\'z}niak}, {Komin}, {Kosack}, {Kostunin}, {Kraus},
  {Lamanna}, {Lau}, {Lemi{\`e}re}, {Lemoine-Goumard}, {Lenain}, {Leser},
  {Levy}, {Lohse}, {Lypova}, {Mackey}, {Majumdar}, {Malyshev}, {Marandon},
  {Marcowith}, {Mares}, {Mariaud}, {Mart{\'\i}-Devesa}, {Marx}, {Maurin},
  {Meintjes}, {Mitchell}, {Moderski}, {Mohamed}, {Mohrmann}, {Muller}, {Moore},
  {Moulin}, {Murach}, {Nakashima}, {de Naurois}, {Ndiyavala}, {Niederwanger},
  {Niemiec}, {Oakes}, {O'Brien}, {Odaka}, {Ohm}, {de O{\~n}a Wilhelmi},
  {Ostrowski}, {Oya}, {Panter}, {Parsons}, {Perennes}, {Petrucci}, {Peyaud},
  {Piel}, {Pita}, {Poireau}, {Priyana Noel}, {Prokhorov}, {Prokoph},
  {P{\"u}hlhofer}, {Punch}, {Quirrenbach}, {Raab}, {Rauth}, {Reimer}, {Reimer},
  {Remy}, {Renaud}, {Rieger}, {Rinchiuso}, {Romoli}, {Rowell}, {Rudak},
  {Ruiz-Velasco}, {Sahakian}, {Saito}, {Sanchez}, {Santangelo}, {Sasaki},
  {Schlickeiser}, {Sch{\"u}ssler}, {Schulz}, {Schutte}, {Schwanke},
  {Schwemmer}, {Seglar-Arroyo}, {Senniappan}, {Seyffert}, {Shafi},
  {Shiningayamwe}, {Simoni}, {Sinha}, {Sol}, {Specovius}, {Spir-Jacob},
  {Stawarz}, {Steenkamp}, {Stegmann}, {Steppa}, {Takahashi}, {Tavernier},
  {Taylor}, {Terrier}, {Tiziani}, {Tluczykont}, {Trichard}, {Tsirou}, {Tsuji},
  {Tuffs}, {Uchiyama}, {van der Walt}, {van Eldik}, {van Rensburg}, {van
  Soelen}, {Vasileiadis}, {Veh}, {Venter}, {Vincent}, {Vink}, {Voisin},
  {V{\"o}lk}, {Vuillaume}, {Wadiasingh}, {Wagner}, {White}, {Wierzcholska},
  {Yang}, {Yoneda}, {Zacharias}, {Zanin}, {Zdziarski}, {Zech}, {Ziegler},
  {Zorn}, {{\.Z}ywucka}, \& {Smith}}]{2020AA...633A.162H}
{H.~E.~S.~S. Collaboration}, {Abdalla}, H., {Adam}, R., {et~al.} 2020, \aap,
  633, A162, \dodoi{10.1051/0004-6361/201935906}

\bibitem[{{Harris} {et~al.}(2020){Harris}, {Millman}, {van der Walt},
  {Gommers}, {Virtanen}, {Cournapeau}, {Wieser}, {Taylor}, {Berg}, {Smith},
  {Kern}, {Picus}, {Hoyer}, {van Kerkwijk}, {Brett}, {Haldane}, {del R{\'\i}o},
  {Wiebe}, {Peterson}, {G{\'e}rard-Marchant}, {Sheppard}, {Reddy}, {Weckesser},
  {Abbasi}, {Gohlke}, \& {Oliphant}}]{2020Natur.585..357H}
{Harris}, C.~R., {Millman}, K.~J., {van der Walt}, S.~J., {et~al.} 2020, \nat,
  585, 357, \dodoi{10.1038/s41586-020-2649-2}

\bibitem[{{Hauser} \& {Dwek}(2001)}]{2001ARA&A..39..249H}
{Hauser}, M.~G., \& {Dwek}, E. 2001, \araa, 39, 249,
  \dodoi{10.1146/annurev.astro.39.1.249}

\bibitem[{{Heitler}(1954)}]{1954qtr..book.....H}
{Heitler}, W. 1954, {Quantum theory of radiation}

\bibitem[{{HESS Collaboration} {et~al.}(2013){HESS Collaboration},
  {Abramowski}, {Acero}, {Aharonian}, {Akhperjanian}, {Ang{\"u}ner}, {Anton},
  {Balenderan}, {Balzer}, {Barnacka}, {Becherini}, {Becker Tjus},
  {Bernl{\"o}hr}, {Birsin}, {Bissaldi}, {Biteau}, {Boisson}, {Bolmont},
  {Bordas}, {Brucker}, {Brun}, {Brun}, {Bulik}, {Carrigan}, {Casanova},
  {Cerruti}, {Chadwick}, {Chalme-Calvet}, {Chaves}, {Cheesebrough},
  {Chr{\'e}tien}, {Colafrancesco}, {Cologna}, {Conrad}, {Couturier}, {Dalton},
  {Daniel}, {Davids}, {Degrange}, {Deil}, {deWilt}, {Dickinson},
  {Djannati-Ata{\"\i}}, {Domainko}, {Drury}, {Dubus}, {Dutson}, {Dyks},
  {Dyrda}, {Edwards}, {Egberts}, {Eger}, {Espigat}, {Farnier}, {Fegan},
  {Feinstein}, {Fernandes}, {Fernandez}, {Fiasson}, {Fontaine}, {F{\"o}rster},
  {F{\"u}{\ss}ling}, {Gajdus}, {Gallant}, {Garrigoux}, {Gast}, {Giebels},
  {Glicenstein}, {G{\"o}ring}, {Grondin}, {Grudzi{\'n}ska}, {H{\"a}ffner},
  {Hague}, {Hahn}, {Harris}, {Heinzelmann}, {Henri}, {Hermann}, {Hervet},
  {Hillert}, {Hinton}, {Hofmann}, {Hofverberg}, {Holler}, {Horns},
  {Jacholkowska}, {Jahn}, {Jamrozy}, {Janiak}, {Jankowsky}, {Jung},
  {Kastendieck}, {Katarzy{\'n}ski}, {Katz}, {Kaufmann}, {Kh{\'e}lifi},
  {Kieffer}, {Klepser}, {Klochkov}, {Klu{\'z}niak}, {Kneiske}, {Kolitzus},
  {Komin}, {Kosack}, {Krakau}, {Krayzel}, {Kr{\"u}ger}, {Laffon}, {Lamanna},
  {Lefaucheur}, {Lemoine-Goumard}, {Lenain}, {Lennarz}, {Lohse}, {Lopatin},
  {Lu}, {Marandon}, {Marcowith}, {Maurin}, {Maxted}, {Mayer}, {McComb},
  {Medina}, {M{\'e}hault}, {Menzler}, {Meyer}, {Moderski}, {Mohamed}, {Moulin},
  {Murach}, {Naumann}, {de Naurois}, {Nedbal}, {Niemiec}, {Nolan}, {Oakes},
  {Ohm}, {de O{\~n}a Wilhelmi}, {Opitz}, {Ostrowski}, {Oya}, {Panter},
  {Parsons}, {Paz Arribas}, {Pekeur}, {Pelletier}, {Perez}, {Petrucci},
  {Peyaud}, {Pita}, {Poon}, {P{\"u}hlhofer}, {Punch}, {Quirrenbach}, {Raab},
  {Raue}, {Reimer}, {Reimer}, {Renaud}, {de los Reyes}, {Rieger}, {Rob},
  {Rosier-Lees}, {Rowell}, {Rudak}, {Rulten}, {Sahakian}, {Sanchez},
  {Santangelo}, {Schlickeiser}, {Sch{\"u}ssler}, {Schulz}, {Schwanke},
  {Schwarzburg}, {Schwemmer}, {Sol}, {Spengler}, {Spie{\ss}}, {Stawarz},
  {Steenkamp}, {Stegmann}, {Stinzing}, {Stycz}, {Sushch}, {Szostek},
  {Tavernet}, {Terrier}, {Tluczykont}, {Trichard}, {Valerius}, {van Eldik},
  {Vasileiadis}, {Venter}, {Viana}, {Vincent}, {V{\"o}lk}, {Volpe}, {Vorster},
  {Wagner}, {Wagner}, {Ward}, {Weidinger}, {White}, {Wierzcholska}, {Willmann},
  {W{\"o}rnlein}, {Wouters}, {Zacharias}, {Zajczyk}, {Zdziarski}, {Zech},
  {Zechlin}, {Perkins}, {Ojha}, {Stevens}, {Edwards}, \&
  {Kadler}}]{2013MNRAS.434.1889H}
{HESS Collaboration}, {Abramowski}, A., {Acero}, F., {et~al.} 2013, \mnras,
  434, 1889, \dodoi{10.1093/mnras/stt1081}

\bibitem[{{Hooper} \& {Serpico}(2007)}]{2007PhRvL..99w1102H}
{Hooper}, D., \& {Serpico}, P.~D. 2007, \prl, 99, 231102,
  \dodoi{10.1103/PhysRevLett.99.231102}

\bibitem[{{Horns} {et~al.}(2012){Horns}, {Maccione}, {Meyer}, {Mirizzi},
  {Montanino}, \& {Roncadelli}}]{2012PhRvD..86g5024H}
{Horns}, D., {Maccione}, L., {Meyer}, M., {et~al.} 2012, \prd, 86, 075024,
  \dodoi{10.1103/PhysRevD.86.075024}

\bibitem[{{Hoyle}(1969)}]{1969Natur.223..936H}
{Hoyle}, F. 1969, \nat, 223, 936, \dodoi{10.1038/223936a0}

\bibitem[{{Kneiske} {et~al.}(2004){Kneiske}, {Bretz}, {Mannheim}, \&
  {Hartmann}}]{2004A&A...413..807K}
{Kneiske}, T.~M., {Bretz}, T., {Mannheim}, K., \& {Hartmann}, D.~H. 2004, \aap,
  413, 807, \dodoi{10.1051/0004-6361:20031542}

\bibitem[{{Kronberg}(1994)}]{1994RPPh...57..325K}
{Kronberg}, P.~P. 1994, Reports on Progress in Physics, 57, 325,
  \dodoi{10.1088/0034-4885/57/4/001}

\bibitem[{Kronberg {et~al.}(1999)Kronberg, Lesch, \& Hopp}]{Kronberg_1999}
Kronberg, P.~P., Lesch, H., \& Hopp, U. 1999, The Astrophysical Journal, 511,
  56, \dodoi{10.1086/306662}

\bibitem[{{MAGIC Collaboration} {et~al.}(2008){MAGIC Collaboration}, {Albert},
  {Aliu}, {Anderhub}, {Antonelli}, {Antoranz}, {Backes}, {Baixeras}, {Barrio},
  {Bartko}, {Bastieri}, {Becker}, {Bednarek}, {Berger}, {Bernardini},
  {Bigongiari}, {Biland}, {Bock}, {Bonnoli}, {Bordas}, {Bosch-Ramon}, {Bretz},
  {Britvitch}, {Camara}, {Carmona}, {Chilingarian}, {Commichau}, {Contreras},
  {Cortina}, {Costado}, {Covino}, {Curtef}, {Dazzi}, {De Angelis}, {de Cea del
  Pozo}, {de los Reyes}, {De Lotto}, {De Maria}, {De Sabata}, {Delgado Mendez},
  {Dominguez}, {Dorner}, {Doro}, {Errando}, {Fagiolini}, {Ferenc},
  {Fern{\'a}ndez}, {Firpo}, {Fonseca}, {Font}, {Galante}, {Garc{\'\i}a
  L{\'o}pez}, {Garczarczyk}, {Gaug}, {Goebel}, {Hayashida}, {Herrero},
  {H{\"o}hne}, {Hose}, {Hsu}, {Huber}, {Jogler}, {Kneiske}, {Kranich}, {La
  Barbera}, {Laille}, {Leonardo}, {Lindfors}, {Lombardi}, {Longo}, {L{\'o}pez},
  {Lorenz}, {Majumdar}, {Maneva}, {Mankuzhiyil}, {Mannheim}, {Maraschi},
  {Mariotti}, {Mart{\'\i}nez}, {Mazin}, {Meucci}, {Meyer}, {Miranda},
  {Mirzoyan}, {Mizobuchi}, {Moles}, {Moralejo}, {Nieto}, {Nilsson}, {Ninkovic},
  {Otte}, {Oya}, {Panniello}, {Paoletti}, {Paredes}, {Pasanen}, {Pascoli},
  {Pauss}, {Pegna}, {Perez-Torres}, {Persic}, {Peruzzo}, {Piccioli}, {Prada},
  {Prandini}, {Puchades}, {Raymers}, {Rhode}, {Rib{\'o}}, {Rico}, {Rissi},
  {Robert}, {R{\"u}gamer}, {Saggion}, {Saito}, {Salvati}, {Sanchez-Conde},
  {Sartori}, {Satalecka}, {Scalzotto}, {Scapin}, {Schmitt}, {Schweizer},
  {Shayduk}, {Shinozaki}, {Shore}, {Sidro}, {Sierpowska-Bartosik},
  {Sillanp{\"a}{\"a}}, {Sobczynska}, {Spanier}, {Stamerra}, {Stark}, {Takalo},
  {Tavecchio}, {Temnikov}, {Tescaro}, {Teshima}, {Tluczykont}, {Torres},
  {Turini}, {Vankov}, {Venturini}, {Vitale}, {Wagner}, {Wittek}, {Zabalza},
  {Zandanel}, {Zanin}, \& {Zapatero}}]{2008Sci...320.1752M}
{MAGIC Collaboration}, {Albert}, J., {Aliu}, E., {et~al.} 2008, Science, 320,
  1752, \dodoi{10.1126/science.1157087}

\bibitem[{{MAGIC Collaboration} {et~al.}(2018{\natexlab{a}}){MAGIC
  Collaboration}, {Ahnen}, {Ansoldi}, {Antonelli}, {Arcaro}, {Baack},
  {Babi{\'c}}, {Banerjee}, {Bangale}, {Barres de Almeida}, {Barrio},
  {Bednarek}, {Bernardini}, {Berse}, {Berti}, {Bhattacharyya}, {Biland},
  {Blanch}, {Bonnoli}, {Carosi}, {Carosi}, {Ceribella}, {Chatterjee}, {Colak},
  {Colin}, {Colombo}, {Contreras}, {Cortina}, {Covino}, {Cumani}, {da Vela},
  {Dazzi}, {de Angelis}, {de Lotto}, {Delfino}, {Delgado}, {di Pierro},
  {Dom{\'\i}nguez}, {Dominis Prester}, {Dorner}, {Doro}, {Einecke},
  {Elsaesser}, {Fallah Ramazani}, {Fern{\'a}ndez-Barral}, {Fidalgo}, {Fonseca},
  {Font}, {Fruck}, {Galindo}, {Garc{\'\i}a L{\'o}pez}, {Garczarczyk}, {Gaug},
  {Giammaria}, {Godinovi{\'c}}, {Gora}, {Guberman}, {Hadasch}, {Hahn},
  {Hassan}, {Hayashida}, {Herrera}, {Hose}, {Hrupec}, {Ishio}, {Konno}, {Kubo},
  {Kushida}, {Kuve{\v{z}}di{\'c}}, {Lelas}, {Lindfors}, {Lombardi}, {Longo},
  {L{\'o}pez}, {Maggio}, {Majumdar}, {Makariev}, {Maneva}, {Manganaro},
  {Mannheim}, {Maraschi}, {Mariotti}, {Mart{\'\i}nez}, {Masuda}, {Mazin},
  {Mielke}, {Minev}, {Miranda}, {Mirzoyan}, {Moralejo}, {Moreno}, {Moretti},
  {Nagayoshi}, {Neustroev}, {Niedzwiecki}, {Nievas Rosillo}, {Nigro},
  {Nilsson}, {Ninci}, {Nishijima}, {Noda}, {Nogu{\'e}s}, {Paiano}, {Palacio},
  {Paneque}, {Paoletti}, {Paredes}, {Pedaletti}, {Peresano}, {Persic}, {Prada
  Moroni}, {Prandini}, {Puljak}, {Garcia}, {Reichardt}, {Rhode}, {Rib{\'o}},
  {Rico}, {Righi}, {Rugliancich}, {Saito}, {Satalecka}, {Schweizer}, {Sitarek},
  {{\v{S}}nidari{\'c}}, {Sobczynska}, {Stamerra}, {Strzys}, {Suri{\'c}},
  {Takahashi}, {Takalo}, {Tavecchio}, {Temnikov}, {Terzi{\'c}}, {Teshima},
  {Torres-Alb{\`a}}, {Treves}, {Tsujimoto}, {Vanzo}, {Vazquez Acosta}, {Vovk},
  {Ward}, {Will}, {Zari{\'c}}, {Becerra Gonz{\'a}lez}, {Tanaka}, {Ojha},
  {Finke}, {L{\"a}hteenm{\"a}ki}, {J{\"a}rvel{\"a}}, {Tornikoski},
  {Ramakrishnan}, {Hovatta}, {Jorstad}, {Marscher}, {Larionov}, {Borman},
  {Grishina}, {Kopatskaya}, {Larionova}, {Morozova}, {Savchenko}, {Troitskaya},
  {Troitsky}, {Vasilyev}, {Agudo}, {Molina}, {Casadio}, {Gurwell}, {Carnerero},
  {Protasio}, \& {Acosta Pulido}}]{2018AA...617A..30M}
{MAGIC Collaboration}, {Ahnen}, M.~L., {Ansoldi}, S., {et~al.}
  2018{\natexlab{a}}, \aap, 617, A30, \dodoi{10.1051/0004-6361/201832624}

\bibitem[{{MAGIC Collaboration} {et~al.}(2018{\natexlab{b}}){MAGIC
  Collaboration}, {Acciari}, {Ansoldi}, {Antonelli}, {Arbet Engels}, {Arcaro},
  {Baack}, {Babi{\'c}}, {Banerjee}, {Bangale}, {Barres de Almeida}, {Barrio},
  {Bednarek}, {Bernardini}, {Berti}, {Besenrieder}, {Bhattacharyya},
  {Bigongiari}, {Biland}, {Blanch}, {Bonnoli}, {Carosi}, {Ceribella}, {Cikota},
  {Colak}, {Colin}, {Colombo}, {Contreras}, {Cortina}, {Covino}, {D'Elia}, {da
  Vela}, {Dazzi}, {de Angelis}, {de Lotto}, {Delfino}, {Delgado}, {di Pierro},
  {Do Souto Espi{\~n}era}, {Dom{\'\i}nguez}, {Dominis Prester}, {Dorner},
  {Doro}, {Einecke}, {Elsaesser}, {Fallah Ramazani}, {Fattorini},
  {Fern{\'a}ndez-Barral}, {Ferrara}, {Fidalgo}, {Foffano}, {Fonseca}, {Font},
  {Fruck}, {Galindo}, {Gallozzi}, {Garc{\'\i}a L{\'o}pez}, {Garczarczyk},
  {Gaug}, {Giammaria}, {Godinovi{\'c}}, {Guberman}, {Hadasch}, {Hahn},
  {Hassan}, {Herrera}, {Hoang}, {Hrupec}, {Inoue}, {Ishio}, {Iwamura}, {Kubo},
  {Kushida}, {Kuve{\v{z}}di{\'c}}, {Lamastra}, {Lelas}, {Leone}, {Lindfors},
  {Lombardi}, {Longo}, {L{\'o}pez}, {L{\'o}pez-Oramas}, {Maggio}, {Majumdar},
  {Makariev}, {Maneva}, {Manganaro}, {Mannheim}, {Maraschi}, {Mariotti},
  {Mart{\'\i}nez}, {Masuda}, {Mazin}, {Minev}, {Miranda}, {Mirzoyan}, {Molina},
  {Moralejo}, {Moreno}, {Moretti}, {Munar-Adrover}, {Neustroev}, {Niedzwiecki},
  {Nievas Rosillo}, {Nigro}, {Nilsson}, {Ninci}, {Nishijima}, {Noda},
  {Nogu{\'e}s}, {Paiano}, {Palacio}, {Paneque}, {Paoletti}, {Paredes},
  {Pedaletti}, {Pe{\~n}il}, {Peresano}, {Persic}, {Prada Moroni}, {Prandini},
  {Puljak}, {Garcia}, {Rhode}, {Rib{\'o}}, {Rico}, {Righi}, {Rugliancich},
  {Saha}, {Saito}, {Satalecka}, {Schweizer}, {Sitarek}, {{\v{S}}nidari{\'c}},
  {Sobczynska}, {Somero}, {Stamerra}, {Strzys}, {Suri{\'c}}, {Tavecchio},
  {Temnikov}, {Terzi{\'c}}, {Teshima}, {Torres-Alb{\`a}}, {Tsujimoto}, {van
  Scherpenberg}, {Vanzo}, {Vazquez Acosta}, {Vovk}, {Ward}, {Will},
  {Zari{\'c}}, {Fermi-Lat Collaboration}, {Becerra Gonz{\'a}lez}, {Raiteri},
  {Sandrinelli}, {Hovatta}, {Kiehlmann}, {Max-Moerbeck}, {Tornikoski},
  {L{\"a}hteenm{\"a}ki}, {Tammi}, {Ramakrishnan}, {Thum}, {Agudo}, {Molina},
  {G{\'o}mez}, {Fuentes}, {Casadio}, {Traianou}, {Myserlis}, \&
  {Kim}}]{2018AA...619A.159M}
{MAGIC Collaboration}, {Acciari}, V.~A., {Ansoldi}, S., {et~al.}
  2018{\natexlab{b}}, \aap, 619, A159, \dodoi{10.1051/0004-6361/201833618}

\bibitem[{{MAGIC Collaboration} {et~al.}(2019){MAGIC Collaboration}, {Acciari},
  {Ansoldi}, {Antonelli}, {Arbet Engels}, {Baack}, {Babi{\'c}}, {}, {Banerjee},
  {Barres de Almeida}, {Barrio}, {Becerra Gonz{\'a}lez}, {Bednarek},
  {Bellizzi}, {Bernardini}, {Berti}, {Besenrieder}, {Bhattacharyya},
  {Bigongiari}, {Biland}, {Blanch}, {Bonnoli}, {Bo{\v{s}}njak}, {Busetto},
  {Carosi}, {Ceribella}, {Cerruti}, {Chai}, {Chilingaryan}, {Cikota}, {Colak},
  {Colin}, {Colombo}, {Contreras}, {Cortina}, {Covino}, {D'Elia}, {da Vela},
  {Dazzi}, {de Angelis}, {de Lotto}, {Delfino}, {Delgado}, {Depaoli}, {di
  Pierro}, {di Venere}, {Do Souto Espi{\~n}eira}, {Dominis Prester}, {Donini},
  {Dorner}, {Doro}, {Elsaesser}, {Fallah Ramazani}, {Fattorini}, {Ferrara},
  {Fidalgo}, {Foffano}, {Fonseca}, {Font}, {Fruck}, {Fukami}, {Garc{\'\i}a
  L{\'o}pez}, {Garczarczyk}, {Gasparyan}, {Gaug}, {Giglietto}, {Giordano},
  {Godinovi{\'c}}, {}, {Green}, {Guberman}, {Hadasch}, {Hahn}, {Herrera},
  {Hoang}, {Hrupec}, {H{\"u}tten}, {Inada}, {Inoue}, {Ishio}, {Iwamura},
  {Jouvin}, {Kerszberg}, {Kubo}, {Kushida}, {Lamastra}, {Lelas}, {Leone},
  {Lindfors}, {Lombardi}, {Longo}, {L{\'o}pez}, {L{\'o}pez-Coto},
  {L{\'o}pez-Oramas}, {Loporchio}, {Machado de Oliveira Fraga}, {Maggio},
  {Majumdar}, {Makariev}, {Mallamaci}, {Maneva}, {Manganaro}, {Mannheim},
  {Maraschi}, {Mariotti}, {Mart{\'\i}nez}, {Mazin}, {Mi{\'c}}, {Anovi{\'c}},
  {}, {Miceli}, {Minev}, {Miranda}, {Mirzoyan}, {Molina}, {Moralejo},
  {Morcuende}, {Moreno}, {Moretti}, {Munar-Adrover}, {Neustroev}, {Nigro},
  {Nilsson}, {Ninci}, {Nishijima}, {Noda}, {Nogu{\'e}s}, {Nozaki}, {Paiano},
  {Palacio}, {Palatiello}, {Paneque}, {Paoletti}, {Paredes}, {Pe{\~n}il},
  {Peresano}, {Persic}, {Prada Moroni}, {Prandini}, {Puljak}, {Rhode},
  {Rib{\'o}}, {Rico}, {Righi}, {Rugliancich}, {Saha}, {Sahakyan}, {Saito},
  {Sakurai}, {Satalecka}, {Schmidt}, {Schweizer}, {Sitarek},
  {{\v{S}}nidari{\'c}}, {}, {Sobczynska}, {Somero}, {Stamerra}, {Strom},
  {Strzys}, {Suda}, {Suri{\'c}}, {}, {Takahashi}, {Tavecchio}, {Temnikov},
  {Terzi{\'c}}, {}, {Teshima}, {Torres-Alb{\`a}}, {Tosti}, {Vagelli}, {van
  Scherpenberg}, {Vanzo}, {Vazquez Acosta}, {Vigorito}, {Vitale}, {Vovk},
  {Will}, {Zari{\'c}}, {}, {Asano}, {D'Ammando}, \&
  {Clavero}}]{2019MNRAS.490.2284M}
---. 2019, \mnras, 490, 2284, \dodoi{10.1093/mnras/stz2725}

\bibitem[{{Malik} {et~al.}(2022){Malik}, {Sahayanathan}, {Shah}, {Iqbal},
  {Manzoor}, \& {Bhatt}}]{2022MNRAS.511..994M}
{Malik}, Z., {Sahayanathan}, S., {Shah}, Z., {et~al.} 2022, \mnras, 511, 994,
  \dodoi{10.1093/mnras/stab3173}

\bibitem[{Masaki {et~al.}(2017)Masaki, Aoki, \& Soda}]{PhysRevD.96.043519}
Masaki, E., Aoki, A., \& Soda, J. 2017, Phys. Rev. D, 96, 043519,
  \dodoi{10.1103/PhysRevD.96.043519}

\bibitem[{{Meyer} {et~al.}(2022){Meyer}, {Davies}, \&
  {Kuhlmann}}]{2022icrc.confE.557M}
{Meyer}, M., {Davies}, J., \& {Kuhlmann}, J. 2022, in 37th International Cosmic
  Ray Conference, 557, \dodoi{10.22323/1.395.0557}

\bibitem[{{Mirizzi} {et~al.}(2008){Mirizzi}, {Raffelt1}, \&
  {Serpico}}]{2008LNP...741..115M}
{Mirizzi}, A., {Raffelt1}, G.~G., \& {Serpico}, P.~D. 2008, in Axions, ed.
  M.~{Kuster}, G.~{Raffelt}, \& B.~{Beltr{\'a}n}, Vol. 741, 115

\bibitem[{{O'Brien}(2017)}]{2017arXiv170802160O}
{O'Brien}, S. 2017, arXiv e-prints, arXiv:1708.02160.
\newblock \doarXiv{1708.02160}

\bibitem[{{Petry} {et~al.}(2002){Petry}, {Bond}, {Bradbury}, {Buckley},
  {Carter-Lewis}, {Cui}, {Duke}, {de la Calle Perez}, {Falcone}, {Fegan},
  {Fegan}, {Finley}, {Gaidos}, {Gibbs}, {Gammell}, {Hall}, {Hall}, {Hillas},
  {Holder}, {Horan}, {Jordan}, {Kertzman}, {Kieda}, {Kildea}, {Knapp},
  {Kosack}, {Krennrich}, {LeBohec}, {Moriarty}, {M{\"u}ller}, {Nagai}, {Ong},
  {Page}, {Pallassini}, {Power-Mooney}, {Quinn}, {Reay}, {Reynolds}, {Rose},
  {Schroedter}, {Sembroski}, {Sidwell}, {Stanton}, {Swordy}, {Vassiliev},
  {Wakely}, {Walker}, \& {Weekes}}]{2002ApJ...580..104P}
{Petry}, D., {Bond}, I.~H., {Bradbury}, S.~M., {et~al.} 2002, \apj, 580, 104,
  \dodoi{10.1086/343102}

\bibitem[{{Protheroe} \& {Meyer}(2000)}]{2000PhLB..493....1P}
{Protheroe}, R.~J., \& {Meyer}, H. 2000, Physics Letters B, 493, 1,
  \dodoi{10.1016/S0370-2693(00)01113-8}

\bibitem[{Pshirkov {et~al.}(2016)Pshirkov, Tinyakov, \&
  Urban}]{PhysRevLett.116.191302}
Pshirkov, M.~S., Tinyakov, P.~G., \& Urban, F.~R. 2016, Phys. Rev. Lett., 116,
  191302, \dodoi{10.1103/PhysRevLett.116.191302}

\bibitem[{{Raffelt} \& {Stodolsky}(1988)}]{1988PhRvD..37.1237R}
{Raffelt}, G., \& {Stodolsky}, L. 1988, \prd, 37, 1237,
  \dodoi{10.1103/PhysRevD.37.1237}

\bibitem[{{Raue} \& {Mazin}(2008)}]{2008IJMPD..17.1515R}
{Raue}, M., \& {Mazin}, D. 2008, International Journal of Modern Physics D, 17,
  1515, \dodoi{10.1142/S0218271808013091}

\bibitem[{{Rees}(1968)}]{1968Natur.219..127R}
{Rees}, M.~J. 1968, \nat, 219, 127, \dodoi{10.1038/219127a0}

\bibitem[{{Rees} {et~al.}(1968){Rees}, {Sciama}, \&
  {Setti}}]{1968Natur.217..326R}
{Rees}, M.~J., {Sciama}, D.~W., \& {Setti}, G. 1968, \nat, 217, 326,
  \dodoi{10.1038/217326a0}

\bibitem[{Roncadelli {et~al.}(2009)Roncadelli, Angelis, \&
  Mansutti}]{2009Evidence}
Roncadelli, M., Angelis, A.~D., \& Mansutti, O. 2009, American Institute of
  Physics

\bibitem[{Rubinstein(2001)}]{2001Magnetic}
Rubinstein, G. 2001, Physics Reports

\bibitem[{{Sanchez} {et~al.}(2013){Sanchez}, {Fegan}, \&
  {Giebels}}]{2013A&A...554A..75S}
{Sanchez}, D.~A., {Fegan}, S., \& {Giebels}, B. 2013, \aap, 554, A75,
  \dodoi{10.1051/0004-6361/201220631}

\bibitem[{{Simet} {et~al.}(2008){Simet}, {Hooper}, \&
  {Serpico}}]{2008PhRvD..77f3001S}
{Simet}, M., {Hooper}, D., \& {Serpico}, P.~D. 2008, \prd, 77, 063001,
  \dodoi{10.1103/PhysRevD.77.063001}

\bibitem[{{Singh} \& {Meintjes}(2020)}]{2020JAsGe...9..309S}
{Singh}, K.~K., \& {Meintjes}, P.~J. 2020, NRIAG Journal of Astronomy and
  Geophysics, 9, 309, \dodoi{10.1080/20909977.2020.1743468}

\bibitem[{{Sinha} {et~al.}(2014){Sinha}, {Sahayanathan}, {Misra}, {Godambe}, \&
  {Acharya}}]{2014ApJ...795...91S}
{Sinha}, A., {Sahayanathan}, S., {Misra}, R., {Godambe}, S., \& {Acharya},
  B.~S. 2014, \apj, 795, 91, \dodoi{10.1088/0004-637X/795/1/91}

\bibitem[{{Stecker} {et~al.}(2007){Stecker}, {Baring}, \&
  {Summerlin}}]{2007ApJ...667L..29S}
{Stecker}, F.~W., {Baring}, M.~G., \& {Summerlin}, E.~J. 2007, \apjl, 667, L29,
  \dodoi{10.1086/522005}

\bibitem[{{Stecker} \& {de Jager}(1998)}]{1998A&A...334L..85S}
{Stecker}, F.~W., \& {de Jager}, O.~C. 1998, \aap, 334, L85.
\newblock \doarXiv{astro-ph/9804196}

\bibitem[{Tavecchio \& Mazin(2009)}]{2009Intrinsic}
Tavecchio, F., \& Mazin, D. 2009, Blackwell Publishing Ltd, 392, 0

\bibitem[{{Urry} \& {Padovani}(1995)}]{1995PASP..107..803U}
{Urry}, C.~M., \& {Padovani}, P. 1995, \pasp, 107, 803, \dodoi{10.1086/133630}

\bibitem[{{Virtanen} {et~al.}(2020){Virtanen}, {Gommers}, {Oliphant},
  {Haberland}, {Reddy}, {Cournapeau}, {Burovski}, {Peterson}, {Weckesser},
  {Bright}, {van der Walt}, {Brett}, {Wilson}, {Millman}, {Mayorov}, {Nelson},
  {Jones}, {Kern}, {Larson}, {Carey}, {Polat}, {Feng}, {Moore}, {VanderPlas},
  {Laxalde}, {Perktold}, {Cimrman}, {Henriksen}, {Quintero}, {Harris},
  {Archibald}, {Ribeiro}, {Pedregosa}, {van Mulbregt}, \& {SciPy 1. 0
  Contributors}}]{2020NatMe..17..261V}
{Virtanen}, P., {Gommers}, R., {Oliphant}, T.~E., {et~al.} 2020, Nature
  Methods, 17, 261, \dodoi{10.1038/s41592-019-0686-2}

\bibitem[{{Wakely} \& {Horan}(2008)}]{2008ICRC....3.1341W}
{Wakely}, S.~P., \& {Horan}, D. 2008, in International Cosmic Ray Conference,
  Vol.~3, International Cosmic Ray Conference, 1341--1344

\bibitem[{{Wouters} \& {Brun}(2014)}]{2014JCAP...01..016W}
{Wouters}, D., \& {Brun}, P. 2014, \jcap, 2014, 016,
  \dodoi{10.1088/1475-7516/2014/01/016}

\bibitem[{{Zheng} \& {Kang}(2013)}]{2013ApJ...764..113Z}
{Zheng}, Y.~G., \& {Kang}, T. 2013, \apj, 764, 113,
  \dodoi{10.1088/0004-637X/764/2/113}

\end{thebibliography}
\bibliographystyle{aasjournal}
\end{document}